\documentclass{article}

\usepackage[final, nonatbib]{neurips_2022}

\usepackage[utf8]{inputenc} 
\usepackage[T1]{fontenc}    
\usepackage{hyperref}       
\usepackage{url}            
\usepackage{booktabs}       
\usepackage{amsfonts}       
\usepackage{nicefrac}       
\usepackage{microtype}      
\usepackage{xcolor}         

\usepackage{caption}

\usepackage{subfig}
\usepackage{amsmath,amsfonts}
\usepackage{algorithmic}
\usepackage{algorithm}
\usepackage{array}
\usepackage{textcomp}
\usepackage{stfloats}
\usepackage{url}
\usepackage{verbatim}
\usepackage{graphicx}
\usepackage{multirow}
\usepackage{bm}
\newcommand{\ie}{{\emph{i.e.}}}
\newcommand{\eg}{{\emph{e.g.}}}
\newcommand{\etal}{{\emph{et al.}}}
\makeatletter
\newcommand*{\rom}[1]{\expandafter\@slowromancap\romannumeral #1@}
\makeatother

\title{Hiding Images in Deep Probabilistic Models}

\author{%
    Haoyu Chen\\
    Department of Computer Science\\
    City University of Hong Kong\\
    \texttt{haoychen3-c@my.cityu.edu.hk} \\
    \And
    Linqi Song \\
    Department of Computer Science\\
    City University of Hong Kong \\
    \texttt{linqi.song@cityu.edu.hk} \\
    \AND
    Zhenxing Qian \\
    School of Computer Science \\
    Fudan University \\
    \texttt{zxqian@fudan.edu.cn} \\
    \And
    Xinpeng Zhang \\
    School of Computer Science \\
    Fudan University \\
    \texttt{zhangxinpeng@fudan.edu.cn} \\
    \And
    Kede Ma\thanks{Corresponding author.} \\
    Department of Computer Science \\
    City University of Hong Kong\\
    \texttt{kede.ma@cityu.edu.hk} \\
}

\hypersetup{colorlinks = true,
            allcolors  = black, 
            citecolor  = blue,
            linkcolor = red,
            }

\begin{document}

\maketitle

\begin{abstract}
Data hiding with deep neural networks (DNNs) has experienced impressive successes in recent years. A prevailing scheme is to train an autoencoder, consisting of an \textit{encoding} network to embed (or transform) secret messages in (or into) a carrier, and a \textit{decoding} network to extract the hidden messages. This scheme may 
suffer from several limitations regarding practicability, security, and embedding capacity.
In this work, we describe a different computational framework to hide images in deep probabilistic models. Specifically, we use a DNN to model the probability density of cover images, and hide a secret image in one particular location of the learned distribution. As an instantiation, we adopt a SinGAN, a pyramid of generative adversarial networks (GANs), to learn the patch distribution of one cover image. We hide the secret image by fitting a deterministic mapping from a fixed set of noise maps (generated by an embedding key) to the secret image during patch distribution learning. The stego SinGAN, behaving 
as the original SinGAN, is publicly communicated; only the receiver with the embedding key is able to extract the secret image. We demonstrate the feasibility of our SinGAN approach in terms of extraction accuracy and model security. Moreover, we show the flexibility of the proposed method in terms of hiding multiple images for different receivers and obfuscating the secret image.

\end{abstract}

\section{Introduction}
Data hiding generally refers to the process of hiding a form of secret message in another form of cover media, while minimizing the introduced distortions to the cover media \cite{cheddad2010digital, provos2003hide}. For human eavesdroppers, the measured distortion should be consistent with human judgments, penalizing errors that are most perceptually or cognitively noticeable \cite{wang2004image,adi2018turning}; for machine eavesdroppers, the distortion should be ``invisible'' in a way that bypasses digital steganalysis tools such as StegExpose \cite{boehm2014stegexpose} and more recent deep learning-based ones \cite{boroumand2018deep}. Only the informed receiver typically with a shared embedding key (through a secure subliminal channel \cite{simmons1984prisoners}) is able to extract the secret message. The form of the secret 
message can be encrypted bit streams \cite{cox2007digital}, texts \cite{jassim2013novel}, audio signals \cite{li2000transparent}, images \cite{baluja2017}, and videos\cite{swanson1997data}. Similarly, the cover media can also be texts \cite{bennett2004linguistic}, audio signals \cite{jayaram2011information}, images \cite{baluja2017}, videos \cite{swanson1997data}, neural networks \cite{adi2018turning,wang2021data}, and even human behaviours\cite{behave}. 

\begin{figure}[!t]
	\centering
	\subfloat[Secret-in-image hiding. The preparation network is optional.]{%
	    \centering
		\includegraphics[width=\textwidth]{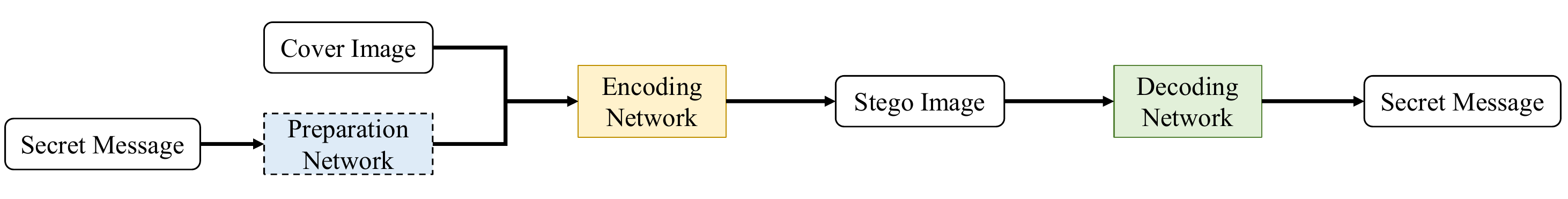}
	}
	
	\subfloat[Secret-in-network hiding. The secret message is usually in the form of a watermark.]{%
	    \centering
		\includegraphics[width=\textwidth]{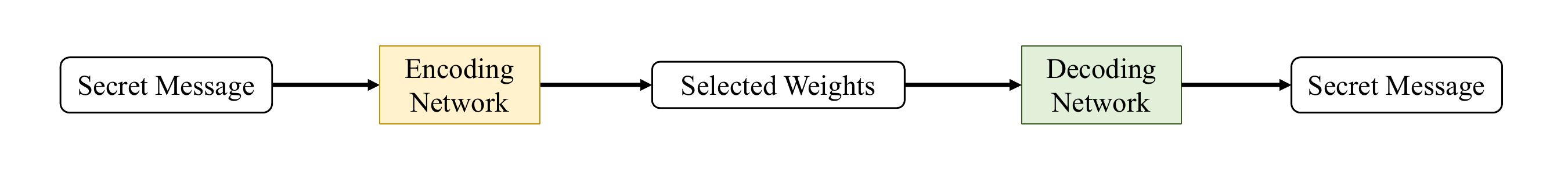}
	}
	
	\subfloat[Constructive (or generative) image hiding.]{%
	    \centering
		\includegraphics[width=\textwidth]{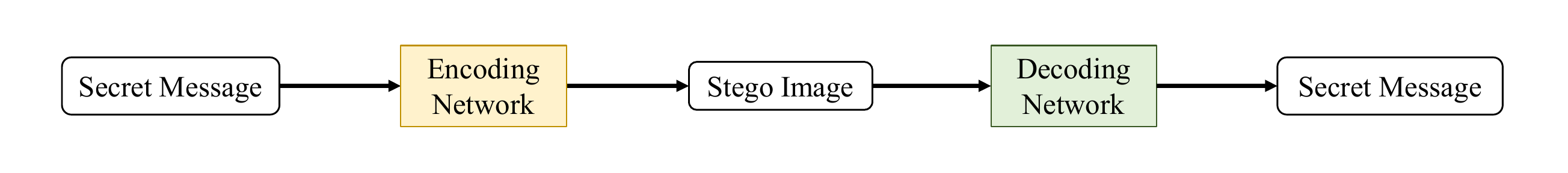}
	}

	\subfloat[Proposed framework for hiding images in deep probabilistic models.]{%
	    \centering
		\includegraphics[width=\textwidth]{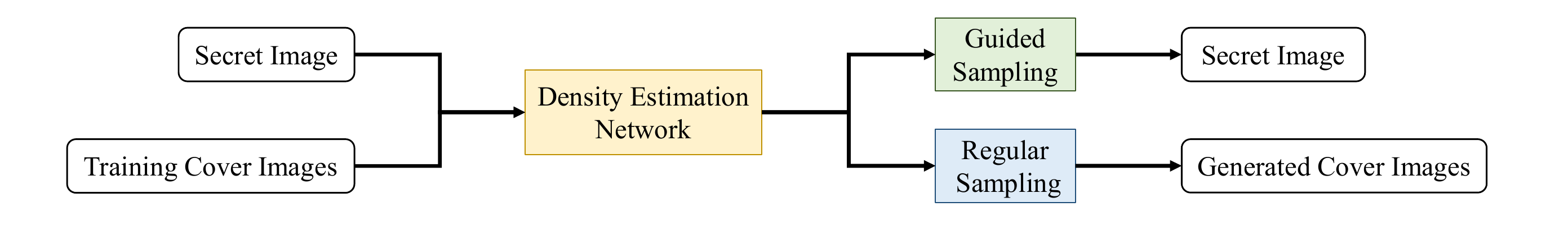}
	}
	\caption{Paradigms for hiding data using DNNs.}\label{fig:paradigm}
\end{figure}

Like many problems in signal and image processing,  data hiding has been revolutionized by the remarkable development of DNNs \cite{baluja2017,zhu2018,hu2018}. These methods typically follow an autoencoder approach with two key components: an \textit{encoding} network
and a \textit{decoding} network. For \textit{secret-in-image} hiding \cite{baluja2017,zhu2018}, the encoding network takes the cover image and the secret message as inputs, and generates a stego image with the hidden message (see Fig. \ref{fig:paradigm} (a)). For \textit{secret-in-network} hiding \cite{adi2018turning,wang2021data}, the cover media becomes some pre-selected weight layers of the encoding network, and the secret message usually serves as a watermark (see Fig.~\ref{fig:paradigm} (b)). For \textit{constructive} (or generative) image hiding \cite{hu2018}, the encoding network directly maps the secret message to the stego image without reliance on 
any cover image (see Fig.~\ref{fig:paradigm} (c)). In all cases, the decoding network is responsible for  extracting the secret message. Despite demonstrated success, the autoencoder scheme may suffer from three main drawbacks. First, the decoding network, whose size may be significantly larger than that of the secret message, must be sent to the receiver side via the subliminal channel\footnote{The decoding network, if shared via a public channel, will be suspectable, as 
it is only trained to extract the secret message rather than performing typical machine learning tasks.}, making the paradigm less practical. Second, it is not hard to re-train existing DNN-based steganalysis methods \cite{boroumand2018deep,wang2019} to identify stego images (or stego weight matrices), making the paradigm less secure. Third, it is difficult to hide multiple images for different receivers via the same encoding and decoding networks, making the paradigm less flexible. 

\noindent\textbf{Our Contributions}. In this paper, we propose to hide images in deep probabilistic models, which is substantially different from the previous autoencoder scheme (see Fig. \ref{fig:paradigm} (d)). The key idea is to use a DNN to model the high-dimensional probability density of training cover images, and hide the secret image in one particular location of the learned distribution. The stego DNN for density estimation is publicly communicated, from which we may draw samples that look like training cover images. Only guided sampling by the embedding key (shared between the sender and receiver) is able to reproduce the secret image. 

We construct a specific example under the proposed probabilistic image hiding framework. Specifically, we adopt a SinGAN \cite{singan}, a pyramid of generative adversarial networks (GANs) \cite{gan}, to implicitly learn the patch distribution of a single cover image. During distribution learning, we use the same SinGAN to fit a deterministic mapping from a fixed set of noise maps (generated by the shared embedding key) to the secret image, which completes the image hiding process. The stego SinGAN that behaves like the original one is publicly communicated. A single forward propagation suffices to extract the secret image by the receiver with the embedding key, and no decoding network is trained and transmitted. Experiments demonstrate that the proposed method is 1) feasible,  extracting the secret image with improved accuracy (compared to the autoencoder-based methods), 2) secure, behaving normally as the original SinGAN in several aspects, and 3) flexible, hiding multiple images for different receivers and obfuscating the secret image with graceful performance degradation.

\section{Related Work}
Depending on the applications, data hiding can be broadly divided into two subtopics: watermarking and steganography. Watermarking~\cite{water} aims to embed a  digital watermark in a multimedia file for copyright protection and content management on social networks. Thus, watermarking techniques focus primarily on the robustness and perceptibility aspects of the embedded watermarks. Steganography \cite{johnson1998exploring,provos2003hide} aims to conceal a secret message within a cover media mainly for covert communication. Thus, steganography pays more attention to the trade-off among embedding capacity, extraction accuracy, and model security. Here we provide a concise review of 
image hiding techniques, and refer  the interested readers to \cite{provos2003hide,cheddad2010digital,zhang2021brief} for more comprehensive surveys of the field. 

\noindent\textbf{Secret-in-Image Hiding}. The most common image steganography modifies the least significant bits (LSBs) of images, either uniformly \cite{chan2004hiding,mielikainen2006lsb} or adaptively \cite{pevny2010using,holub2012designing}, guided by the design of novel distortion functions. Other representative techniques  include pixel value differencing \cite{wu2003steganographic}, histogram shifting \cite{ni2006reversible,zhang2013recursive}, and recursive code construction \cite{zhang2014optimal}. Transform domain steganography   \cite{ruanaidh1996phase,hsu1999hidden,barni2001improved}) have also been proposed with improved capacity and security. Often, these methods leave traces in the form of certain statistical irregularities, which can be easily revealed by simple steganalysis algorithms as countermeasures \cite{fridrich2001detecting,lyu2006steganalysis,holub2013random}. Zhu \etal~\cite{zhu2018} proposed one of the first DNN-based autoencoder for unified watermarking and steganography with an optional noise layer. Baluja \cite{baluja2017} extended it to ``image-in-image'' steganography with a preparation network to preprocess the secret image. Weng \etal~\cite{weng2019high} further extended it to ``video-in-video'' steganography via temporal residual modeling. Normalizing flow-based invertible architectures~\cite{dinh2014nice} have also been investigated for multiple image hiding~\cite{lu2021large,jing2021,DeepMIH}.  Here, we describe a different framework to accomplish a similar but more challenging goal - multiple image hiding for different users - with several other advantages.

\noindent\textbf{Secret-in-Network Hiding}. A prerequisite for secret-in-network hiding is that the hidden message should not affect the network performance on the given machine learning task. As a result, the message (mostly the watermark for intellectual property protection of the neural network) is commonly embedded during network training. Typical strategies for this purpose include parameter regularization \cite{uchida2017embedding},  backdooring \cite{adi2018turning}, output watermarking \cite{wu2020watermarking}, and weight selection \cite{wang2021data}. The proposed framework can be seen as a form of secret-in-network hiding, but with different goals (steganography instead of watermarking).

\noindent\textbf{Constructive (Generative) Image Hiding}. Traditional methods hide secret messages during the construction of some specific types of images, such as textures \cite{wu2015steganography} and fingerprints \cite{li2019}. Recent DNN-based constructive image hiding methods mainly aim to construct the mapping between secret messages and stego images of more unconstrained content types \cite{liu2017coverless,wei2022generative}. The proposed framework can be seen as a form of constructive image hiding, where we hide a secret image during the ``construction'' of a probability density function, but with larger embedding capacity and improved model security.

\section{Hiding Images in Deep Probabilistic Models}
\noindent\textbf{General Framework}. Without loss of generality, we describe the general framework of hiding a single image in deep probabilistic models. The straightforward extension to multiple image hiding is described in Sec. \ref{subsec:fe}. We assume a cover image dataset $\mathcal{D} = \{\bm x^{(1)}, \bm x^{(2)},\ldots, \bm x^{(M)}\}$, where each image is drawn independently from an underlying image distribution $p(\bm x)$. $\mathcal{D}$ may include a single image (where $\bm x^{(i)}$ becomes the $i$-th image patch), images of a specific class (\eg, textures, fingerprints, and faces), and images on the natural image manifold. Also given is a secret image $\bm x^{(s)}$ of arbitrary content and an embedding key $\bm k$ shared by the sender and the receiver, which is  a typical setting in the prisoner's problem formulated by Simmons \cite{simmons1984prisoners}. 

The proposed probabilistic image hiding framework consists of two main steps. \underline{First}, we learn a probability density function $p_s(\bm x)$ over $\mathcal{D}_s = \mathcal{D}\bigcup \{\bm x^{(s)}\}$, \ie, the combination of the cover image dataset and the secret image either explicitly via (approximate) maximum likelihood \cite{larochelle2011neural,dinh2014nice,song2021train,kingma2013auto} or implicitly via likelihood-free inference (\eg, density estimation by comparison) \cite{mohamed2016learning}. The learned probability model $p_s(\bm x)$ is publicly communicated in the proposed framework, which is usually  in the form of a DNN (as the scoring function \cite{gneiting2007strictly} or the sample generating function \cite{mohamed2016learning}). We may as well learn a \textit{reference} probability density function $p_c(\bm x)$ on the cover image dataset $\mathcal{D}$ solely. To guarantee the security of our framework, $p_s(\bm x)$ should be as close to $p_c(\bm x)$ as possible in some statistical distance (\ie, distribution-preserving \cite{cachin1998information}). Provided that the adopted  probability density estimator is robust (to the outlier image $\bm x^{(s)}$), which assigns an infinitesimal probability mass to $\bm x^{(s)}$, the proposed framework is perfectly secure in any statistical distance sense. 
\underline{Second}, we design a guided sampling procedure (with the help of the embedding key $\bm k$) to draw a sample image $\hat{\bm x}^{(s)}$ from $p_s(\bm x)$ that looks identical to the secret image $\bm x^{(s)}$. A third party without the embedding key is only able to generate samples that resemble cover images. It is noteworthy that the proposed framework does not require the learned $p_s(\bm x)$ (or equivalently $p_c(\bm x)$) to be sufficiently close to the underlying $p(\bm x)$, a daunting (if not impossible) task to complete as digital images reside in a very high-dimensional space. In other words, it is perfectly fine that the samples drawn from $p_s(\bm x)$ (without guidance) contain visually noticeable distortions as long as such distortions are shared by the samples drawn from $p_c(\bm x)$.

\noindent\textbf{Specific Example}. Within the general framework of probabilistic image hiding, we provide a specific example that relies on GANs \cite{gan}, a family of \textit{implicit} probability models, which are represented by a stochastic procedure of data sampling. Although more natural to work with in our context compared to \textit{prescribed} probabilistic models \cite{diggle1984monte}, unconditional GANs have the notorious reputation of being difficult to train, especially when modeling image sets with diverse content complexities and rich semantics. Thus, to make our work easily reproducible, we opt for a SinGAN \cite{singan} to learn the internal patch distribution of a single cover image $\bm x^{(c)}_0$ at various scales. 

\begin{figure}[t]
	\centering
	\includegraphics[width=\textwidth]{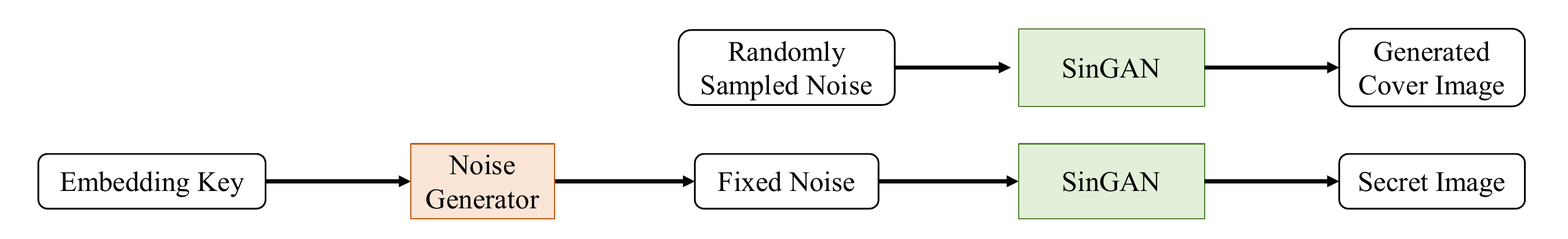}
	\caption{Hiding images in a SinGAN.}
	\label{fig:sin}
\end{figure}

A SinGAN consists of a pyramid of generators $\left\{\bm G_{0}, \bm G_{1} \ldots, \bm G_{N}\right\}$, where $\bm G_n$ takes the upsampled version of the generated image by $\bm G_{n+1}$ as well as an additive white Gaussian noise map $\bm z_n$ of the same size as inputs, and produces an image sample  at the $n$-th scale:
\begin{align}\label{eq:ns}
	\hat{\bm x}^{(c)}_{n}=\bm G_{n}\left(\bm z_{n},\left(\hat{\bm x}^{(c)}_{n+1}\right) \uparrow^{r}\right), \quad n<N,
\end{align}
where $r>1$ is the pre-defined upsampling ratio. The generation process starts at the coarsest scale (\ie, the $N$-th scale), where the input is purely noise:
\begin{align}
	\hat{\bm x}^{(c)}_{N}=\bm G_{N}\left(\bm z_{N}\right),
\end{align}
and progressively makes use of all generators to produce the finest scale image $\hat{\bm x}^{(c)}_0$ with possibly different size and aspect ratio of $\bm x^{(c)}_0$ \cite{singan}. Coupled with the generators is a pyramid of discriminators $\left\{D_0,D_1,\ldots, D_N\right\}$, and each $D_n$ is trained to discriminate between patches extracted from $\hat{\bm x}^{(c)}_n$ (in Eq. \eqref{eq:ns}) and $\bm x^{(c)}_n$, which is a downsampled version of $\bm x^{(c)}_0$ by a factor of $r^n$ (\ie, downsampled $n$ times by a factor of $r$).

The training objective for each scale is a weighted combination of an adversarial term and a reconstruction term:
\begin{align}\label{eq:lossc}
	\min _{\bm G_{n}} \max _{D_{n}}\, \ell_{\mathrm{adv}}\left(\bm G_{n}, D_{n};\bm x^{(c)}\right)+ \lambda \ell_{\mathrm{rec}}\left(\bm G_{n};\bm x^{(c)}\right),
\end{align}
where $\lambda$ is the trade-off parameter.
The adversarial loss $ \ell_{\mathrm{adv}}$ is for penalizing the statistical difference between the patch distribution of the generated $\hat{\bm x}^{(c)}_n$ and that of the (downsampled) cover image $\bm x^{(c)}_n$. The reconstruction loss $\ell_{\mathrm{rec}}$ is for stabilizing the training process by ensuring that $\bm x^{(c)}_n$ can be reconstructed from a specific set of input noise maps. Following the original SinGAN paper \cite{singan}, we use the WGAN-GP loss \cite{wgan} and the mean squared error (MSE) to implement $\ell_\mathrm{adv}$ and $\ell_\mathrm{rec}$, respectively.

After training, the SinGAN is capable of generating new image samples that preserve the patch distribution of $\bm x^{(c)}$, with novel and plausible scene configurations and structures. Once the learning procedure of the SinGAN is clear, hiding the secret image $\bm x^{(s)}$ during the patch distribution learning can be straightforwardly done by modifying the training objective from Eq. \eqref{eq:lossc} to
\begin{align}\label{eq:losss}
	\min _{\bm G_{n}} \max _{D_{n}}\, \ell_{\mathrm{adv}}\left(\bm G_{n}, D_{n};\bm x^{(c)}\right)+ \lambda \ell_{\mathrm{rec}}\left(\bm G_{n};\bm x^{(s)}\right).
\end{align}
That is, a reconstruction loss is replaced to enforce that a specific set of input noise maps $\bm z^{(s)} = \{\bm z^{(s)}_0, \bm z^{(s)}_1, \ldots, \bm z^{(s)}_N \}$ is mapped to the secret image $\bm x^{(s)}$ instead of the cover image $\bm x^{(c)}$. $\bm z^{(s)}$ can be generated by a standard Gaussian pseudo-random number generator \cite{brent1974algorithm} using the embedding key $\bm k$ as the seed. We may as well put back the reconstruction term for $\bm x^{(c)}$, but we find this makes little difference during training and testing. We conjecture that the reconstruction loss 
in Eq. \eqref{eq:losss} not only enables hiding of the secret image, but also plays a similar role in  improving the training stability and convergence. For ease of description, we refer to models optimized for Eq. \eqref{eq:lossc} and Eq. \eqref{eq:losss} as the \textit{original} and \textit{stego} SinGANs, respectively.
  
To extract the secret image $\bm x^{(s)}$, the receiver uses the shared embedding key $\bm k$ to re-generate the noise maps $\bm z^{(s)}$, which is fed to the publicly transmitted stego SinGAN for secret image extraction via a single forward propagation (see Fig. \ref{fig:sin}).

\section{Experiments}
In this section, we perform a series of experiments to verify the promise of our SinGAN approach. First, we evaluate secret image extraction accuracy both quantitatively and qualitatively in comparison to image-in-image hiding methods based on autoencoders. Second, we probe the security of the stego SinGAN by comparing it to the original one in terms of 1) quality and diversity of generated cover images, 2) marginal distribution similarity of model parameters \cite{wang2021data}, and 3) possibility of secret image leakage. Third, we 
experiment with our method in two more challenging scenarios: 1) hiding multiple images within a SinGAN for different users and 2) hiding the content-obfuscated image. We implement
 the generators and discriminators of the SinGAN by medium-size DNNs, whose specifications and training details are given in the Appendix. Our quantitative experiments make use of $200$ test image pairs, whose details are also given in the Appendix.

\begin{table}[tb]
    \centering
    \begin{minipage}{.6\textwidth}
    \captionsetup{width=.9\textwidth}
	\caption{Extraction accuracy comparison when hiding one image. "$\uparrow$": larger is better, and vice versa.}
	\label{tab:ea}
	\centering
	\begin{tabular}[t!]{lcccc}
		\toprule
		Method           &  \#params & PSNR$\uparrow$  & SSIM$\uparrow$   & DISTS$\downarrow$\\
		\midrule
		LSB & --- & 23.06 & 0.785 & 0.095\\
		Baluja17 & 0.48M & 25.91 & 0.874 & 0.102\\
		HiDDeN  & 0.38M & 27.99 & 0.897 & 0.096\\
		Weng19& 42.6M & 35.64 & 0.942 & 0.055\\
		HiNet & 4.05M & 35.59 & 0.952 & 0.047\\
	\midrule
		Ours   & 0.67M &\textbf{36.84} & \textbf{0.958} & \textbf{0.038}\\

		\bottomrule
	\end{tabular}
	\end{minipage}%
    \begin{minipage}{.4\textwidth}
	\caption{Cover image quality, diversity, and weight distribution similarity between the original and stego SinGANs in terms of SIFID, DS, and KLD.}
	\label{tab:qdd}

	\centering
	\begin{tabular}{llll}
		\toprule
		\#images  & SIFID$\downarrow$  & DS$\uparrow$ & KLD$\downarrow$\\
		\midrule
		0 (ref) & 0.041 & 0.407 & 0\\
		1 & 0.046 & 0.430 & 0.001\\
		2 & 0.045 & 0.415 & 0.004\\
		3 & 0.047 & 0.427 & 0.006\\
		4 & 0.051 & 0.438 & 0.008\\
		\bottomrule
	\end{tabular}
	\end{minipage}%
\end{table}

\begin{figure}[t]
    \label{fig:ve}
    \centering
	\subfloat{%
		\includegraphics[width=0.17\textwidth]{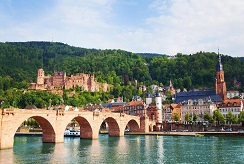}
	}
	\subfloat{%
		\includegraphics[width=0.17\textwidth]{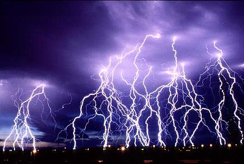}
	}
	\subfloat{%
		\includegraphics[width=0.17\textwidth]{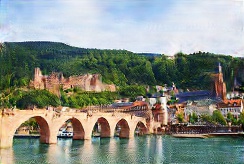}
	}
	\subfloat{%
		\includegraphics[width=0.17\textwidth]{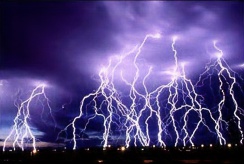}
	}
	\subfloat{%
		\includegraphics[width=0.204\textwidth]{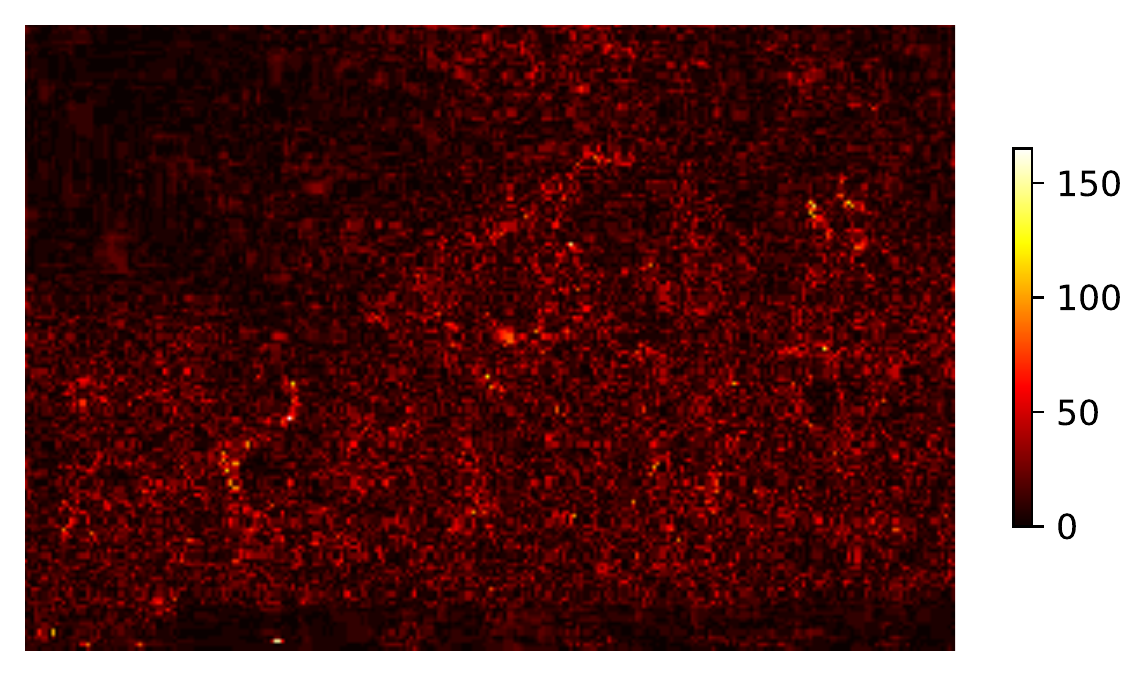}
	}
	\vspace{-0.35cm}
	\subfloat{%
		\includegraphics[width=0.17\textwidth]{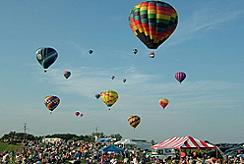}
	}
	\subfloat{%
		\includegraphics[width=0.17\textwidth]{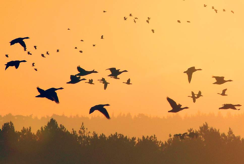}
	}
	\subfloat{%
		\includegraphics[width=0.17\textwidth]{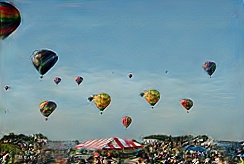}
	}
	\subfloat{%
		\includegraphics[width=0.17\textwidth]{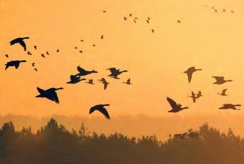}
	}
	\subfloat{%
		\includegraphics[width=0.204\textwidth]{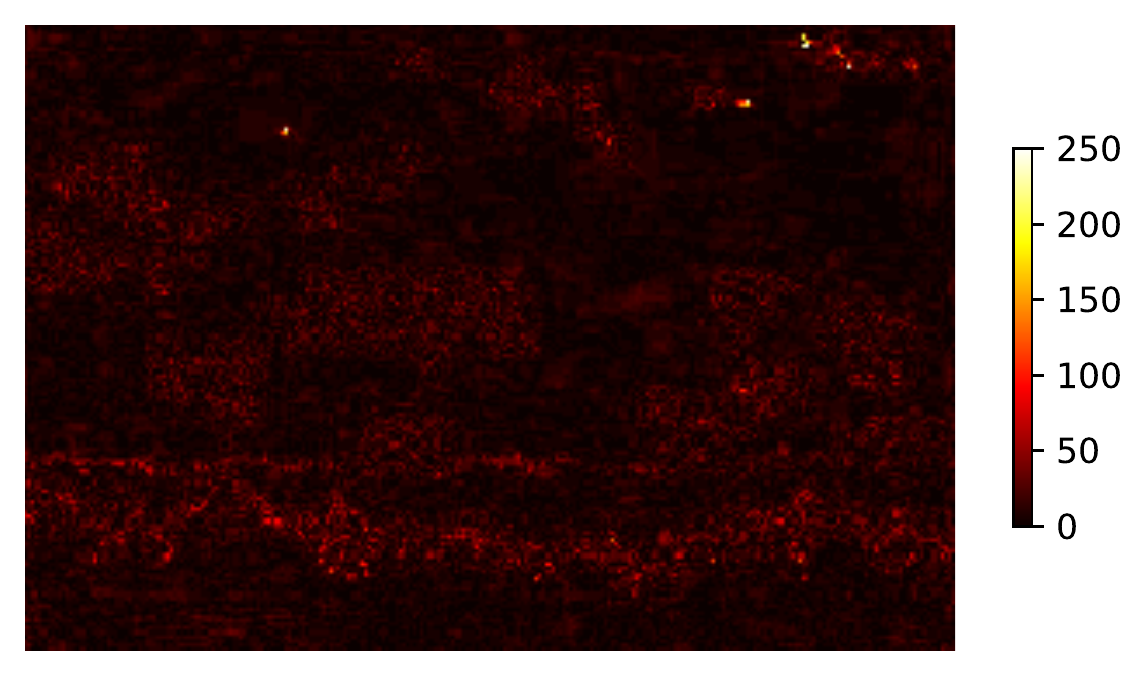}
	}
	\vspace{-0.35cm}
	\subfloat{%
		\includegraphics[width=0.17\textwidth]{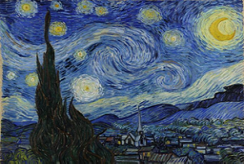}
	}
	\subfloat{%
		\includegraphics[width=0.17\textwidth]{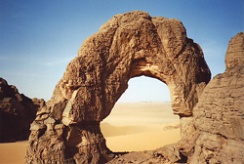}
	}
	\subfloat{%
		\includegraphics[width=0.17\textwidth]{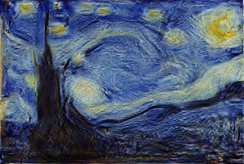}
	}
	\subfloat{%
		\includegraphics[width=0.17\textwidth]{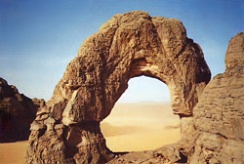}
	}
	\subfloat{%
		\includegraphics[width=0.204\textwidth]{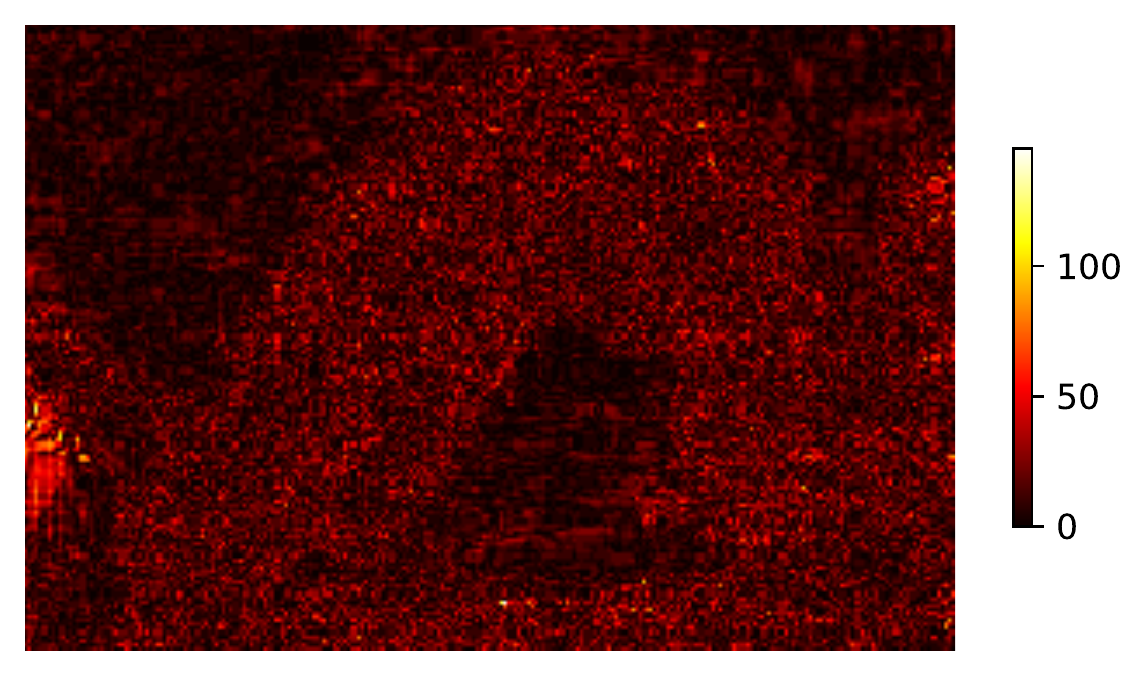}
	}
	\vspace{-0.35cm}
	\subfloat{%
		\includegraphics[width=0.17\textwidth]{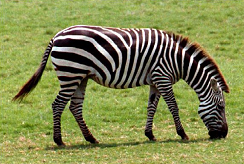}
	}
	\subfloat{%
		\includegraphics[width=0.17\textwidth]{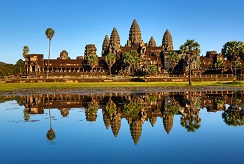}
	}
	\subfloat{%
		\includegraphics[width=0.17\textwidth]{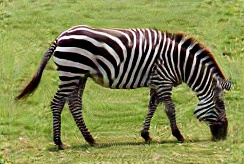}
	}
	\subfloat{%
		\includegraphics[width=0.17\textwidth]{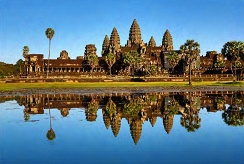}
	}
	\subfloat{%
		\includegraphics[width=0.204\textwidth]{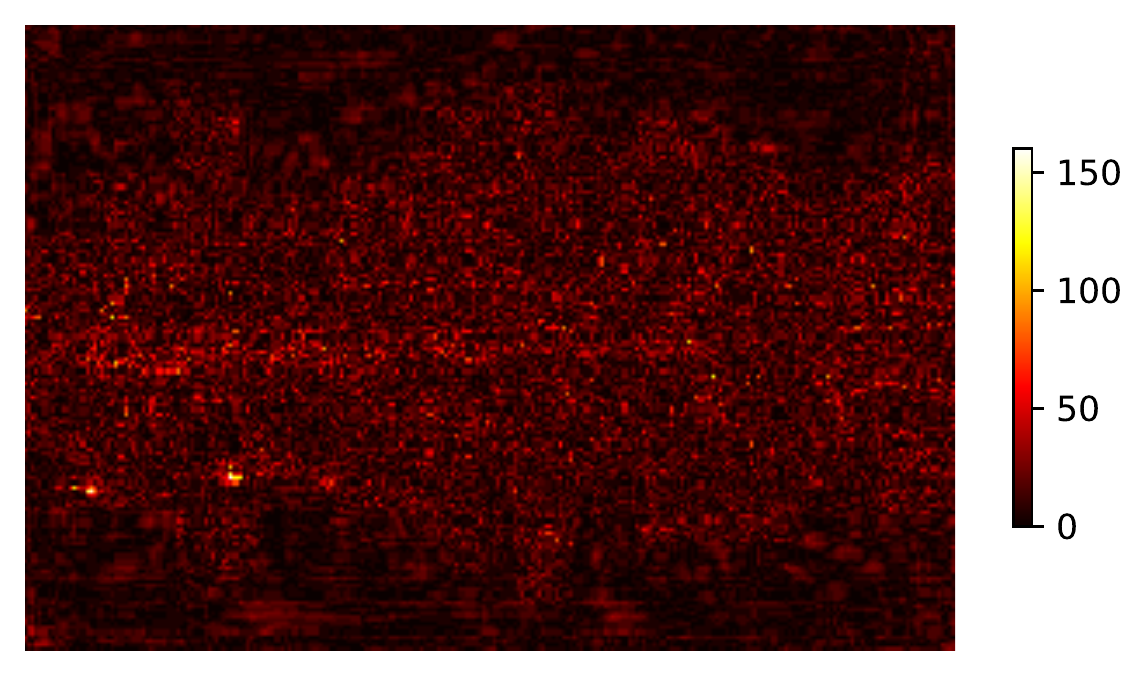}
	}
	\vspace{-0.35cm}
	 \addtocounter{subfigure}{-20}
	\subfloat[Cover $\bm{x}^{(c)}$]{%
		\includegraphics[width=0.17\textwidth]{Images-Ours-1secret/real-cover-3.jpg}
	}
	\subfloat[Secret $\bm{x}^{(s)}$]{%
		\includegraphics[width=0.17\textwidth]{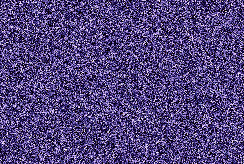}
	}
	\subfloat[Generated $\hat{\bm{x}}^{(c)}$]{%
		\includegraphics[width=0.17\textwidth]{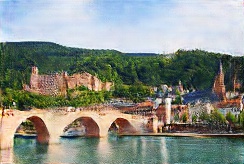}
	}
	\subfloat[Extracted $\hat{\bm{x}}^{(s)}$]{%
		\includegraphics[width=0.17\textwidth]{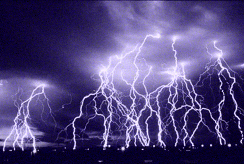}
	}
	\subfloat[$5\times \vert\bm{x}^{(s)} -\hat{\bm{x}}^{(s)}\vert$]{%
		\includegraphics[width=0.204\textwidth]{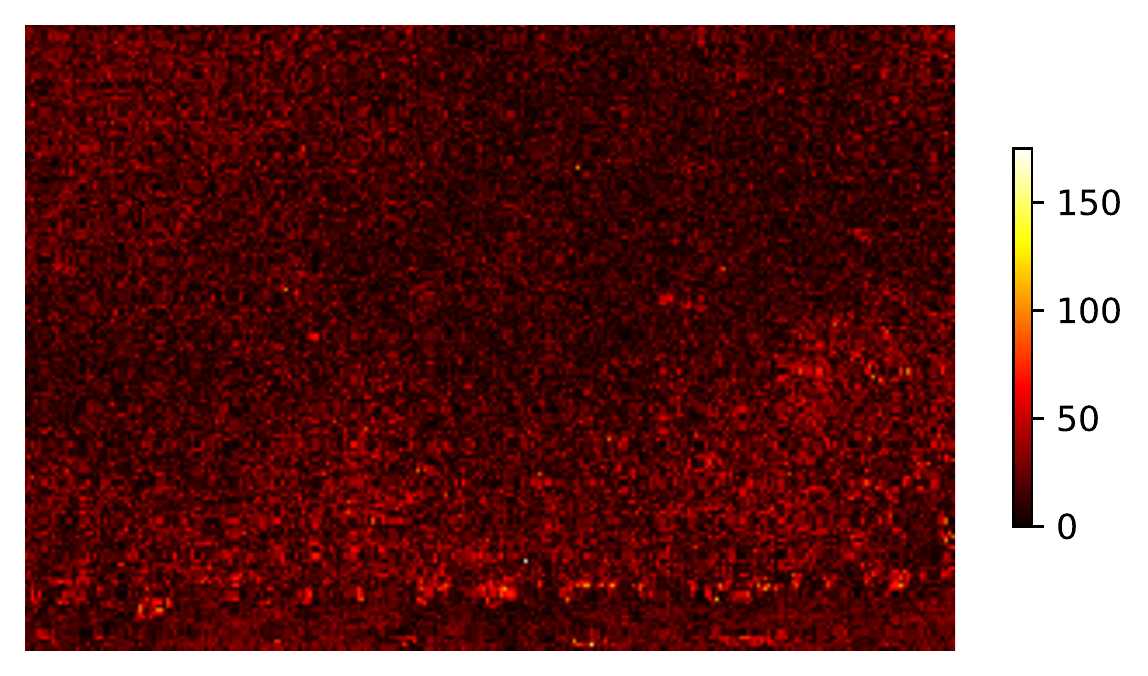}
	}
	\caption{Visual results of our SinGAN approach. The second image of the last row is a pixel-shuffled version of the second image in the first row. Zoom in for improved visibility.}
	\label{fig:ve}
\end{figure}

\subsection{Extraction Accuracy}
In image data hiding, extraction accuracy means how the secret image can be faithfully reproduced at the receiver side. Here, we apply three objective image quality measures to quantitatively evaluate the signal fidelity, perceptual fidelity, and perceptual quality of the extracted secret image: the peak signal-to-noise ratio (PSNR), the structural similarity (SSIM) index \cite{wang2004image}, and the deep image structure and texture similarity (DISTS) measure \cite{ding2020image}. Although our SinGAN approach is the first of its kind in the proposed probabilistic image hiding framework, we compare it with one na\"{i}ve  LSB replacement method, and four image-in-image steganography methods - Baluja17 \cite{baluja2017}, HiDDeN \cite{zhu2018}, Weng19 \cite{weng2019high} and HiNet\cite{jing2021}. The LSB replacement method simply replaces the four LSB planes of the cover image with the four most significant bit (MSB) planes of the secret image.  As the robustness is not our focus, we re-train HiDDeN \cite{zhu2018} for image hiding without the noise layer. Weng19 \cite{weng2019high} is a DNN-based video steganography method, and only the image hiding branch is used for testing.
HiNet \cite{jing2021} relies on a normalizing flow-based invertible DNN to implement the encoder, such that the decoder is simply its inverse. 
We use the publicly available implementations for Baluja17
\cite{dabluja17-github}, HiDDeN\cite{hidden-github}, Weng19\cite{weng19-github} and HiNet\cite{hinet-github} with default training and testing\footnote{It is important to note that the results of autoencoder-based methods in our paper (especially in terms of extraction accuracy in Table~\ref{tab:ea}) may be noticeably different from some
 of the previous publications. This is because we choose to quantize the stego image from the single-precision floating-point format of $32\times 3$ to $8\times 3$ bits per pixel before transmitting it to the receiver side. If such quantization is not properly enforced, trivial hiding solutions may exist because there are just more space to accommodate the cover and secret images by a simple concatenation.} settings, and implement the LSB replacement by ourselves. 

 The extraction accuracy results are shown in Table~\ref{tab:ea}, where we find that the proposed SinGAN approach performs favorably against existing autoencoder-based image-in-image steganography methods.
 We consider the obtained improvements as significant because secret-in-network hiding is generally considered much more difficult than secret-in-image hiding, where the latter has significant spatial redundancy to be reduced. After all, the most recent secret-in-network hiding method \cite{wang2021data} has a limited embedding capacity up to $6,000$ bits. We also show some representative visual results of our method in Fig.~\ref{fig:ve}. It is clear that the stego SinGAN is able to capture the internal patch distribution of the cover image, generating image samples with different but reasonable structures and
configurations of the same natural scene. Moreover, the extracted secret image by guided sampling is visually close to the original, which is adequate for use in the majority of  real-world steganography applications.

\subsection{Model Security} \label{security}
Steganographic analysis (\ie, steganalysis) aims to identify whether there are hidden messages in  suspected media, which constitutes an indispensable part of steganography evaluation. Traditional statistics-based and recent DNN-based steganalysis methods \cite{boehm2014stegexpose,boroumand2018deep} are mostly designed to discriminate between cover and stego images, thus not applicable when the cover media is a DNN. To the best of our knowledge, we are not aware of a steganalysis algorithm that accepts a full DNN (with millions of parameters) as input. Instead, we propose to probe the security of our SinGAN approach from the following three different aspects.

\begin{figure}[t]

    \centering
    \begin{tabular}{c|c}
    \multirow{2}{*}[12pt]{
        \subfloat[Cover Image]{
		\includegraphics[width=0.16\textwidth]{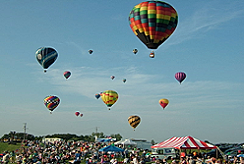}
	}} &

	\subfloat{
		\includegraphics[width=0.14\textwidth]{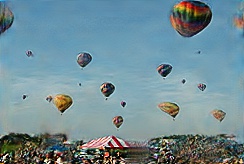}
	}
	\subfloat{
		\includegraphics[width=0.14\textwidth]{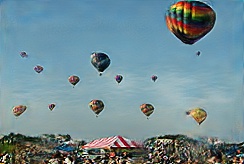}
	}
	\addtocounter{subfigure}{-2} 
	\subfloat[Original]{%
		\includegraphics[width=0.14\textwidth]{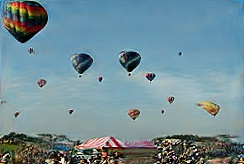}
	}
	\subfloat{
		\includegraphics[width=0.14\textwidth]{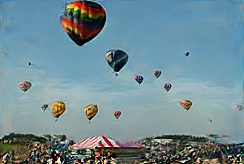}
	}
	\subfloat{
		\includegraphics[width=0.14\textwidth]{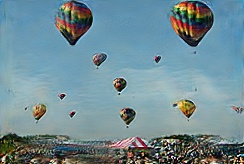}
	}\\ &
	\subfloat{
	    \centering
		\includegraphics[width=0.14\textwidth]{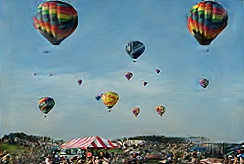}
	}
	\subfloat{
		\includegraphics[width=0.14\textwidth]{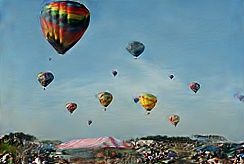}
	}
	\addtocounter{subfigure}{-4}
	\subfloat[Stego]{%
		\includegraphics[width=0.14\textwidth]{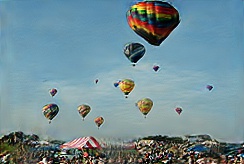}
	}
	\subfloat{
		\includegraphics[width=0.14\textwidth]{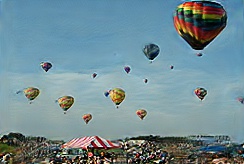}
	}
	\subfloat{
		\includegraphics[width=0.14\textwidth]{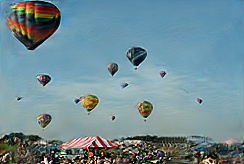}
	}
	\end{tabular}
	
	\caption{Visual comparison of the generated cover images ``Fire balloons'' by the original and stego SinGANs. In our case, the secret is the ``Birds'' image, as shown in the second row of Fig.~\ref{fig:ve}.}
	\label{fig:sampled}
\end{figure}

\noindent\textbf{Quality and Diversity of Generated Cover Images}.
In the proposed framework, the stego SinGAN is publicly transmitted, and thus it must function as the original SinGAN \cite{singan}. As the primary goal of SinGANs is to model internal patch distributions, we examine and compare the quality and diversity of generated cover images by the original and stego SinGANs. To quantify quality, we adopt the single image Fr\'{e}chet inception distance (SIFID) metric, 
as suggested in \cite{singan}. To quantify diversity, for each cover image, we  compute the diversity score (DS) as the standard deviation (std) of each pixel values over $25$ generated samples of the cover content, averaged over all pixels and normalized by the std of pixel intensities of the cover image \cite{singan}. The average quality and diversity results over $200$ test image pairs are shown in Table \ref{tab:qdd}, where we find that both SIFID and DS of the stego SinGAN are statistically indistinguishable from those of the original SinGAN based on a hypothesis testing using $t$-statistics \cite{montgomery2010applied}. Fig. \ref{fig:sampled} shows some randomly sampled examples from the original and stego SinGANs, which provides additional visual evidence that they learn very similar internal patch distributions of the cover image.

\noindent\textbf{Marginal Distribution Similarity of Model Weights}. As the proposed method essentially hides the secret image in the learned weights of the stego SinGAN, it is natural to ask whether its weight distributions significantly deviate from those of the original SinGAN. Following \cite{wang2021data}, we compute the Kullback–Leibler divergence (KLD) between the marginal distributions of model parameters of the original and stego SinGANs, as shown in the last column of Table~\ref{tab:qdd}. We find that the marginal distributions of the two sets of model parameters are almost identical, as evidenced by a KLD close to zero. Similar results can be obtained if the marginal distributions are compared in a  per-stage fashion (see more results in the Appendix with visual comparison of histograms).

\noindent\textbf{Possibility of Secret Image Leakage}. One may also  wonder the possibility of secret image leakage if an adversary constantly draws samples from the stego SinGAN. Such steganalysis arises  naturally from the fact that there is no theoretical guarantee that the mapping between the secret noise (generated by the embedding key) and the secret image is bijective (\ie, one-to-one). In other words,  there might be some other sets of noise that  are also mapped to the secret image, or at least some of its semantically meaningful content. Since the SinGAN is trained to be a black-box sampler, it is difficult to provide a theoretical analysis of the possibility of secret image leakage. Nevertheless, we 
conduct an empirical study, in which we  randomly draw $100,000$ samples from each of the $200$ trained stego SinGANs. Visual inspection of the thumbnails of generated samples indicates that no secret image (or images with similar semantics) is revealed. Therefore, it is safe to empirically conclude that the possibility of secret image leakage is less than $0.001\%$.

In summary, we have empirically proven that the proposed SinGAN approach is secure: hiding a full-size photographic image into a SinGAN does not compromise the quality and diversity of generated cover images, nor skew the weight distribution. It also survives constant sampling by an adversary. Such level of security verifies our claims that the secret image indeed occupies a tiny portion of probability mass of $p_s(\bm x)$, and that $p_s(\bm x)$ and  $p_c(\bm x)$ are statistically close.

\subsection{Further Extensions}\label{subsec:fe}
\noindent\textbf{Hiding Multiple Images for Different Receivers}. Hiding multiple images in a DNN is challenging; doing so for different receivers is even more challenging, and has not been accomplished before. Here the main difficulties lie not only in that more embedding capacity is required, but also in that each receiver must only extract her/his piece of message, and cannot extract (or even affirm the existence of) other messages \cite{wang2021data}. The proposed probabilistic image hiding framework provides a straightforward and elegant extension to hide multiple images for different users, which follows a similar two-step approach. First, learn a probabilistic density function $p_s(
\bm x)$ over $\mathcal{D}_s = \mathcal{D}\bigcup\{\bm x^{(s_1)}, \ldots,\bm x^{(s_T)}\}$, where $T$ is the number of secret images and is considerably smaller than $M$, the number of cover images. Second, design $T$ guided sampling procedures using $T$ different embedding keys $\mathcal{K} = \{\bm k^{(1)}, \ldots,\bm k^{(T)}\}$, shared to $T$ different receivers. The learning goal remains the same: $p_s(\bm x)$ should be close in some statistical distance to the reference distribution $p_c(\bm x)$. The described procedure is more easily understood using the SinGAN instantiation, where we just modify the objective function in Eq.~\eqref{eq:losss} to
\begin{align}\label{eq:lossms}
	\min _{\bm G_{n}} \max _{D_{n}}\, \ell_{\mathrm{adv}}\left(\bm G_{n}, D_{n};\bm x^{(c)}\right)+ \lambda \frac{1}{T}\sum_{t=1}^T\ell_{\mathrm{rec}}\left(\bm G_{n};\bm x^{(s_t)}\right).
\end{align}

After training, the $t$-th receiver is able to re-generate the $t$-th specific set of noise maps $\bm z^{(s_t)}$ using the shared embedding key $\bm k^{(t)}$ for the $t$-th secret image extraction. S/he is, by design, ignorant of the presence (or absence) of other secret images. Even if the receiver is informed in some way that the current SinGAN contains multiple secret images, without extra embedding keys, s/he cannot extract images that are not intended to share with her/him.

\begin{table}[tb]
    \begin{minipage}{.5\textwidth}
	\captionsetup{width=\textwidth}
	\caption{Extraction accuracy of our method when hiding multiple images in one SinGAN.}
	\label{tab:mea}
	\centering
	\begin{tabular}[t!]{lccc}
		\toprule
		\#images    & PSNR$\uparrow$  & SSIM$\uparrow$ & DISTS$\downarrow$\\
		\midrule
		One              & 36.84 & 0.958 & 0.038\\
		Two              & 35.91 & 0.946 & 0.043\\
		Three            & 34.93 & 0.935 & 0.049\\
		Four             & 34.03 & 0.923 & 0.055\\
		\bottomrule
	\end{tabular}
	\end{minipage}%
    \begin{minipage}{.5\textwidth}
    \caption{Extraction accuracy of our method with image obfuscation.}
	\label{tab:obf}
	\centering
	\begin{tabular}{lccc}
		\toprule
		Obfuscation    & PSNR$\uparrow$  & SSIM$\uparrow$  & DISTS$\downarrow$ \\
		\midrule
		No    & 36.84  & 0.958 & 0.038\\
		Yes   & 20.53  & 0.726 & 0.172\\
		\bottomrule
	\end{tabular}
	\end{minipage}%
\end{table}

\begin{figure}[t]
    \centering
    
    \subfloat[1st Original]{%
		\includegraphics[width=0.15\textwidth]{Images-Train/balloons.png}
	}
	\subfloat[Extracted]{%
		\includegraphics[width=0.15\textwidth]{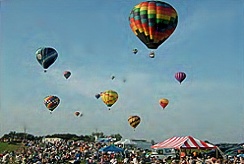}
	}
	\subfloat[Error ($5\times$)]{%
		\includegraphics[width=0.18\textwidth]{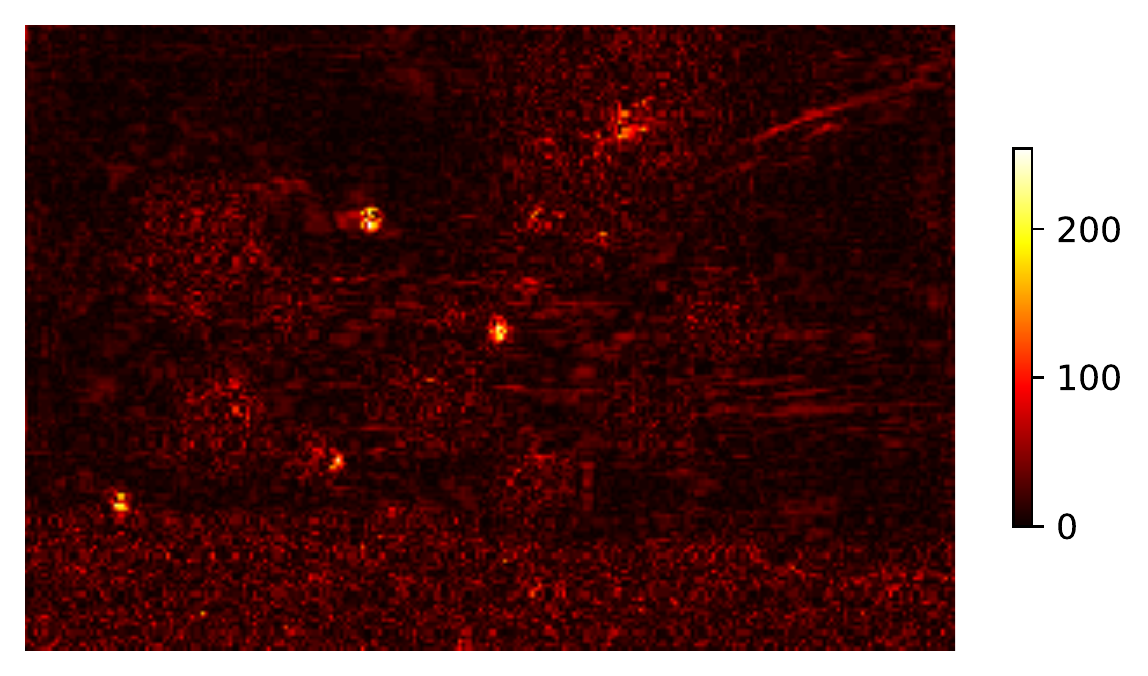}
	}
	\subfloat[2nd Original]{%
		\includegraphics[width=0.15\textwidth]{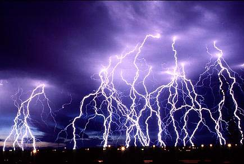}
	}
	\subfloat[Extracted]{%
		\includegraphics[width=0.15\textwidth]{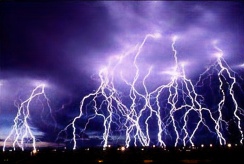}
	}
	\subfloat[Error ($5\times$)]{%
		\includegraphics[width=0.18\textwidth]{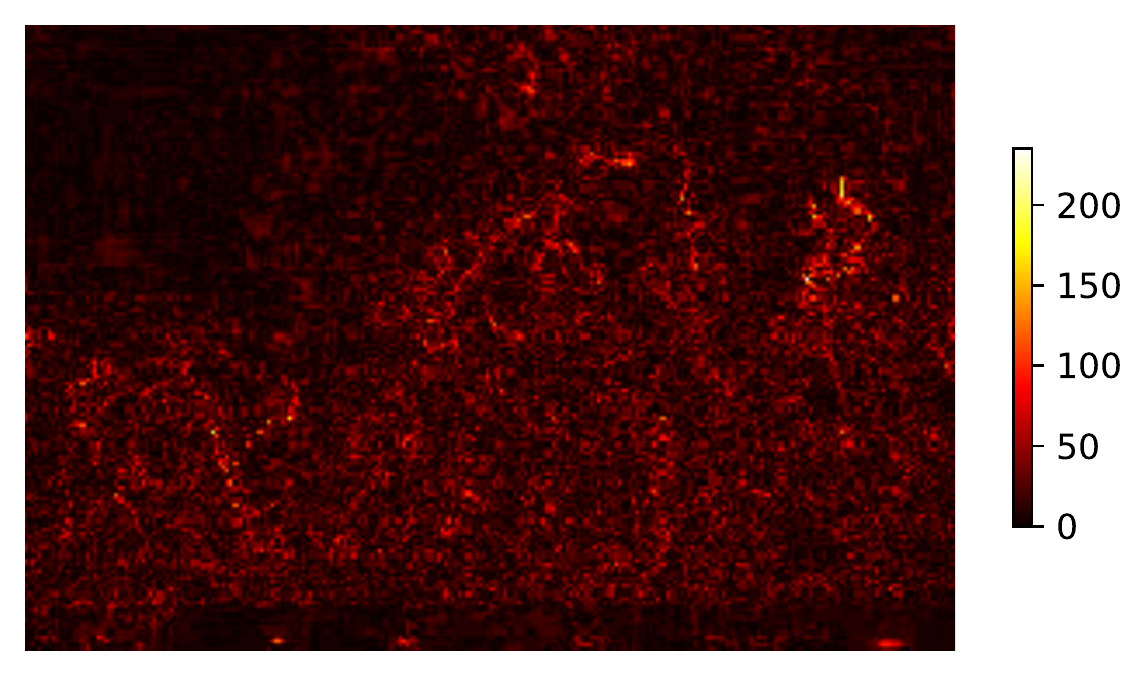}
	}
	
	\vspace{-0.3cm}
	\subfloat[3rd Original]{%
		\includegraphics[width=0.15\textwidth]{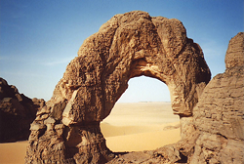}
	}
	\subfloat[Extracted]{%
		\includegraphics[width=0.15\textwidth]{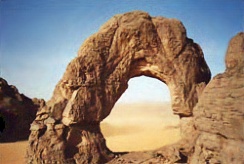}
	}
	\subfloat[Error ($5\times$)]{%
		\includegraphics[width=0.181\textwidth]{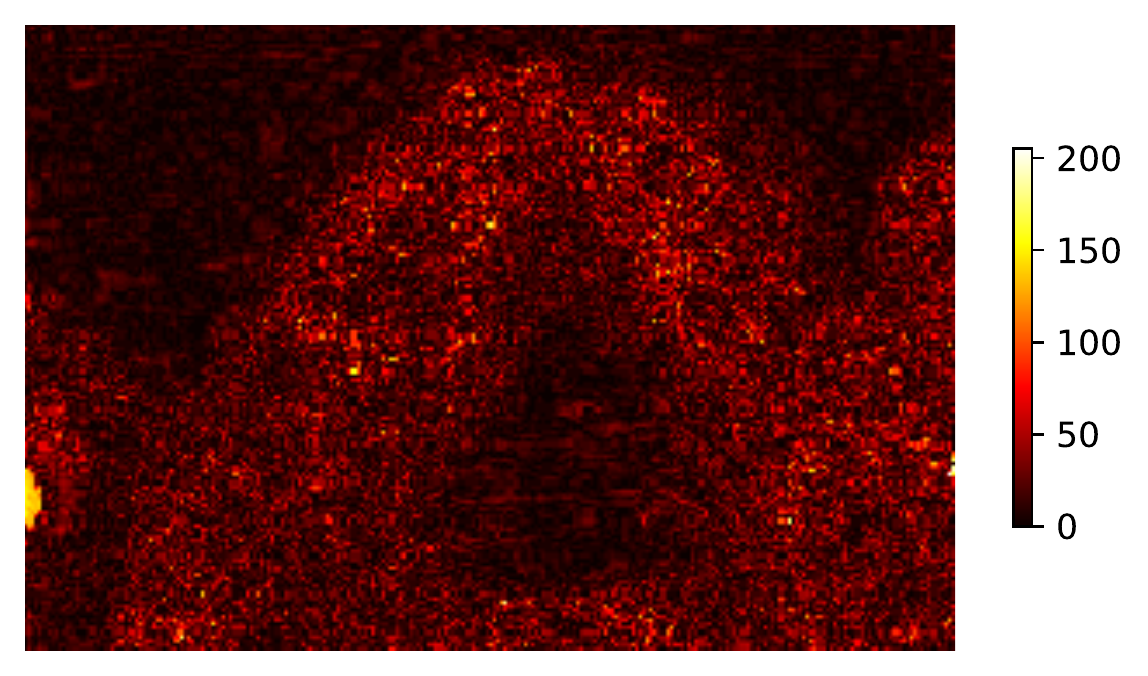}
	}
	\subfloat[4th Original]{%
		\includegraphics[width=0.15\textwidth]{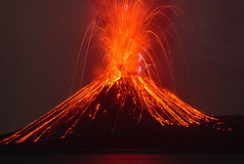}
	}  
	\subfloat[Extracted]{%
		\includegraphics[width=0.15\textwidth]{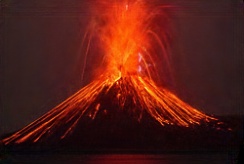}
	}
    \subfloat[Error ($5\times$)]{%
		\includegraphics[width=0.181\textwidth]{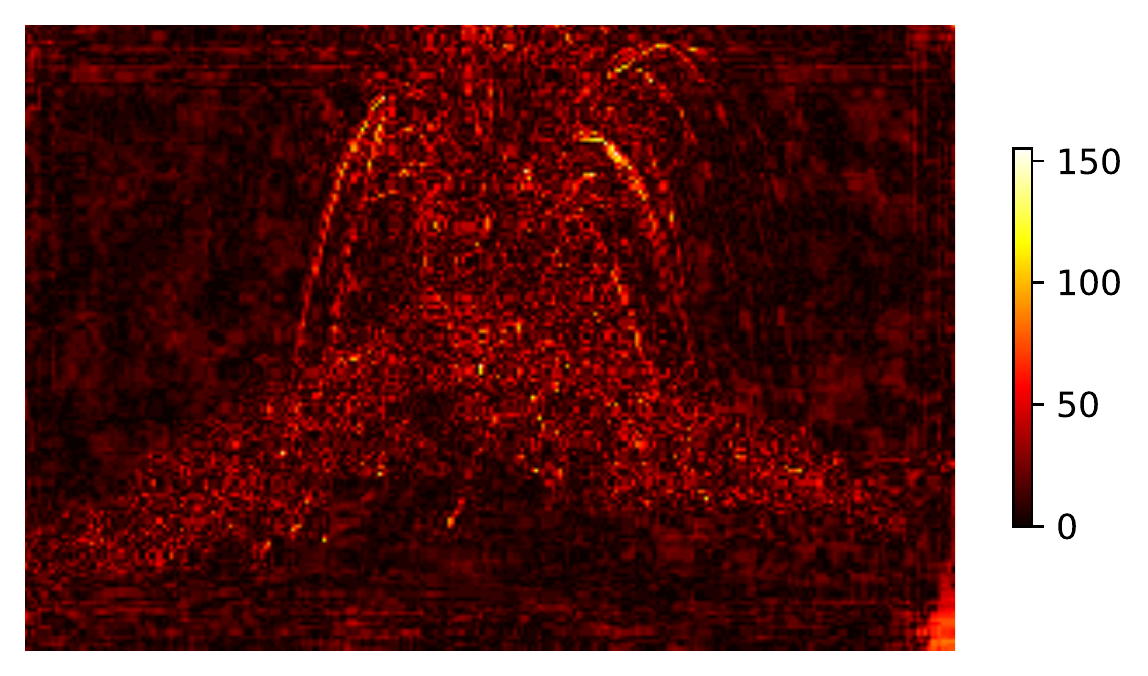}
	}

	\caption{Visual comparison between the original and extracted secret images from a SinGAN with four hidden images. }
	\label{fig:mve}
\end{figure}

We train SinGANs to hide up to four secret images, \ie, $T\in\{2,3,4\}$. For each value of $T$, we train $200$ SinGANs to hide different combinations of cover and secret images. The quality and diversity of the cover image, and the weight distribution similarity between the original and stego SinGANs are shown in Table \ref{tab:qdd}, where we find that hiding multiple images does not seem to compromise the security of the proposed method. The extraction accuracy results and  visual examples are shown in Table \ref{tab:mea} and Fig. \ref{fig:mve}, respectively. With the number of increasing secret images, the extraction accuracy in terms of the three objective metrics degrades gracefully. The visual appearances of the extracted secret images are similar to those of the original, showing great promise of our method in hiding multiple images for different receivers.

\noindent\textbf{Obfuscating the Secret Image}. Inspired by \cite{baluja2017}, we consider obfuscating the secret image by shuffling its pixels (\ie, image scrambling \cite{van2004image}), as a way of strengthening security. In this case, The shuffling key, together with the embedding key, is shared to the receiver. Table~\ref{tab:obf} shows the extraction accuracy results, where we see that pixel shuffling significantly increases the difficulty of image hiding. Nevertheless, from the last row of Fig. \ref{fig:ve}, we observe that the main content is clearly visible, despite somewhat noisy appearance.

\vspace{-0.2cm}
\section{Conclusion and Discussion}
\vspace{-0.2cm}
We have described a new computational framework for hiding images in deep probabilistic models, which is in stark contrast to previous steganography schemes. We provided an instantiation, where we used the SinGAN to build the internal patch distribution of the cover image, and hid the secret image during patch distribution learning. We conducted a series of experiments to demonstrate the feasibility of the proposed SinGAN approach in terms of extraction accuracy and model security. Moreover, our method is readily extended to hide multiple images for different receivers, a challenging task that has not been accomplished before. In addition, it works nicely with pixel shuffling, which adds additional security.

The current work opens the door to a new class of image hiding methods, with many interesting problems to be explored. First, the extraction accuracy of our SinGAN approach still has quite some room for improvement. For applications that require precise recovery of the secret image, it is worth exploring more efficient network structures, optimized for perceptual losses that exploit the physiological properties of the human visual system. Second, our method na\"{i}vely bypasses the steganalysis tools specifically designed for secret-in-image steganography. Currently, we have 
designed three different tests to probe the model security. As SinGANs \cite{singan} can be applied in a much wider range of image manipulation tasks such as super-resolution and paint-to-image translation, the stego SinGAN should be tested in those applications as well for model security. More importantly, we expect future effort to be dedicated to building steganalysis methods that accept a full DNN as input and assess whether it contains secret messages. Third, so far, we have just given the theoretical intuition of the probabilistic image hiding framework, supplied with empirical evidence. It is of mathematical interest to rigorously measure the statistical distance between $p_c(\bm x)$ and $p_s(\bm x)$, in an attempt to answer important questions like 1) where the secret image is hidden in the network (or equivalently the learned distribution) and 2) what the maximum number of secret images is allowed for a given distance constraint (\eg, $\mathrm{KLD}(p_c(\bm x)\parallel  p_s(\bm x)) \le \epsilon$). Fourth, many other generative modeling methods \cite{song2021train} are worthy of deep investigation within the proposed framework to 
circumvent some of the limitations of the SinGAN approach. For example, we may model the (Stein) score function \cite{liu2016kernelized,hyvarinen2005estimation} of $p_s(\bm x)$ (\ie, the gradient of $\log p_s(\bm x)$), and develop guided Langevin-type sampling (by the embedding key) to extract the secret image.

\begin{ack}
The authors would like to thank Song Yang for inspiring discussion. This work was supported in part by the Hong Kong RGC ECS Grants 21213821 (to KDM) and 21212419 (to LQS), the National Natural Science Foundation of China under Grants 62071407, 61936214, U20B2051 and U1936214, the InnoHK initiative, the Government of the HKSAR, Laboratory for AI-Powered Financial Technologies, and the Tencent AI Lab Rhino-Bird Gift Fund.
\end{ack}

\clearpage
\bibliography{arXiv} 
\bibliographystyle{plain}

\clearpage
\appendix

\section{Specifications and Training Details}

\subsection{Model Architecture}
\begin{figure}[h]
    \centering
    \subfloat[$n$-th scale generator.]{
	\includegraphics[width=\textwidth]{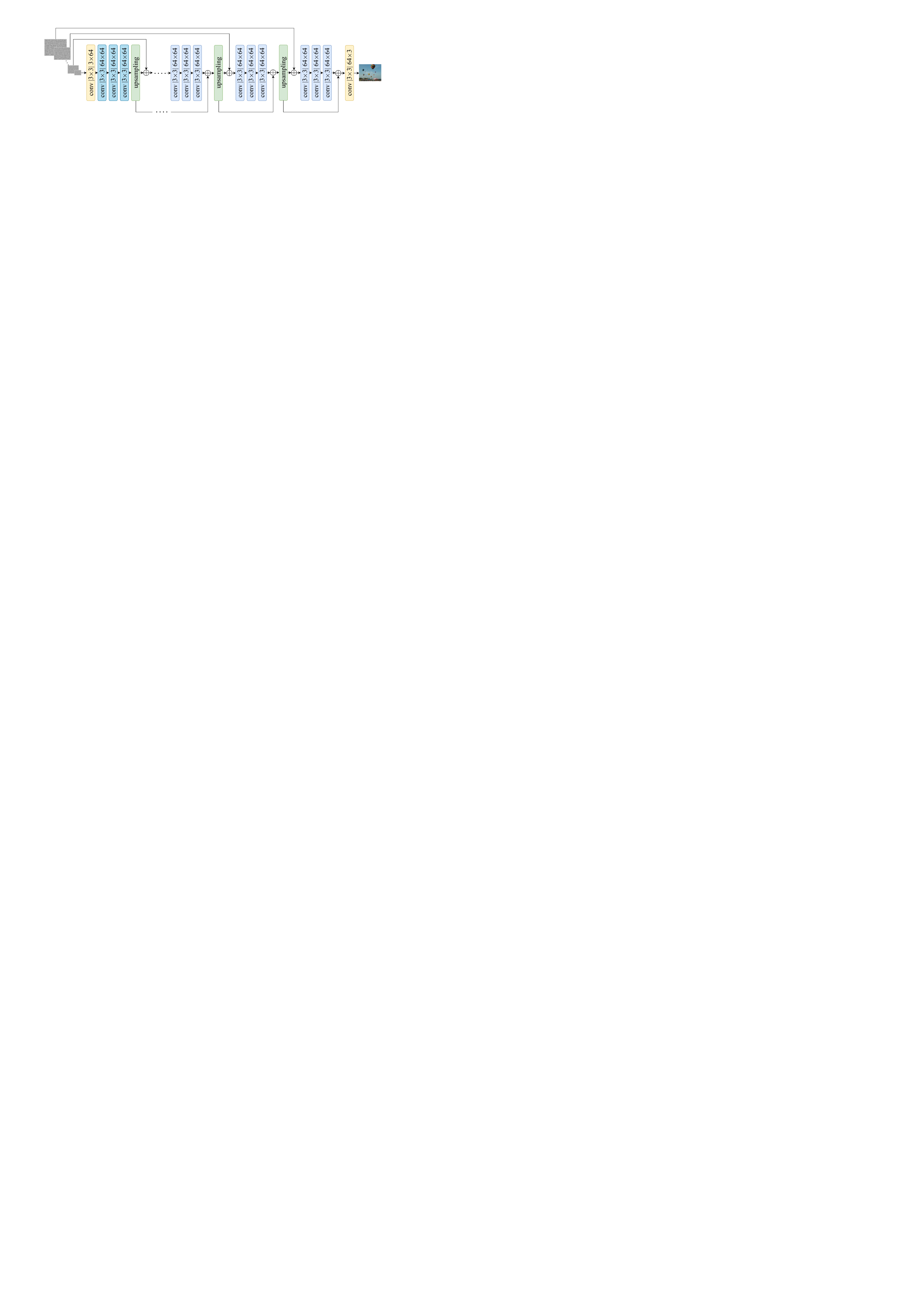}
	}
	
	\subfloat[Discriminator (shared).]{
		\includegraphics[width=0.9\textwidth]{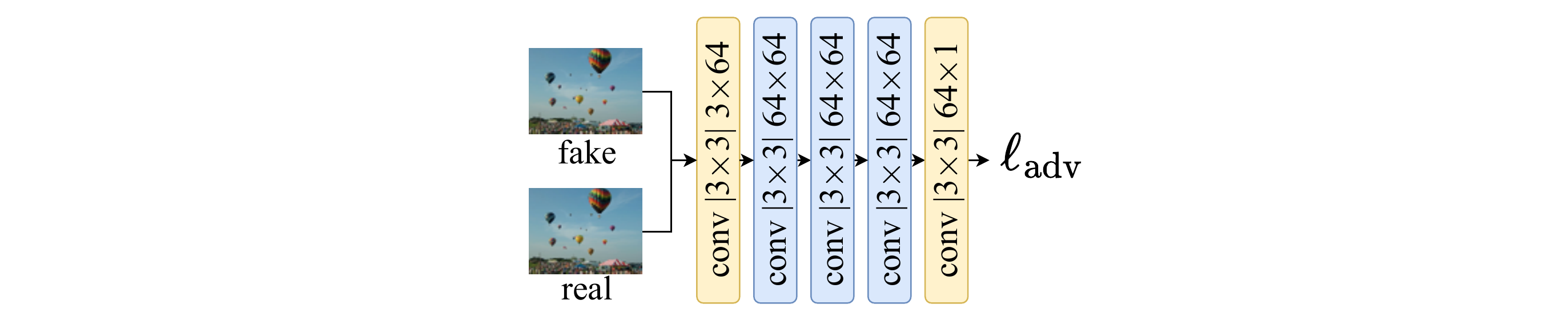}
	}
	
	\caption{Architecture of the SinGAN used in our paper.}
    \label{fig:arc}
\end{figure}

The architecture of the SinGAN used in our paper follows that in~\cite{consingan}. As shown in Fig.~\ref{fig:arc},  the generator at the $n$-th scale consists of a front-end convolution, $N-n+1$ convolution blocks each with three convolution layers. Each convolution block, except for the last one, is followed by an upsampling layer with two residual connections, one for residual learning and one for incorporating a noise map from a higher scale (with a smaller spatial size). The last convolution block is responsible for the $n$-th scale image reconstruction with one residual connection. One back-end convolution with $\mathrm{tanh}()$ nonlinear activation is used to produce a scaled version of the RGB image as output. All convolution layers (except for the back-end one) have $64$ filters with a filter size of $3\times3$, followed by batch normalization and leaky ReLU activation with the negative slope of $0.05$. The weights of the front-end convolution, and the first $N-n$ convolution blocks, and the back-end convolution are inherited from those of the trained generator at the $n+1$-th scale. 

All generators share one discriminator, which is composed of five convolution layers all with a filter size $3\times3$. The first four convolution layers have $64$ filters followed by leaky ReLU activation with the negative slope of $0.05$, while the last convolution layer has a single convolution filter. 

\subsection{Optimization}
We adopt the improved techniques for training SinGANs as recommended in \cite{consingan}. The $\lambda$ in Eq.~\eqref{eq:losss} is set to $10$. The trade-off parameter in WGAN-GP \cite{wgan} is set to  $0.1$ for gradient penalty. For the generator at the $n$-th scale, the newest three convolution blocks along with the front-end and back-end convolutions are jointly trained (or fine-tuned), while holding the older convolution blocks fixed (if any). This training strategy seems effective in preventing mode collapse. Adam\cite{kingma2014adam} is adopted as the stochastic optimizer with an initial learning rate of $0.0005$ and a decay factor of $0.1$ after finishing $80\%$ of iterations, and we set the maximum number of training iterations to $2,000$. The training time is approximately $20$ minutes for one image pair (each with size $244\times 164 \times 3$) on an NVIDIA GeForce RTX3080 GPU.

\section{Test Image Pairs for Quantitative Experiments}
\begin{figure}[h]
     \centering
     \includegraphics[width=\textwidth]{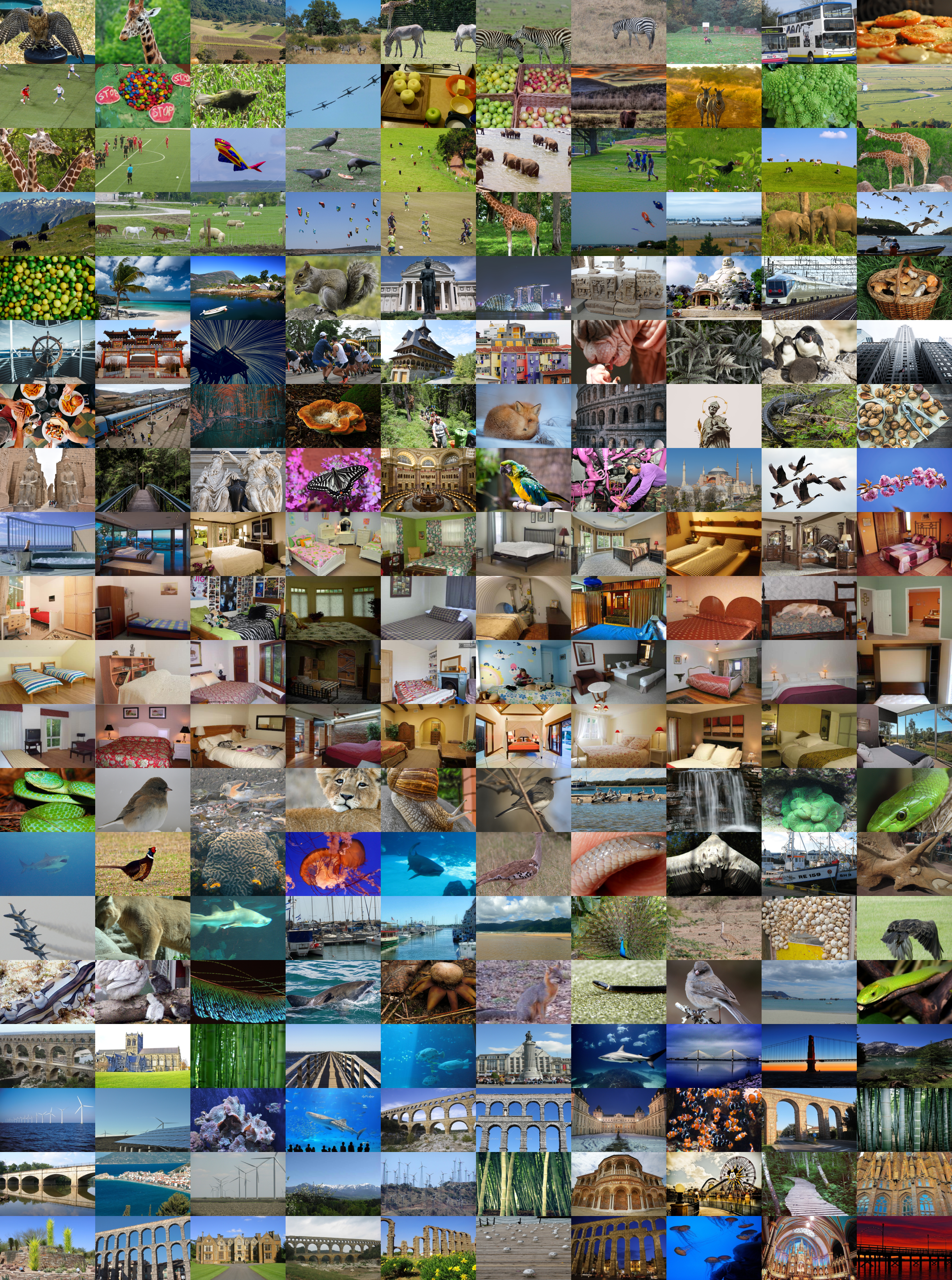}
    \caption{$200$ cover images used for quantitative comparison.}
    \label{fig:cov}
\end{figure}

\begin{figure}[h]
     \centering
     \includegraphics[width=\textwidth]{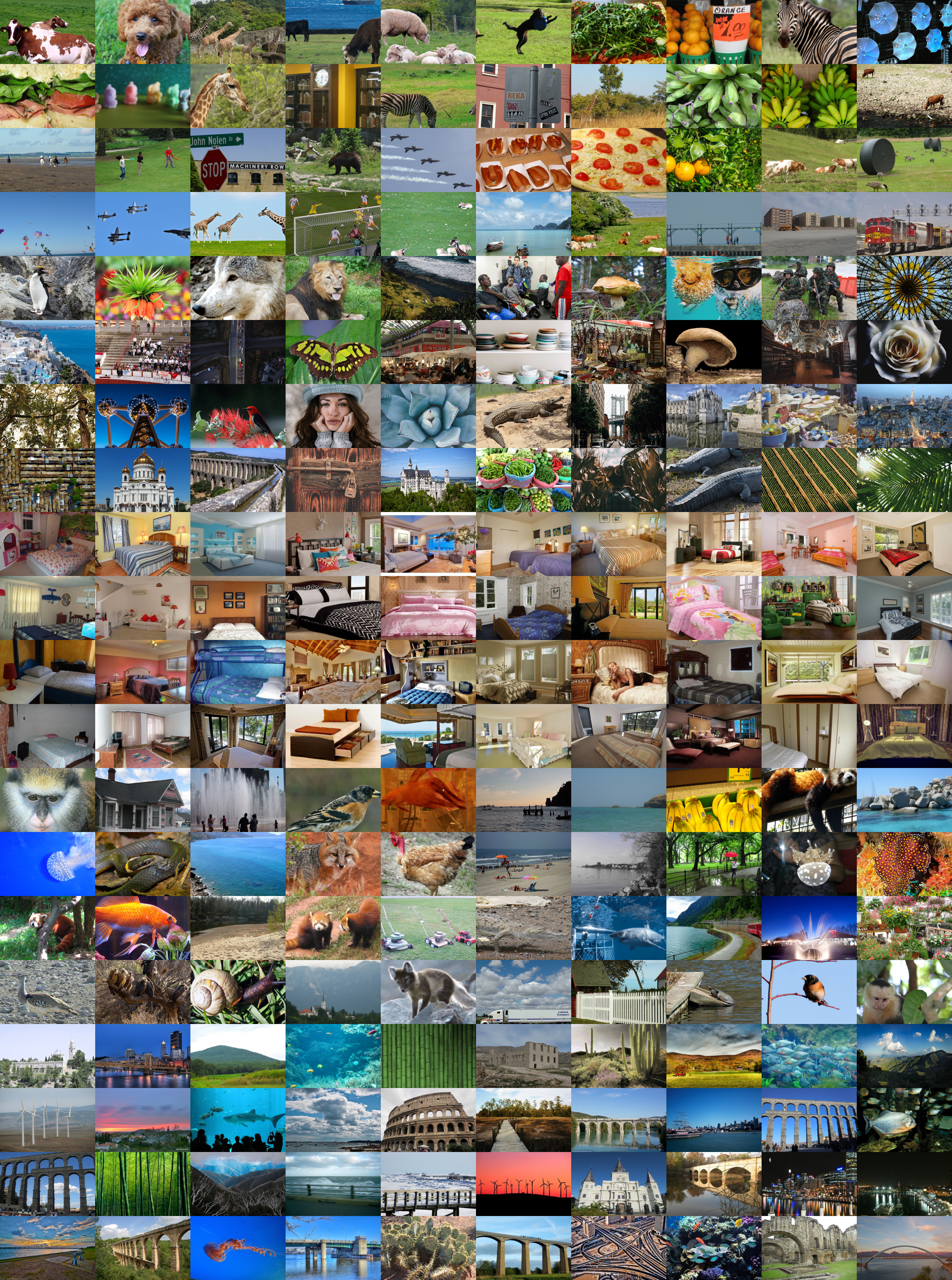}
    \caption{$200$ secret images used for quantitative comparison. The two co-located cover and secret images form one pair.}
    \label{fig:sec}
\end{figure}

The $200$ test image pairs used in our quantitative experiments are randomly drawn from five popular datasets (COCO\cite{lin2014microsoft}, DIV2K\cite{agustsson2017ntire}, LSUN bedroom\cite{yu2015lsun}, ImageNet\cite{deng2009image}, Places\cite{zhou2018places}). Specifically, we sample $80$ images from each dataset to obtain a total of $400$ images, which are randomly partitioned into $200$ cover images and $200$ secret images, as shown in Figs.~\ref{fig:cov} and~\ref{fig:sec} respectively, with co-located images forming one pair.

\clearpage

\section{Histograms of Weight Distribution}
\captionsetup{position=bottom}
\captionsetup[subfloat]{labelformat=parens}
\subsection{Total Weight Distribution}
\begin{figure}[h!]

    \centering

    \subfloat[No secret image (original).]{
	\includegraphics[width=0.5\textwidth]{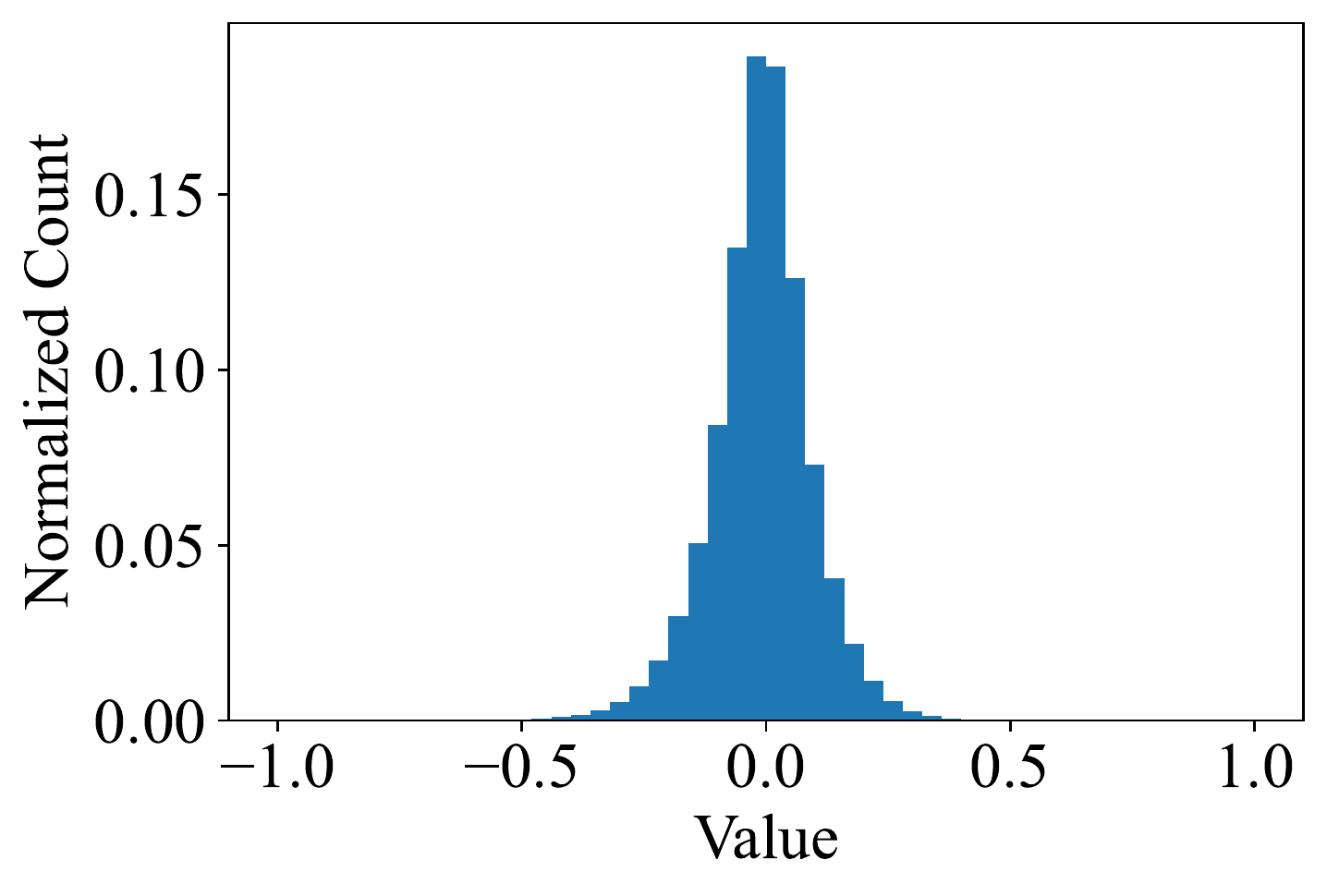}
	}
	
	\subfloat[One secret image.]{
		\includegraphics[width=0.5\textwidth]{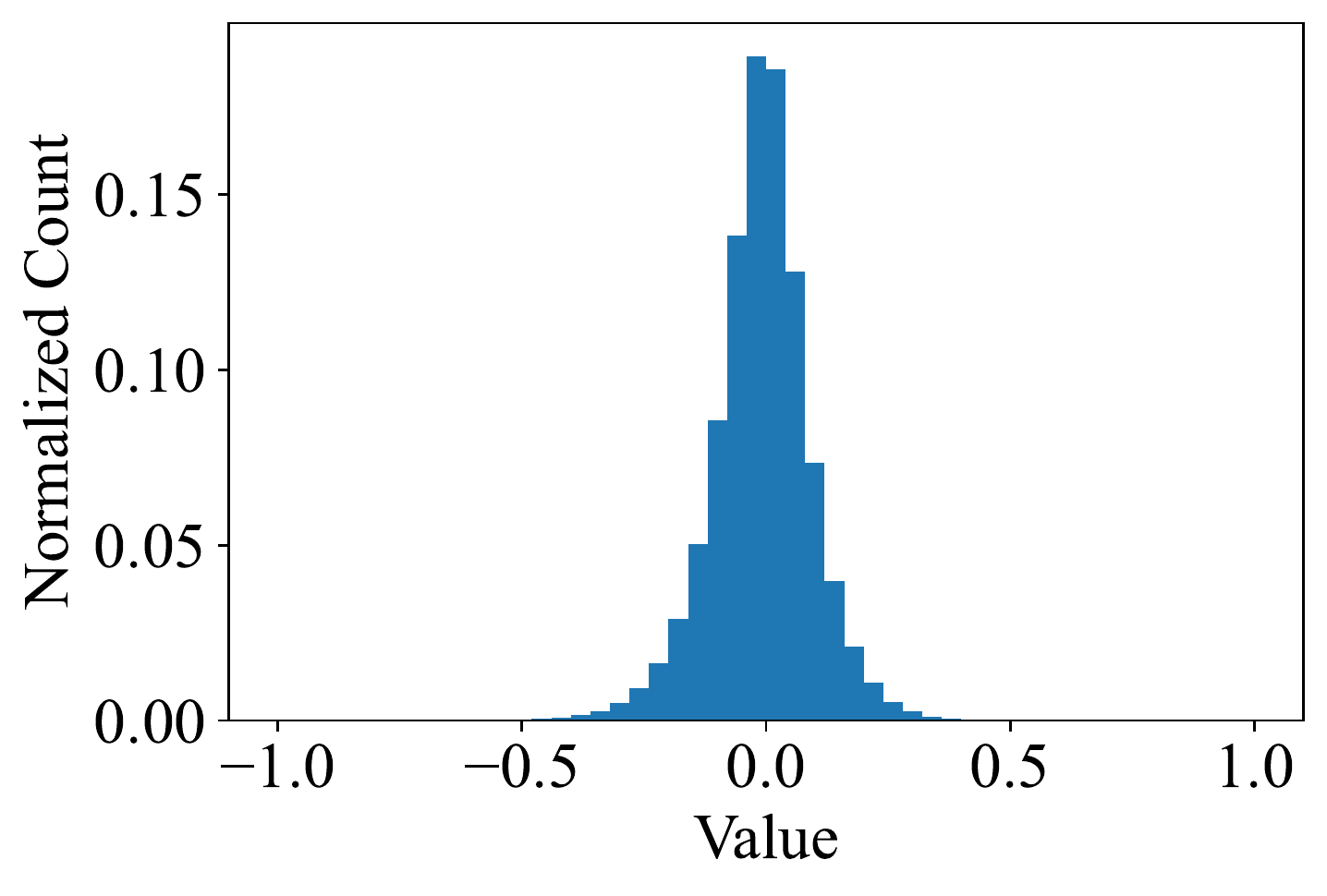}
	}
	\subfloat[Two secret images.]{
		\includegraphics[width=0.5\textwidth]{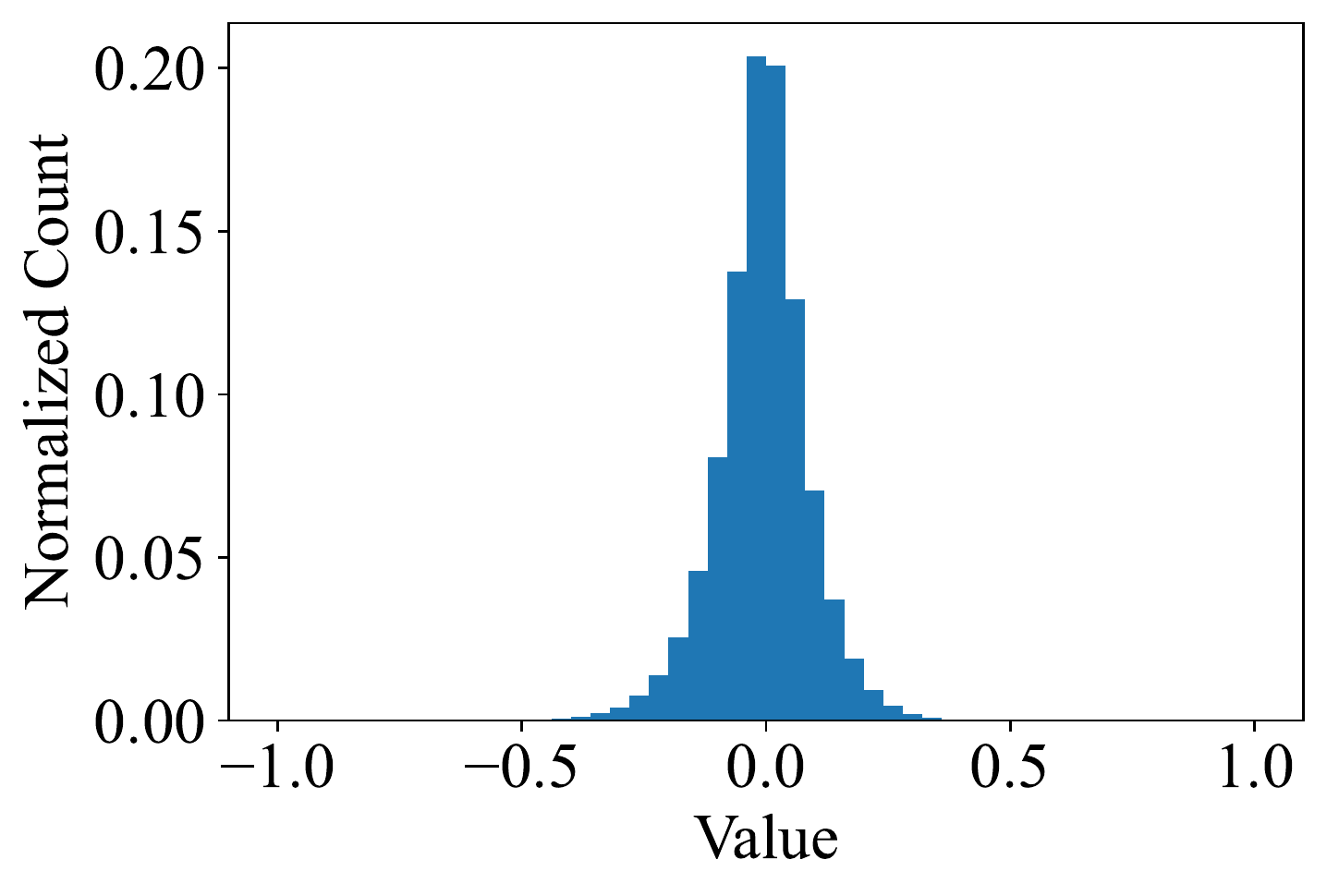}
	}
	
	\subfloat[Three secret images.]{
	    \centering
		\includegraphics[width=0.5\textwidth]{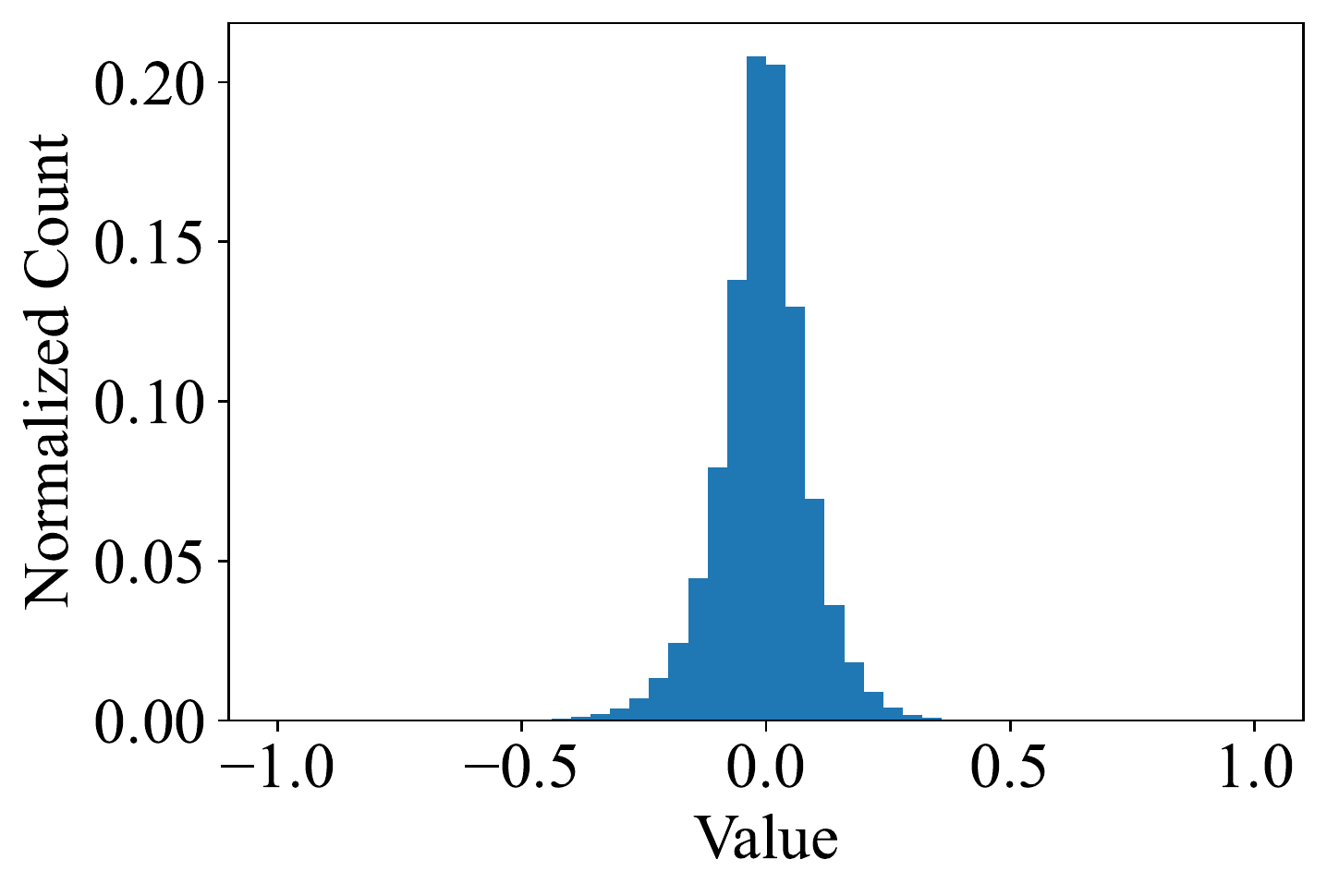}
	}
	\subfloat[Four secret images.]{
		\includegraphics[width=0.5\textwidth]{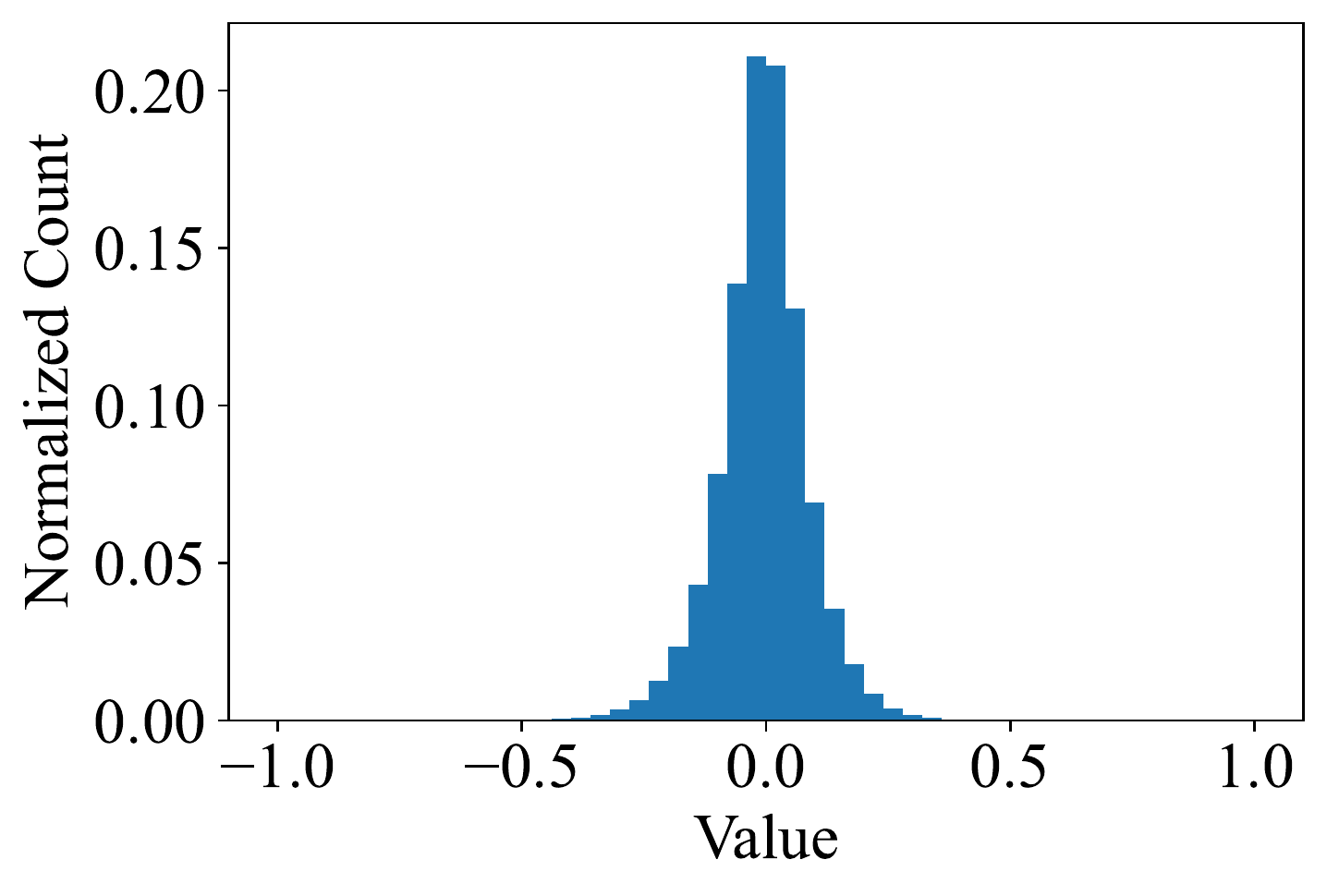}
	}
	
	\caption{Visual comparison of histograms of the total weights.}
	\label{fig:ttw}
\end{figure}

Fig~\ref{fig:ttw} shows the histograms of all weights from $200$ original and stego SinGAN generators. Careful visual inspection shows that the empirical weight distributions of the original and stego SinGAN generators are identical even when we hide up to four images.

\subsection{Per-Stage Weight Distribution}
\begin{figure}[!h]

    \centering
    
    \subfloat[No secret image (original).]{
	\includegraphics[width=0.5\textwidth]{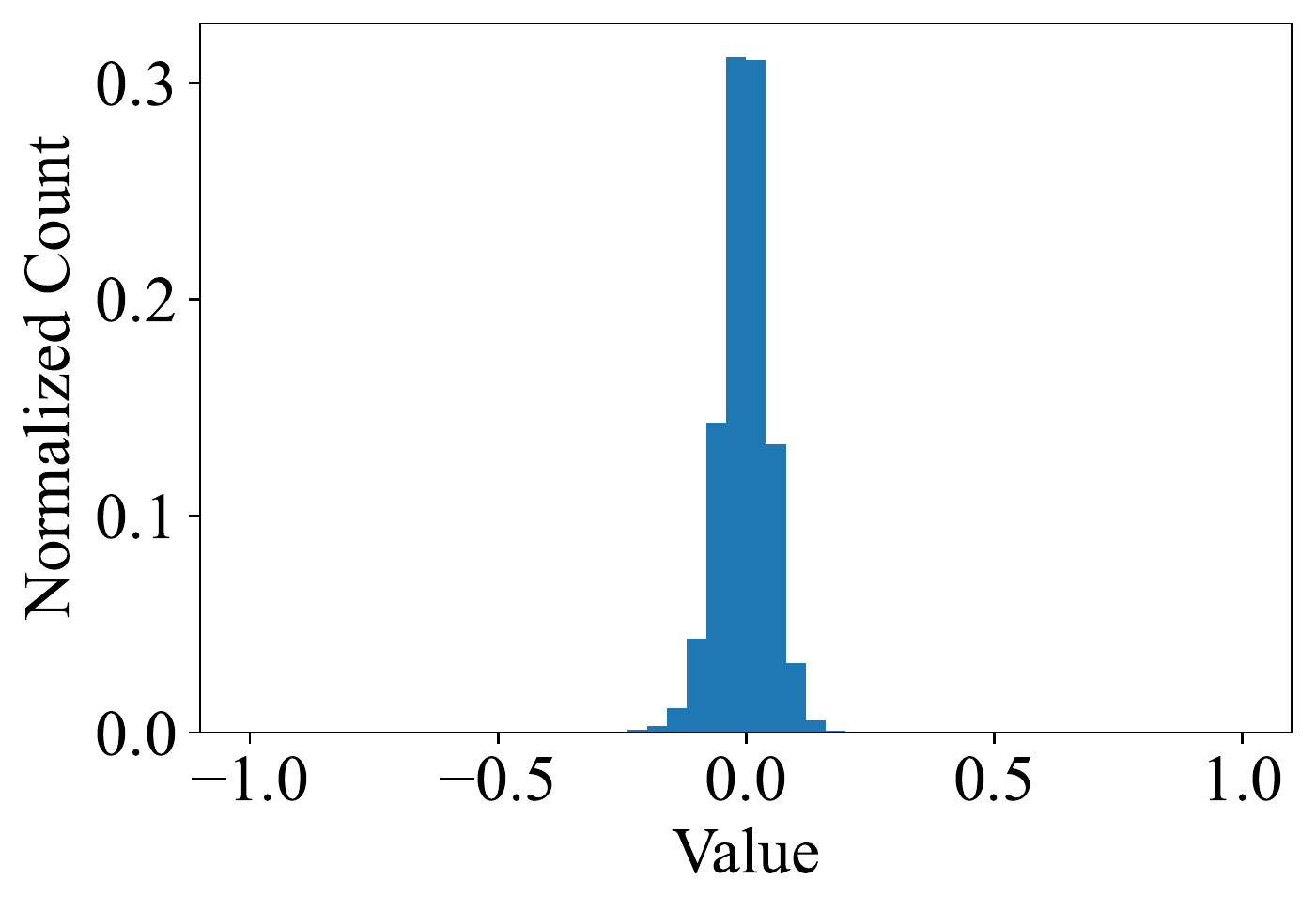}
	} 
	
	\subfloat[One secret image.]{
		\includegraphics[width=0.5\textwidth]{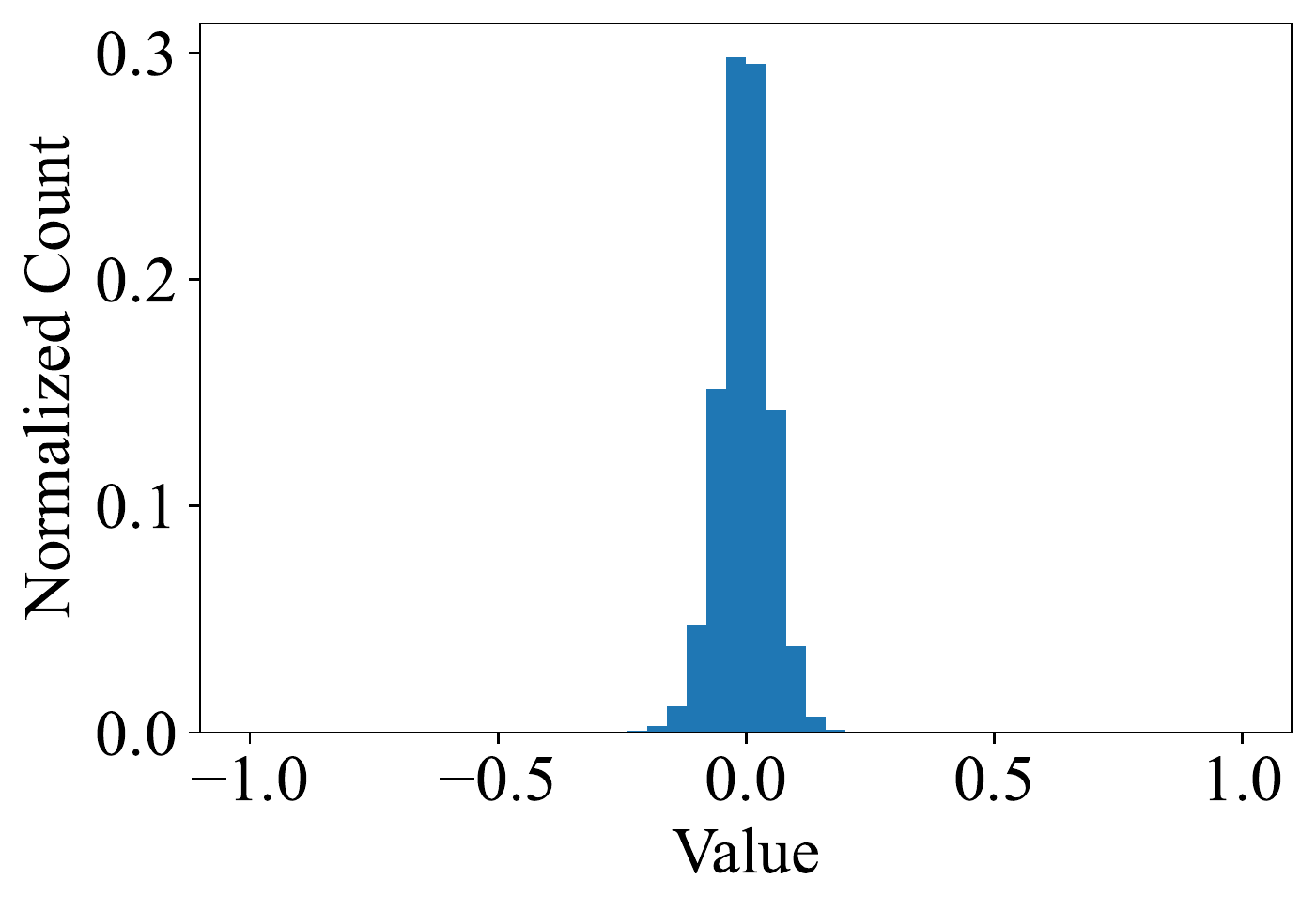}
	}
	\subfloat[Two secret images.]{
		\includegraphics[width=0.5\textwidth]{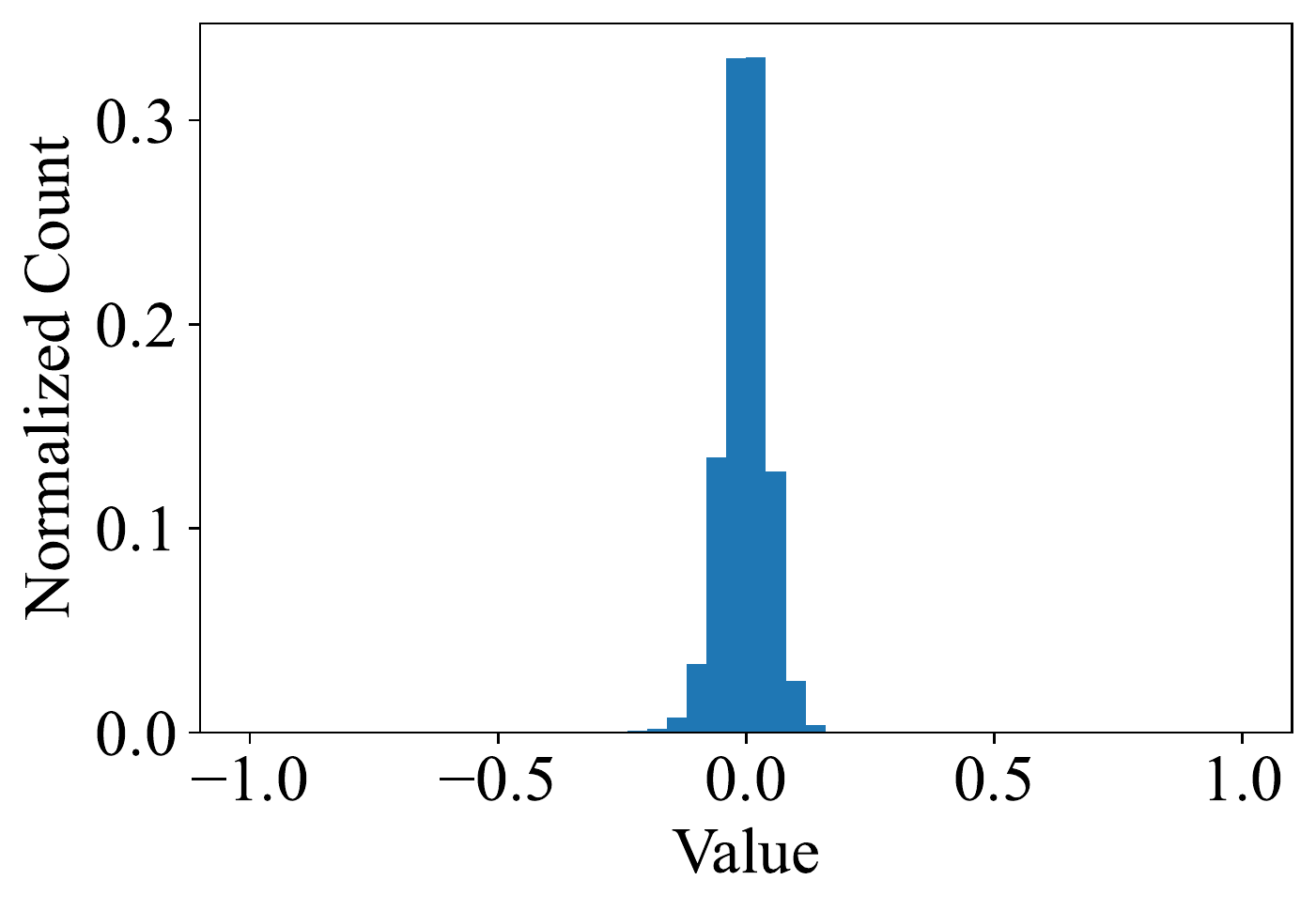}
	}
	
	\subfloat[Three secret images.]{
	    \centering
		\includegraphics[width=0.5\textwidth]{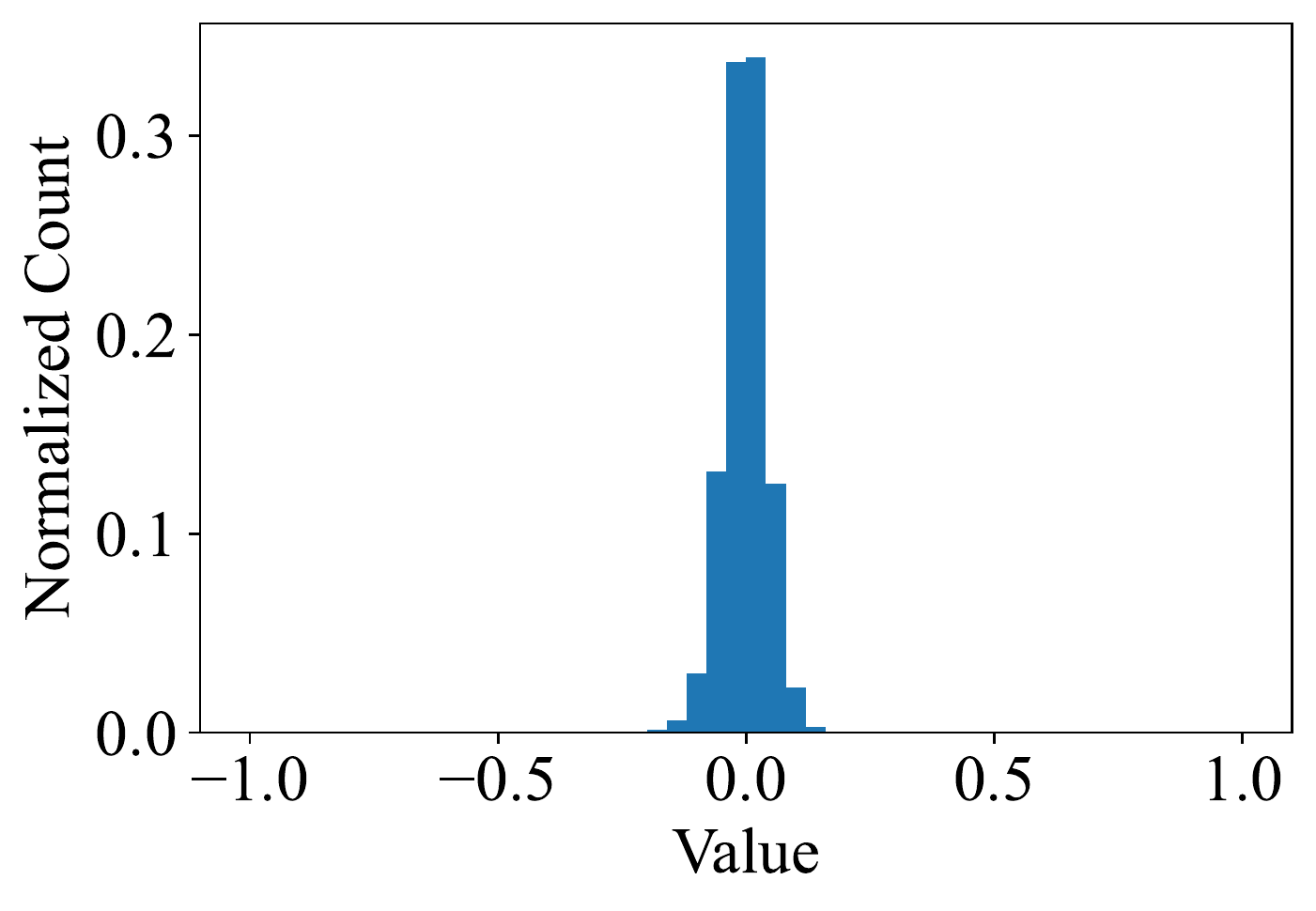}
	}
	\subfloat[Four secret images.]{
		\includegraphics[width=0.5\textwidth]{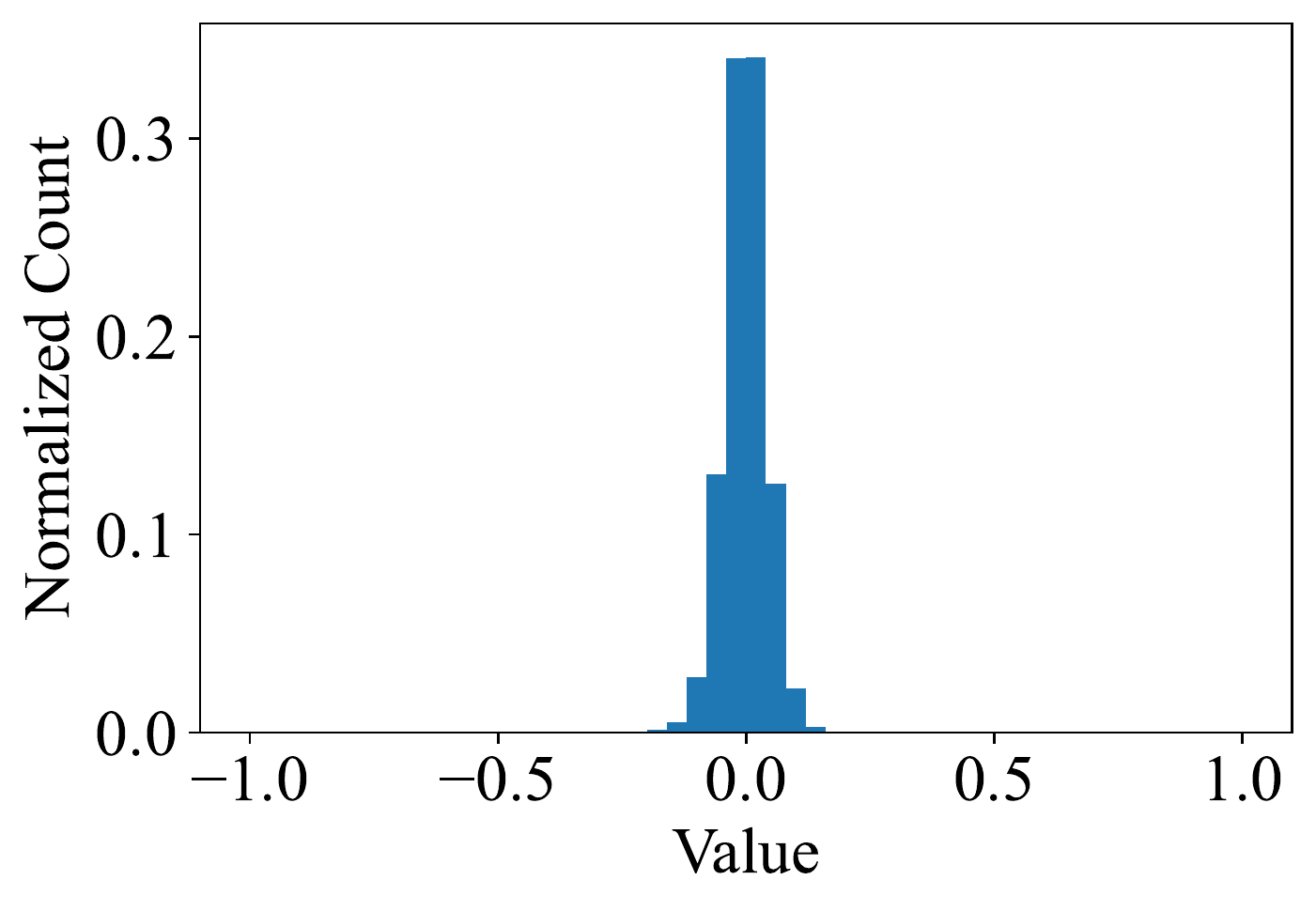}
	}
	
	\caption{Visual comparison of histograms of the first-stage weights.}
	\label{fig:s1}
\end{figure}

\begin{figure}[!h]

    \centering

    \subfloat[No secret image (original).]{
	\includegraphics[width=0.5\textwidth]{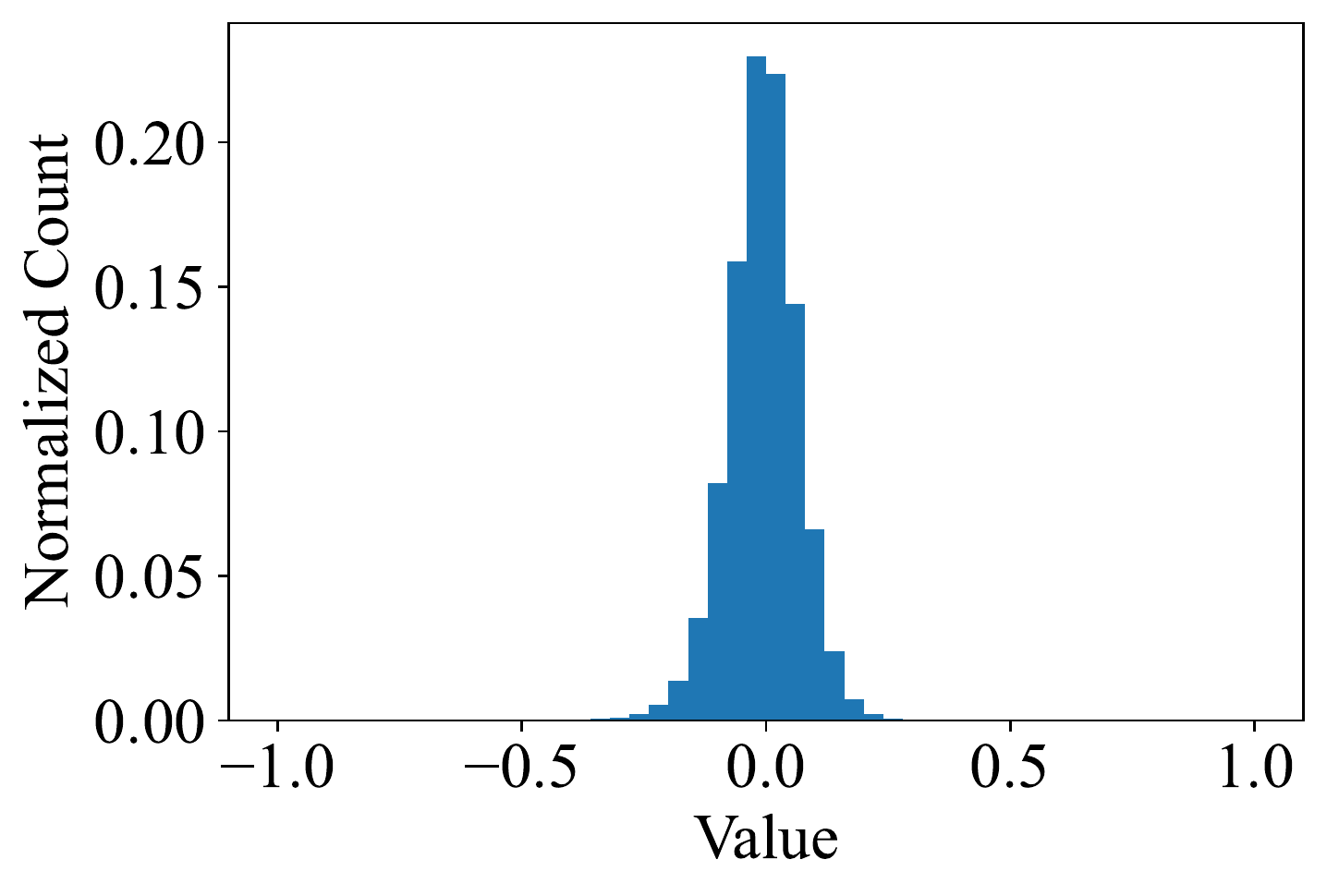}
	}
	
	\subfloat[One secret image.]{
		\includegraphics[width=0.5\textwidth]{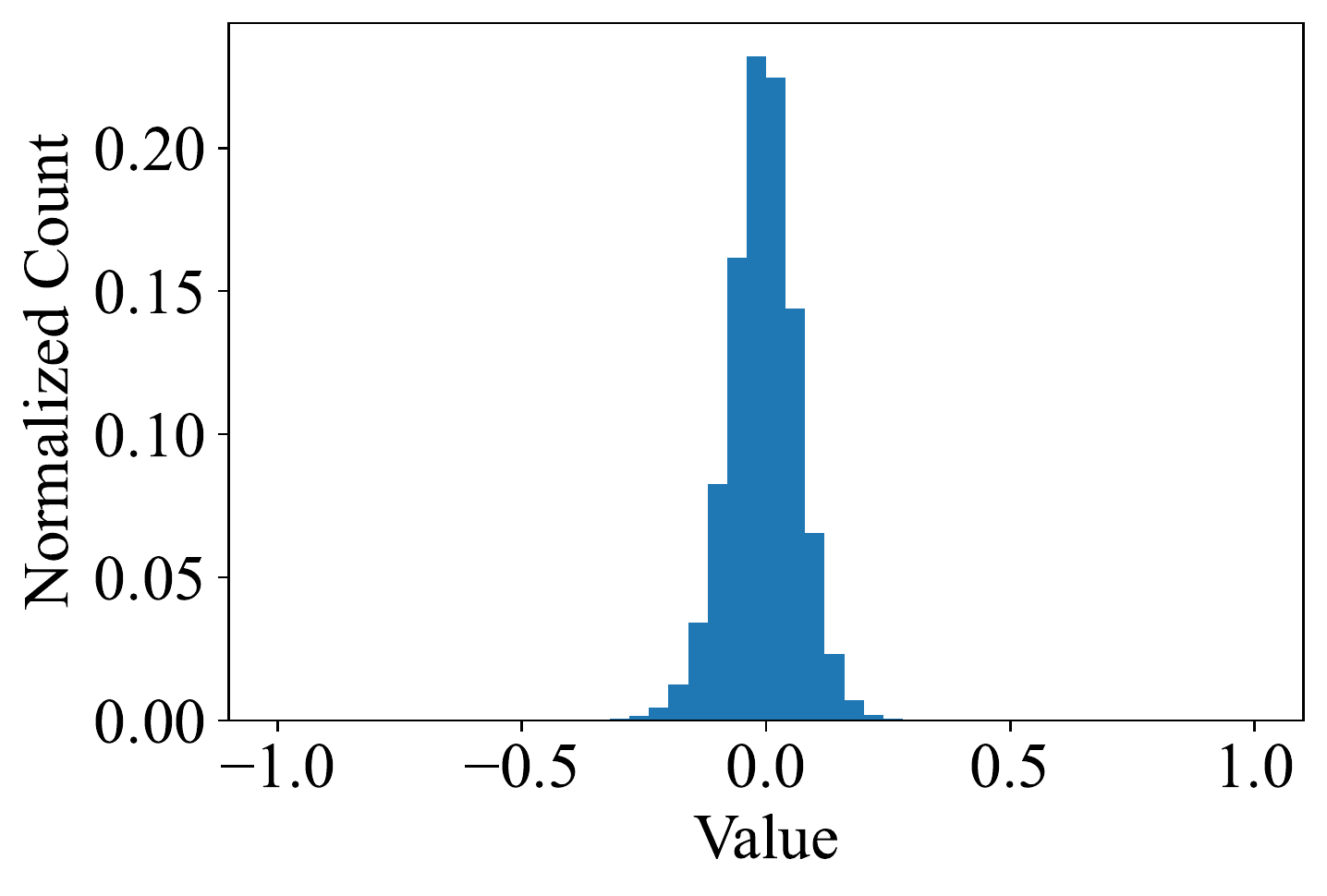}
	}
	\subfloat[Two secret images.]{
		\includegraphics[width=0.5\textwidth]{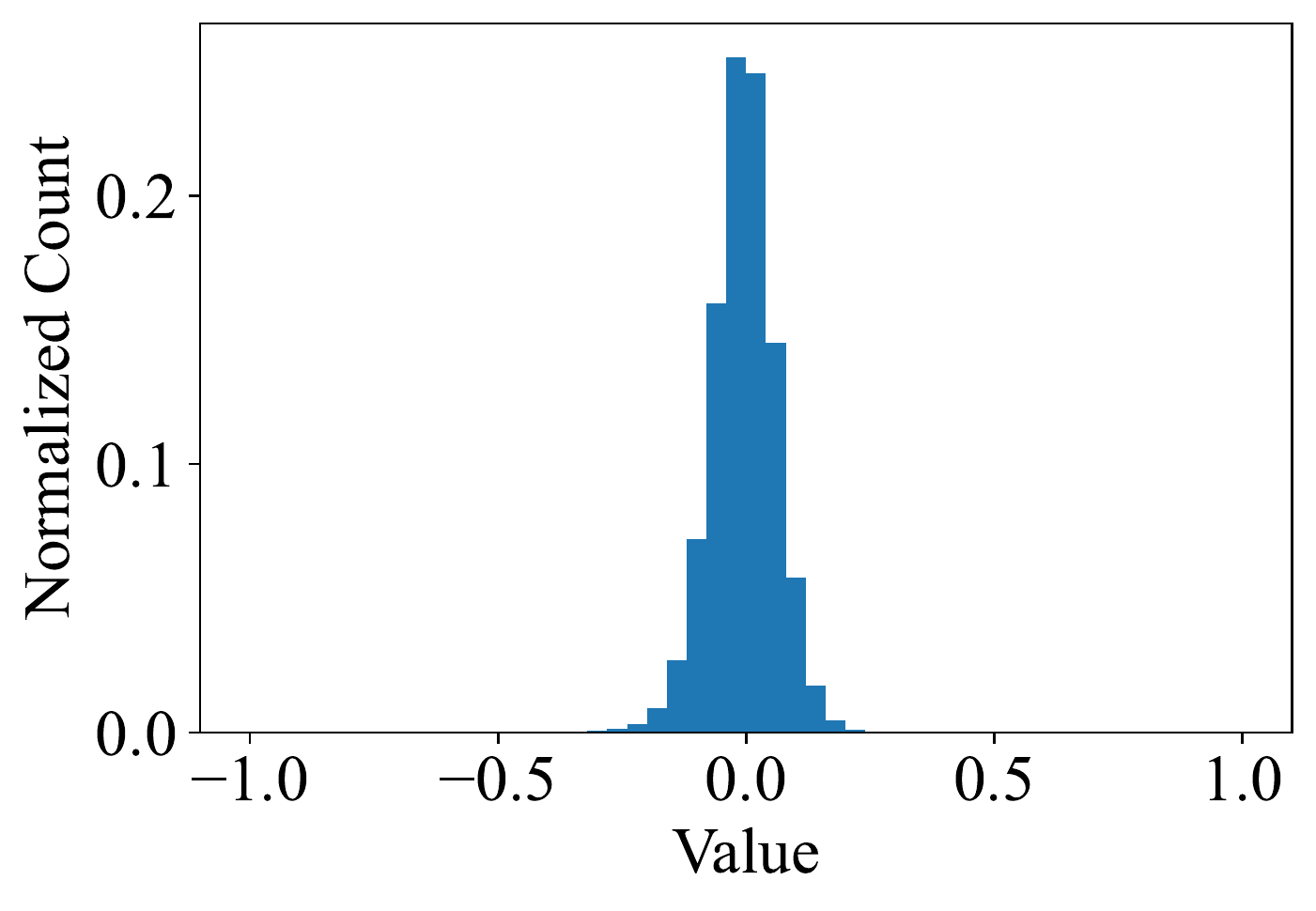}
	}
	
	\subfloat[Three secret images.]{
	    \centering
		\includegraphics[width=0.5\textwidth]{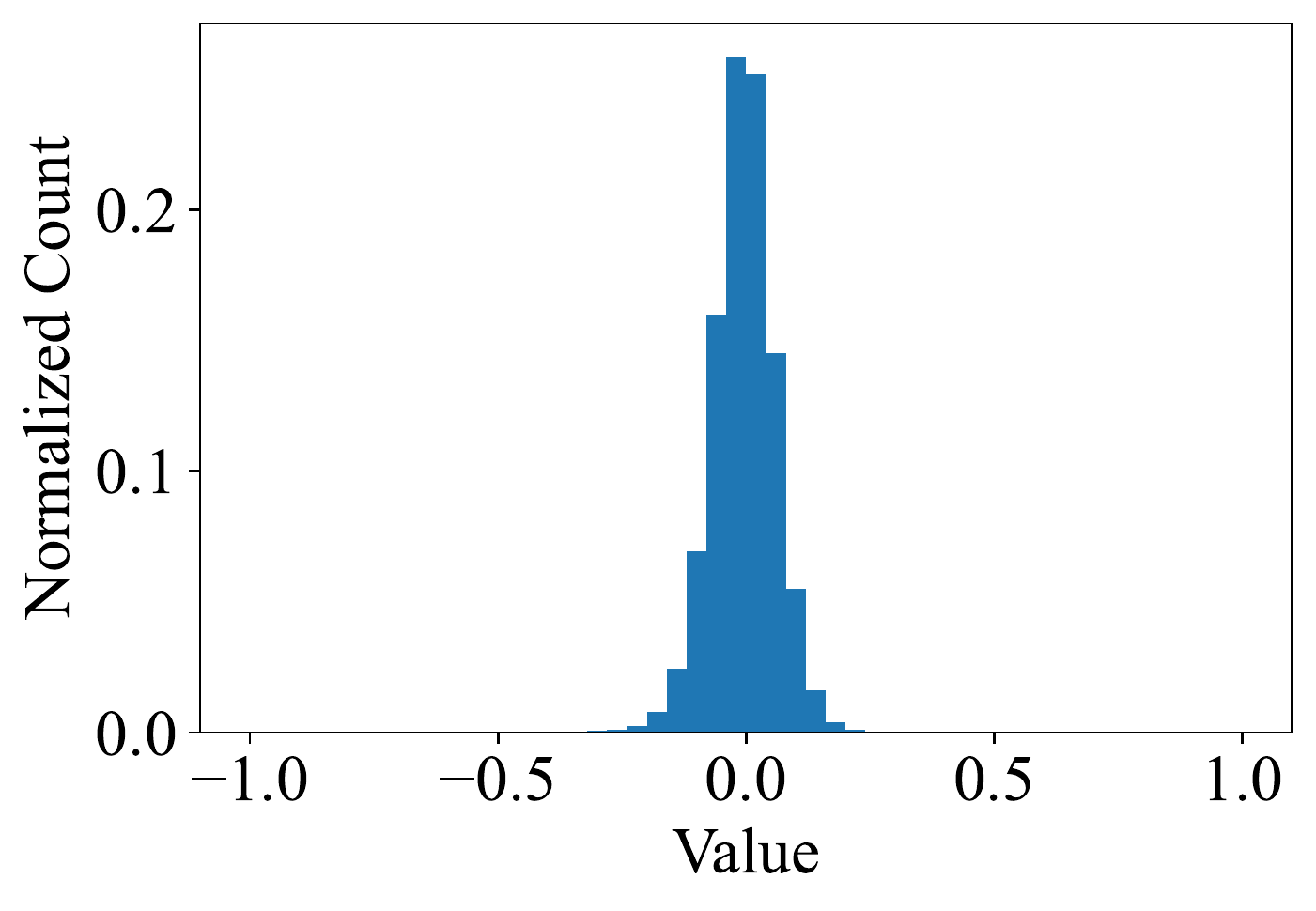}
	}
	\subfloat[Four secret images.]{
		\includegraphics[width=0.5\textwidth]{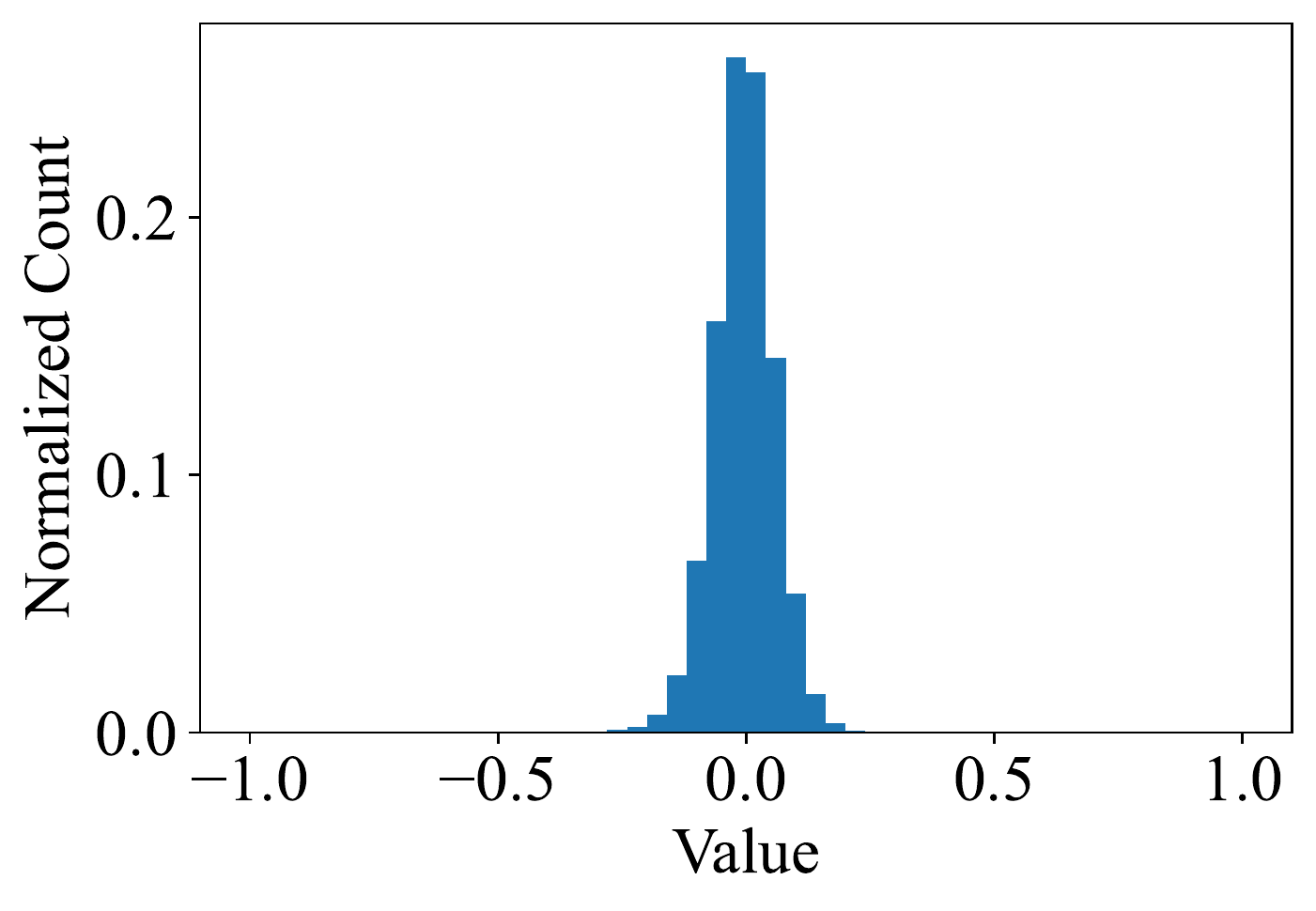}
	}
	
	\caption{Visual comparison of histograms of the second-stage weights.}
	\label{fig:s2}
\end{figure}

\begin{figure}[!h]

    \centering

    \subfloat[No secret image (original).]{
	\includegraphics[width=0.5\textwidth]{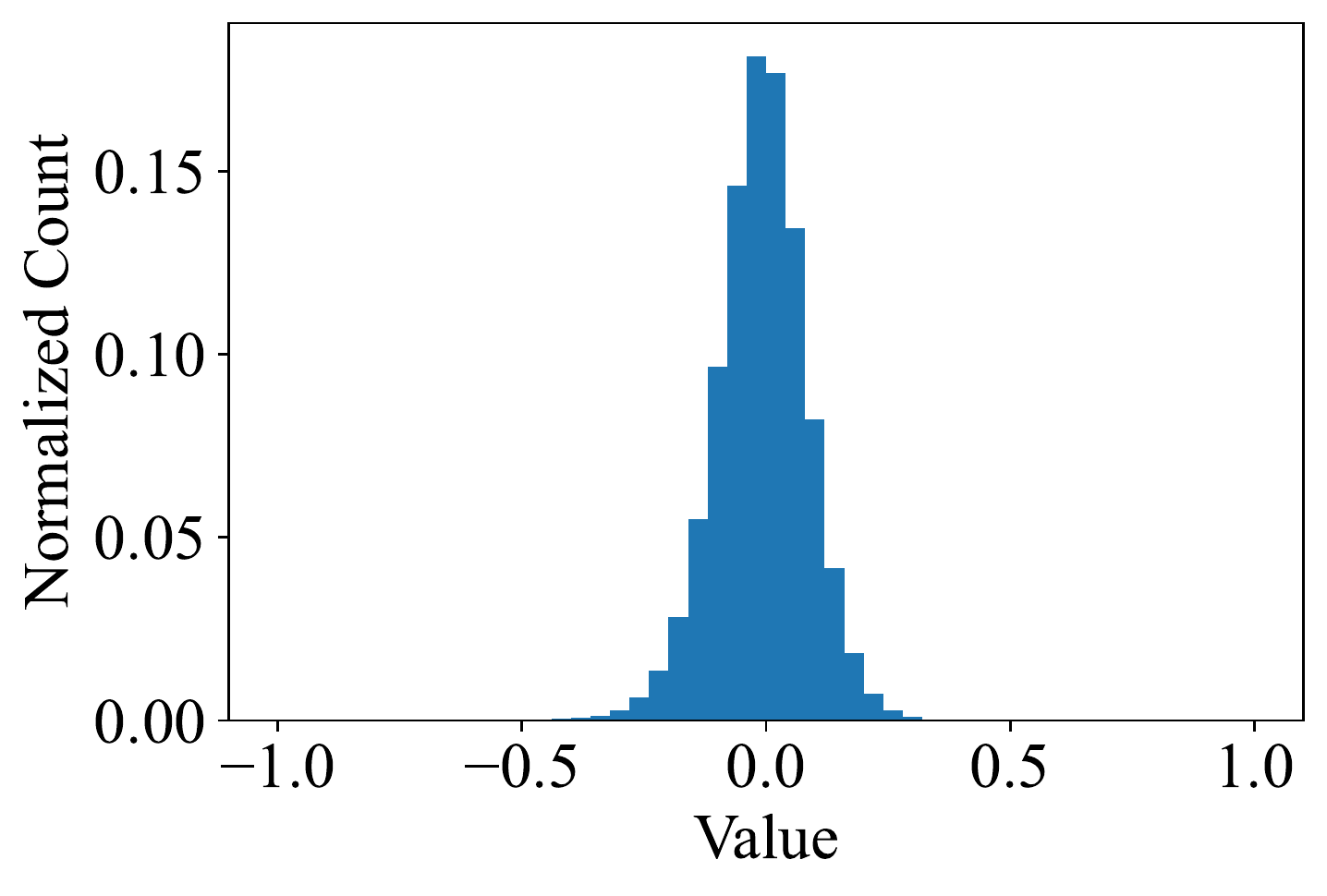}
	}
	
	\subfloat[One secret image.]{
		\includegraphics[width=0.5\textwidth]{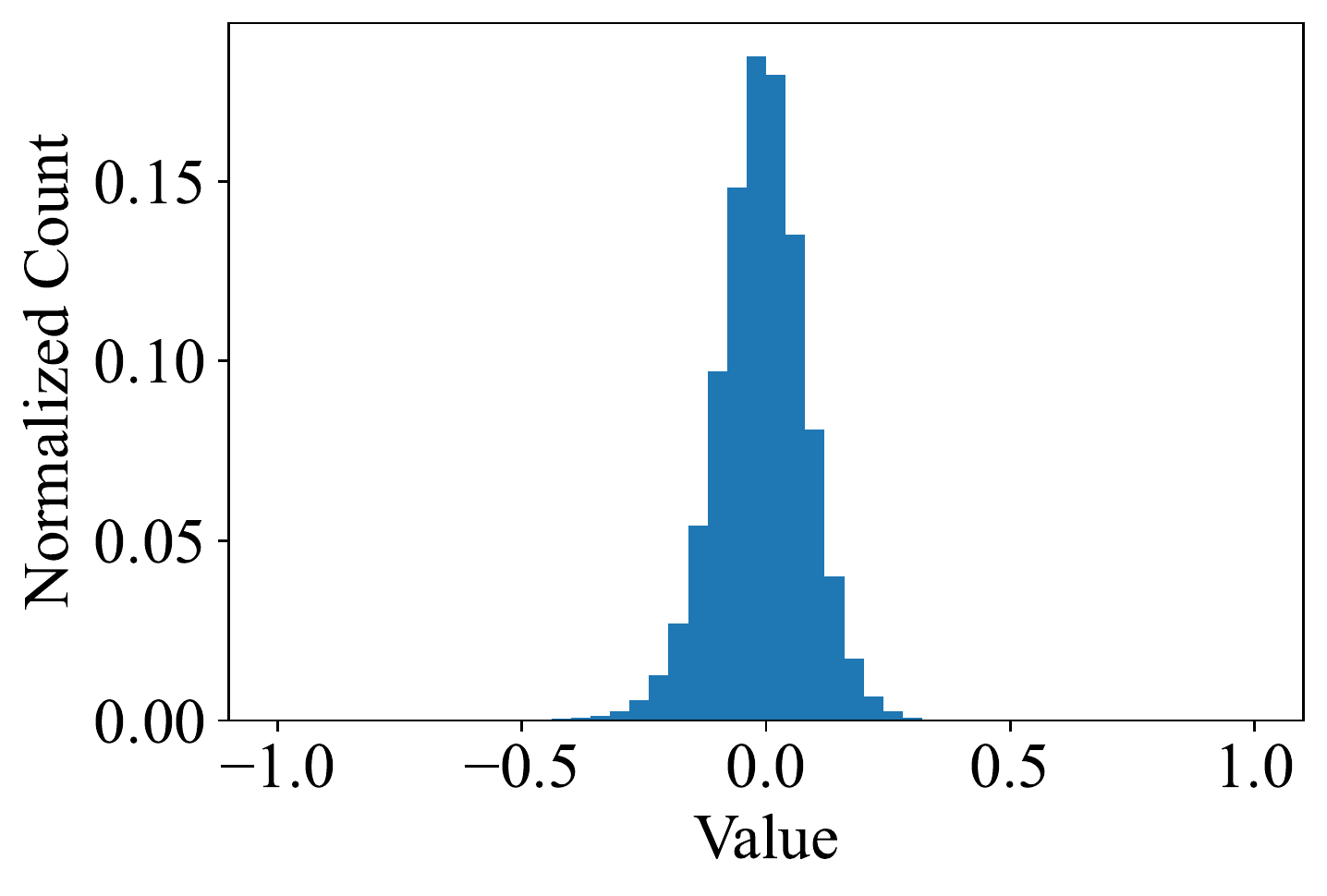}
	}
	\subfloat[Two secret images.]{
		\includegraphics[width=0.5\textwidth]{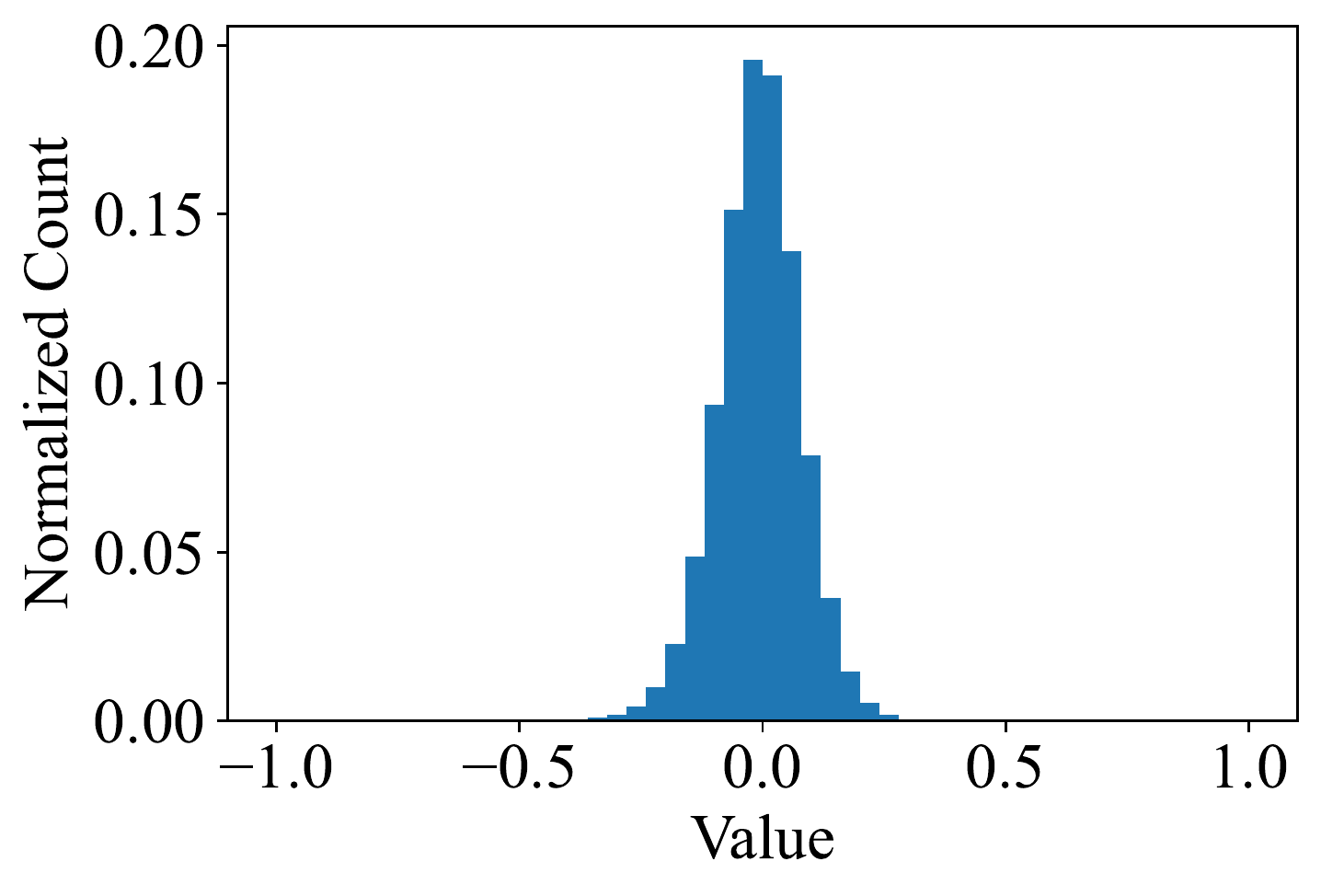}
	}
	
	\subfloat[Three secret images.]{
	    \centering
		\includegraphics[width=0.5\textwidth]{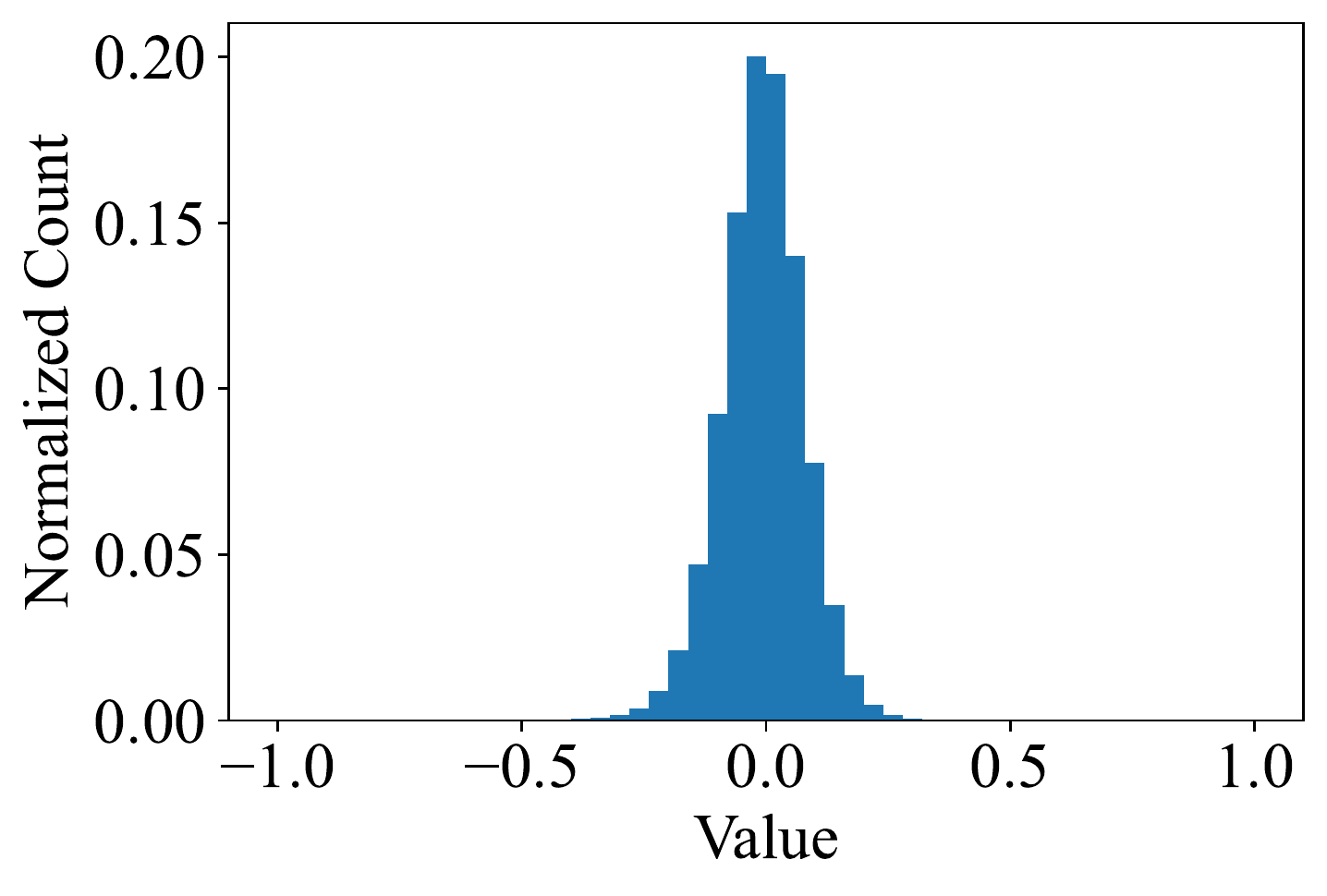}
	}
	\subfloat[Four secret images.]{
		\includegraphics[width=0.5\textwidth]{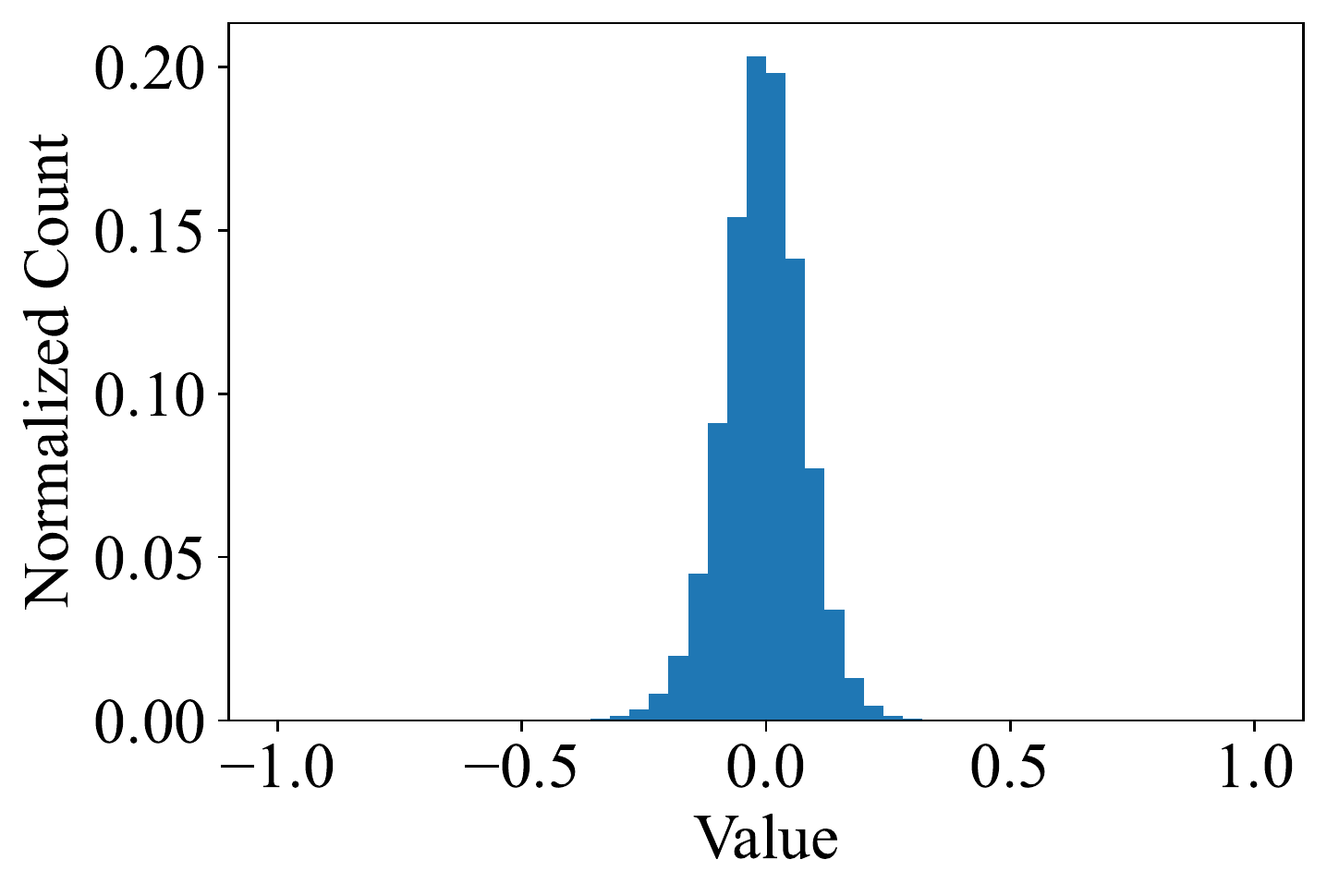}
	}

	\caption{Visual comparison of histograms of the third-stage weights.}
	\label{fig:s3}
\end{figure}

\begin{figure}[!h]
    \centering

    \subfloat[No secret image (original).]{
	\includegraphics[width=0.5\textwidth]{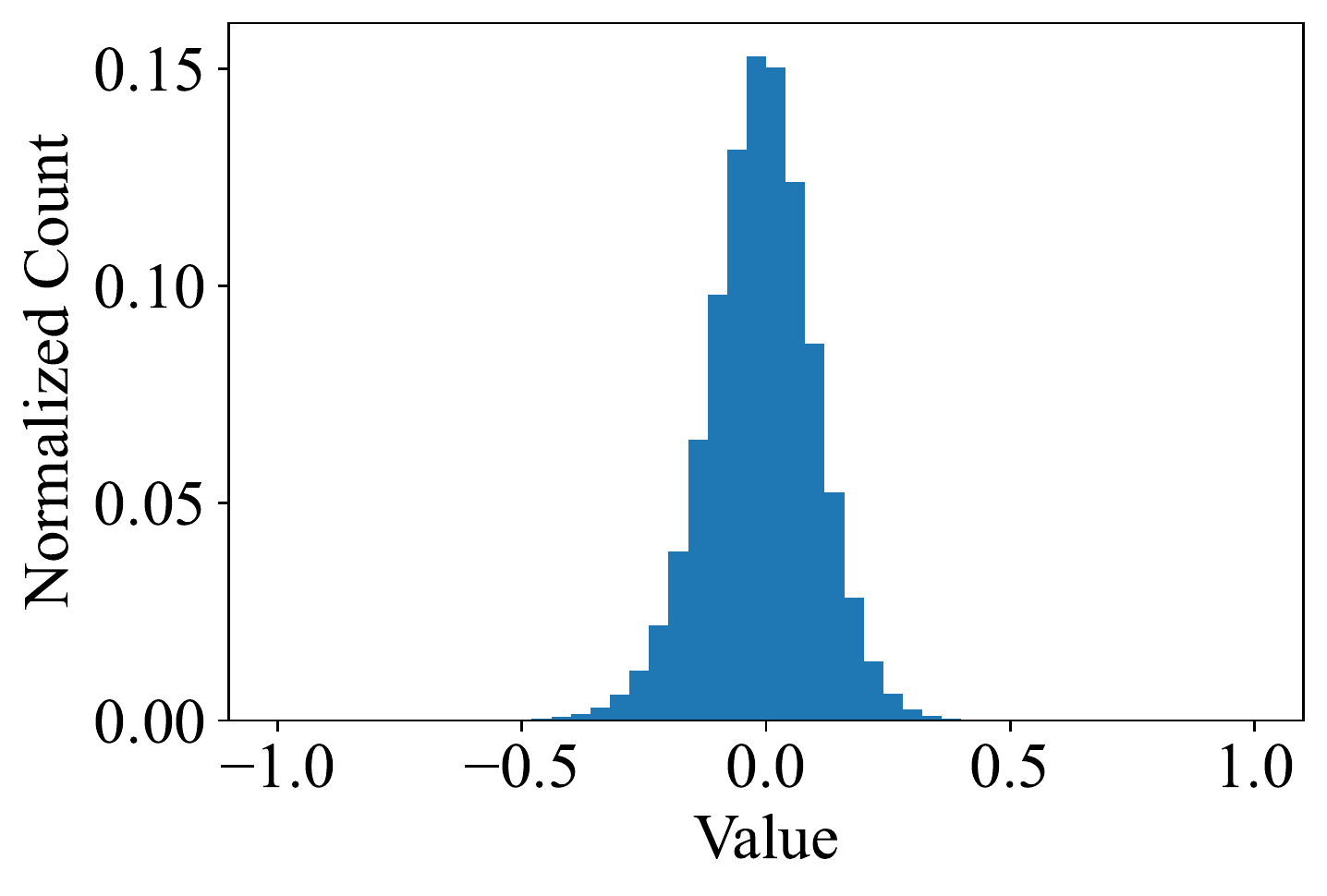}
	}
	
	\subfloat[One secret image.]{
		\includegraphics[width=0.5\textwidth]{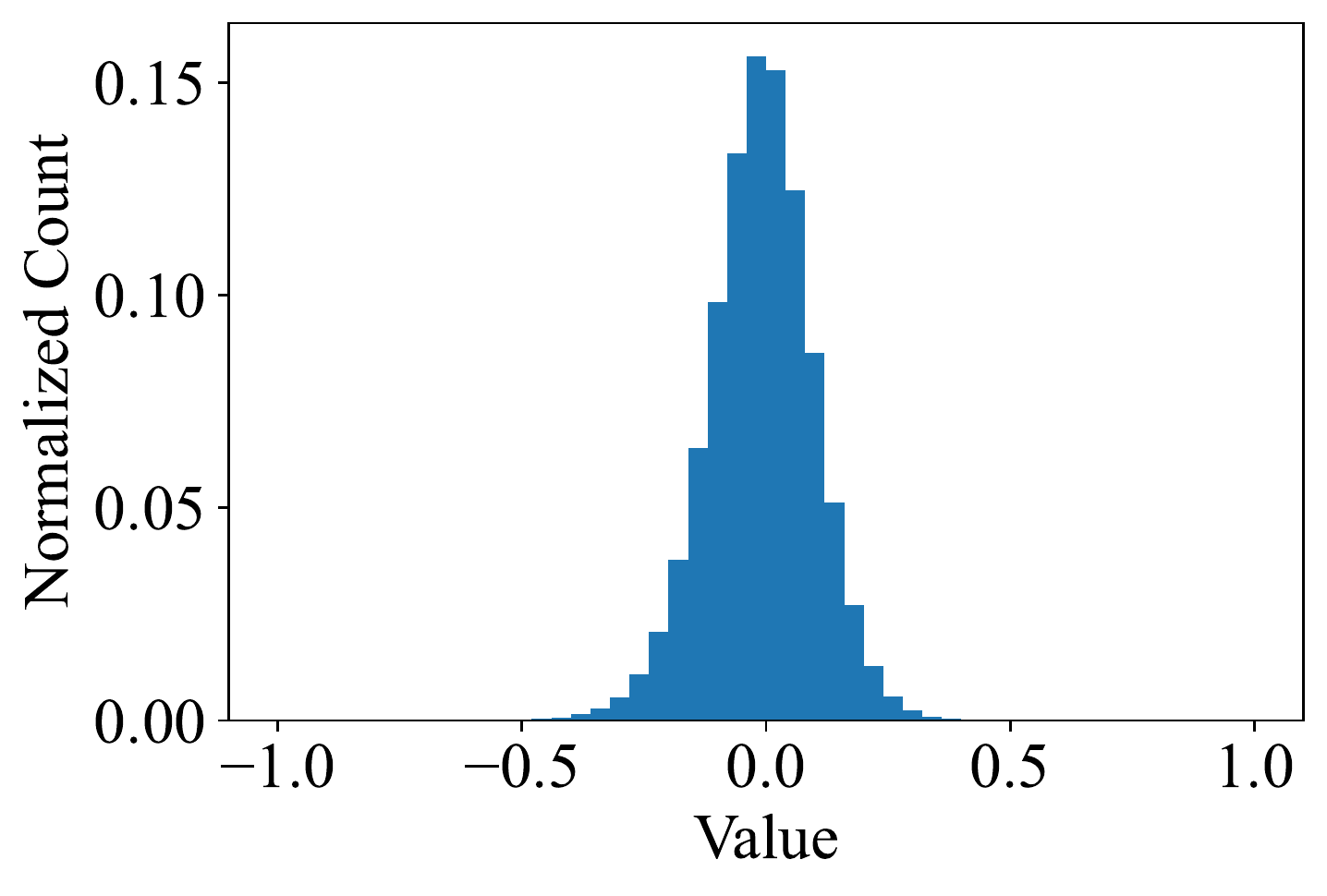}
	}
	\subfloat[Two secret images.]{
		\includegraphics[width=0.5\textwidth]{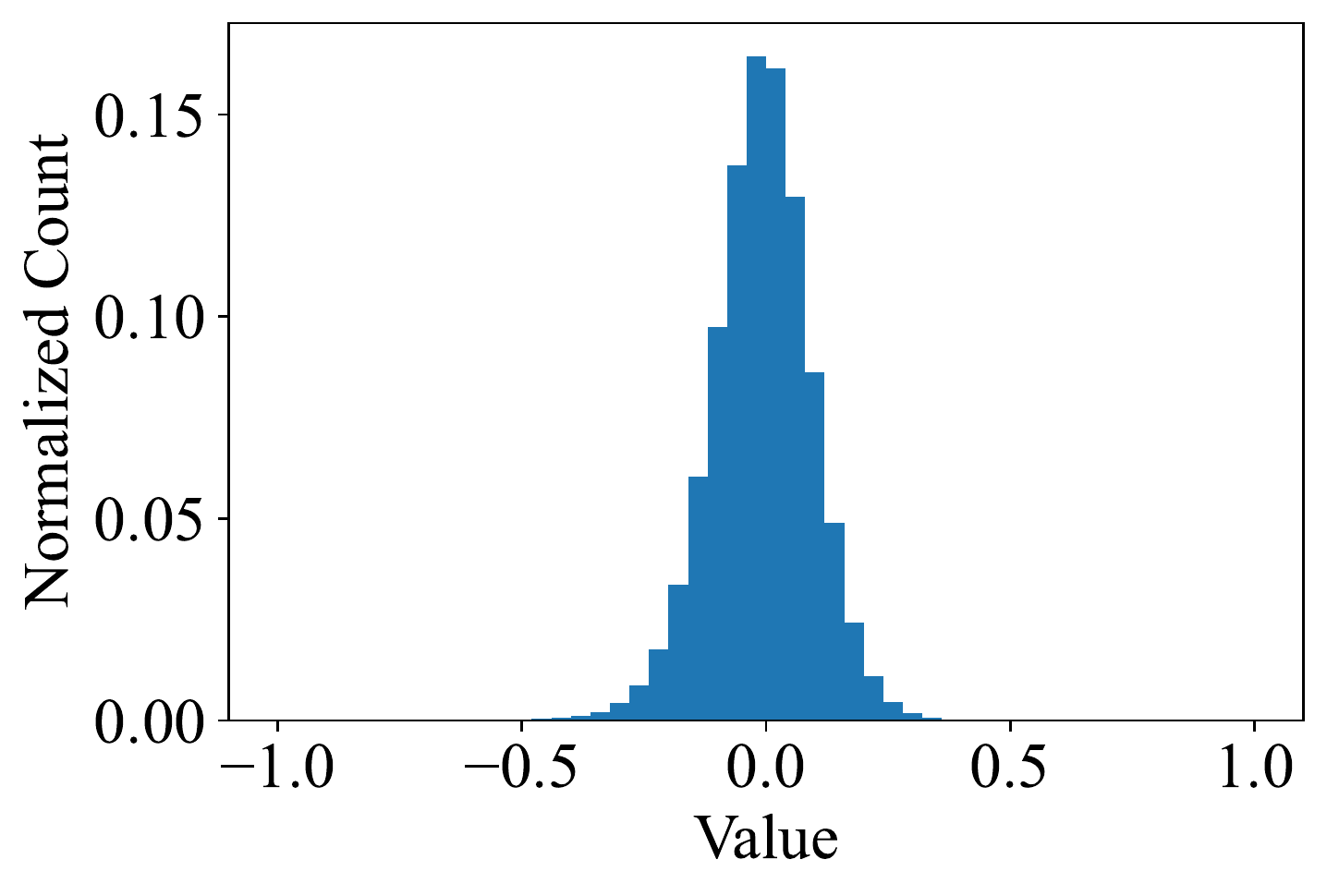}
	}
	
	\subfloat[Three secret images.]{
	    \centering
		\includegraphics[width=0.5\textwidth]{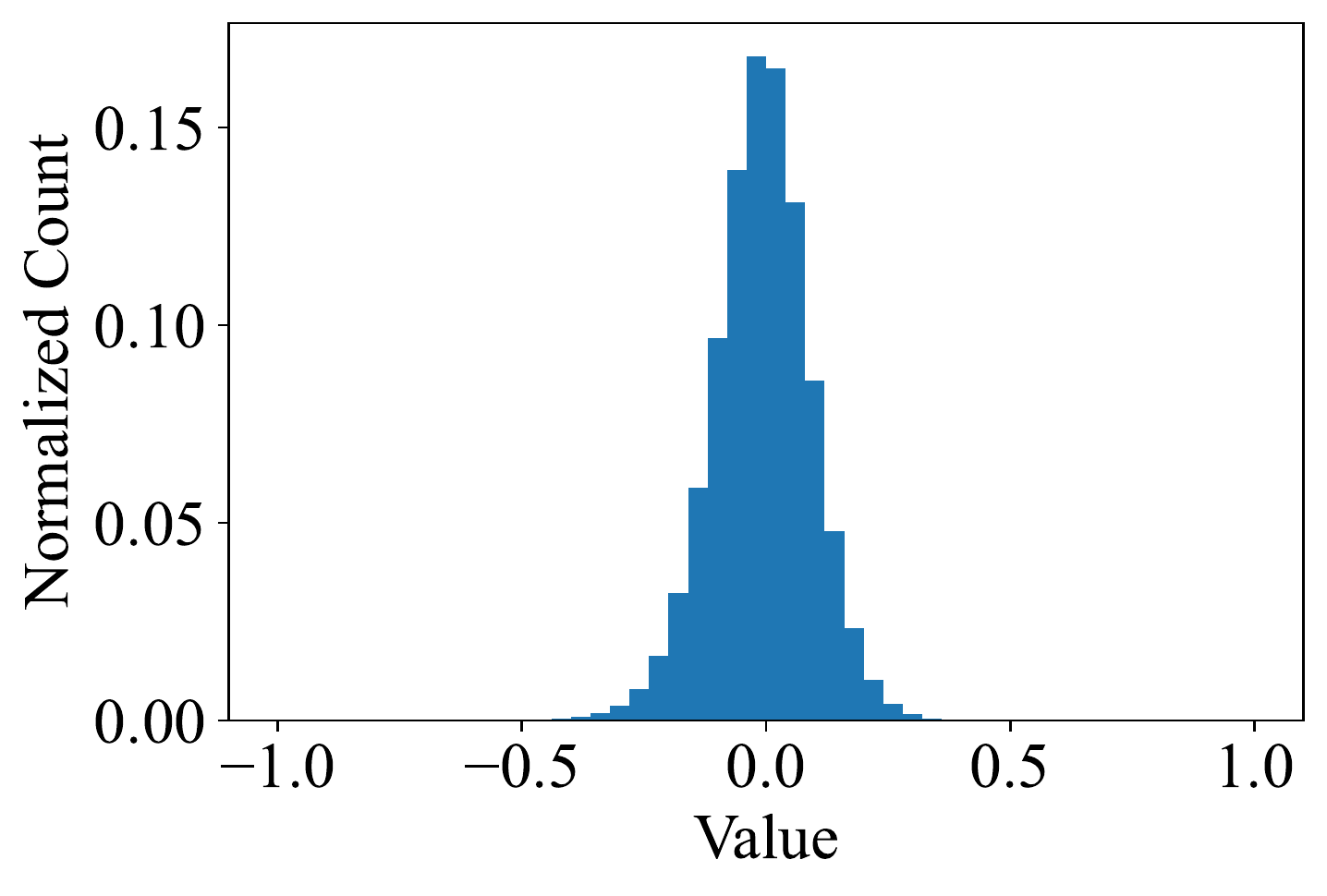}
	}
	\subfloat[Four secret images.]{
		\includegraphics[width=0.5\textwidth]{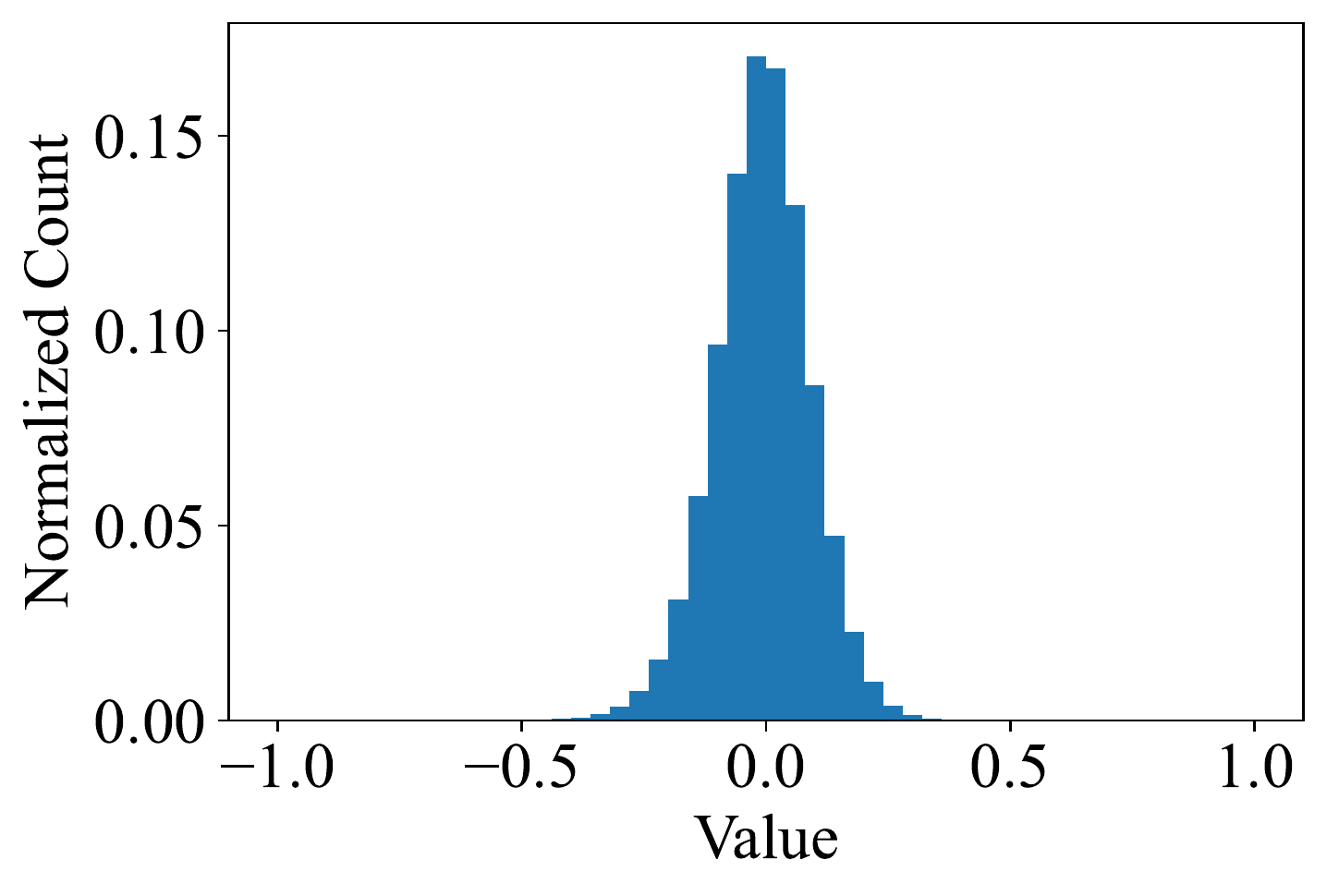}
	}
	
	\caption{Visual comparison of histograms of the fourth-stage weights.}
	\label{fig:s4}
\end{figure}

\begin{figure}[!h]

    \centering

    \subfloat[No secret image (original).]{
	\includegraphics[width=0.5\textwidth]{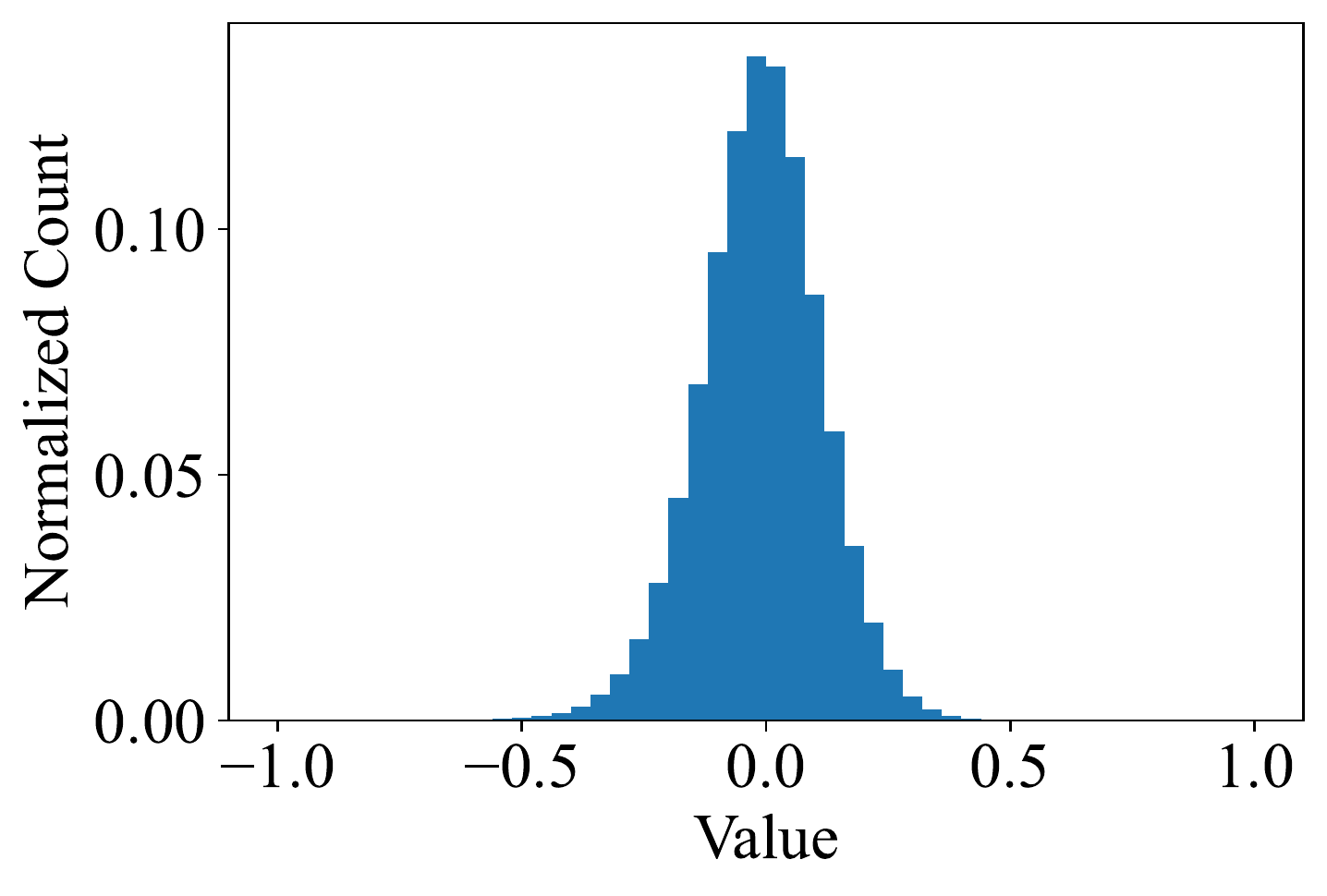}
	}
	
	\subfloat[One secret image.]{
		\includegraphics[width=0.5\textwidth]{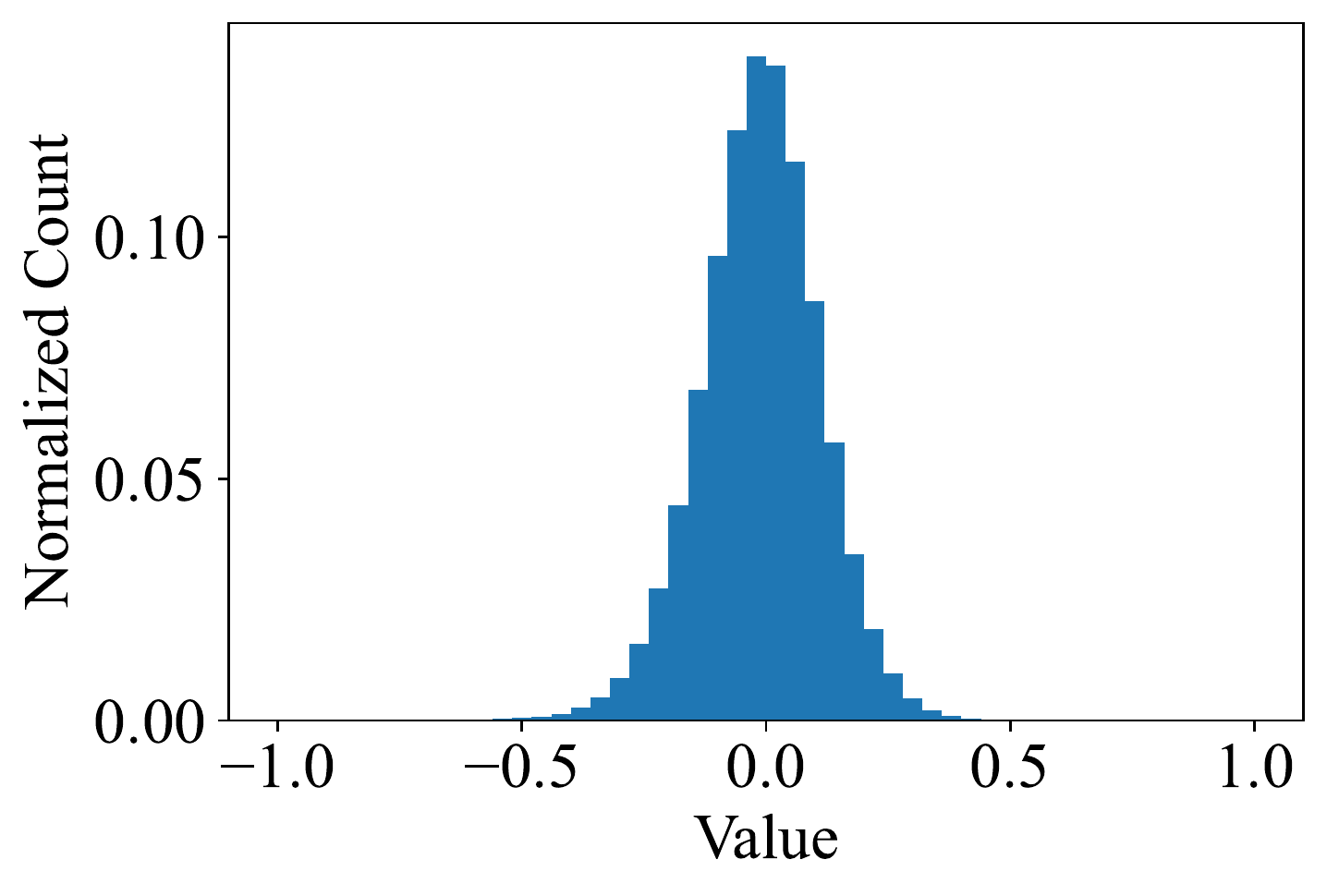}
	}
	\subfloat[Two secret images.]{
		\includegraphics[width=0.5\textwidth]{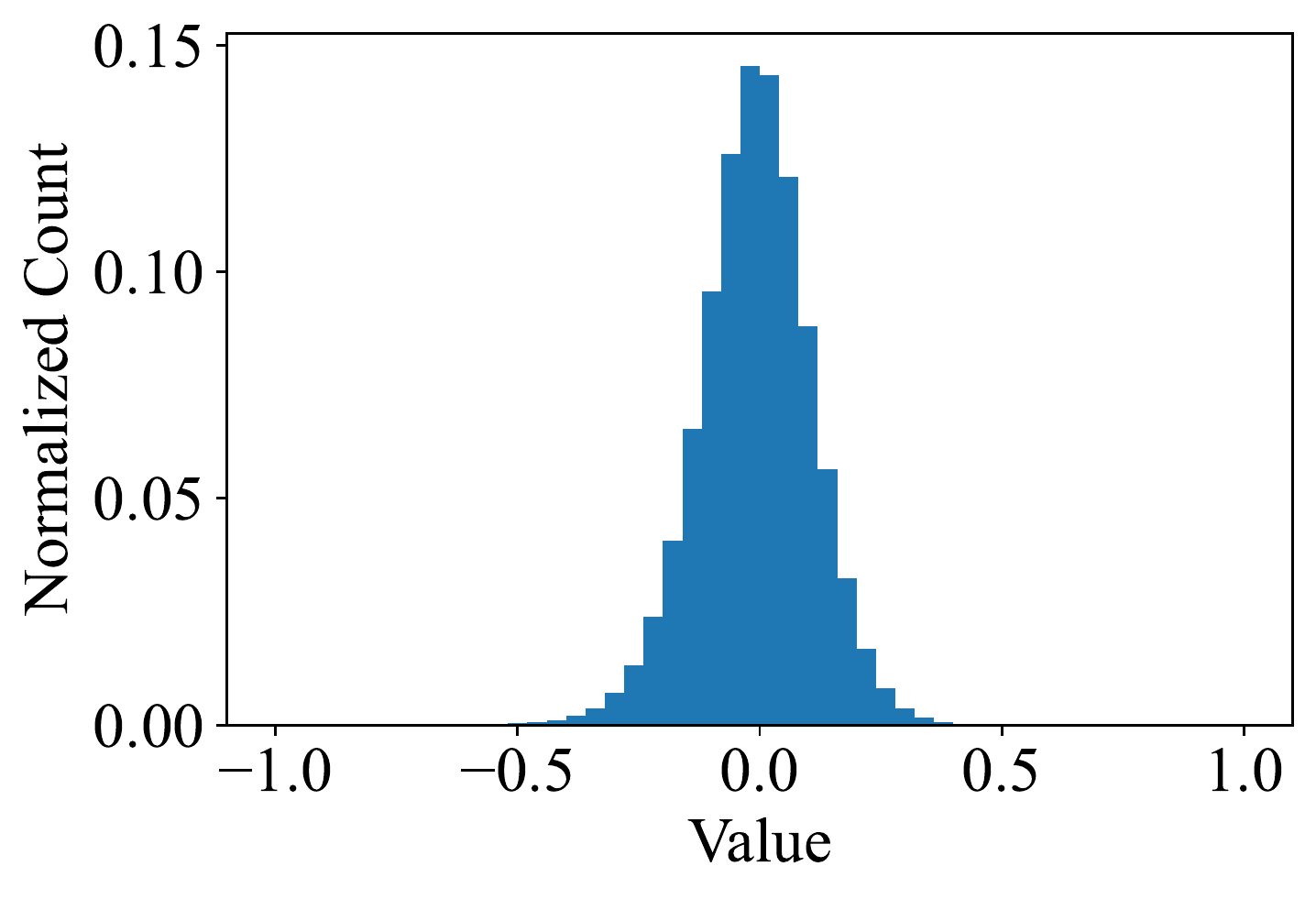}
	}
	
	\subfloat[Three secret images.]{
	    \centering
		\includegraphics[width=0.5\textwidth]{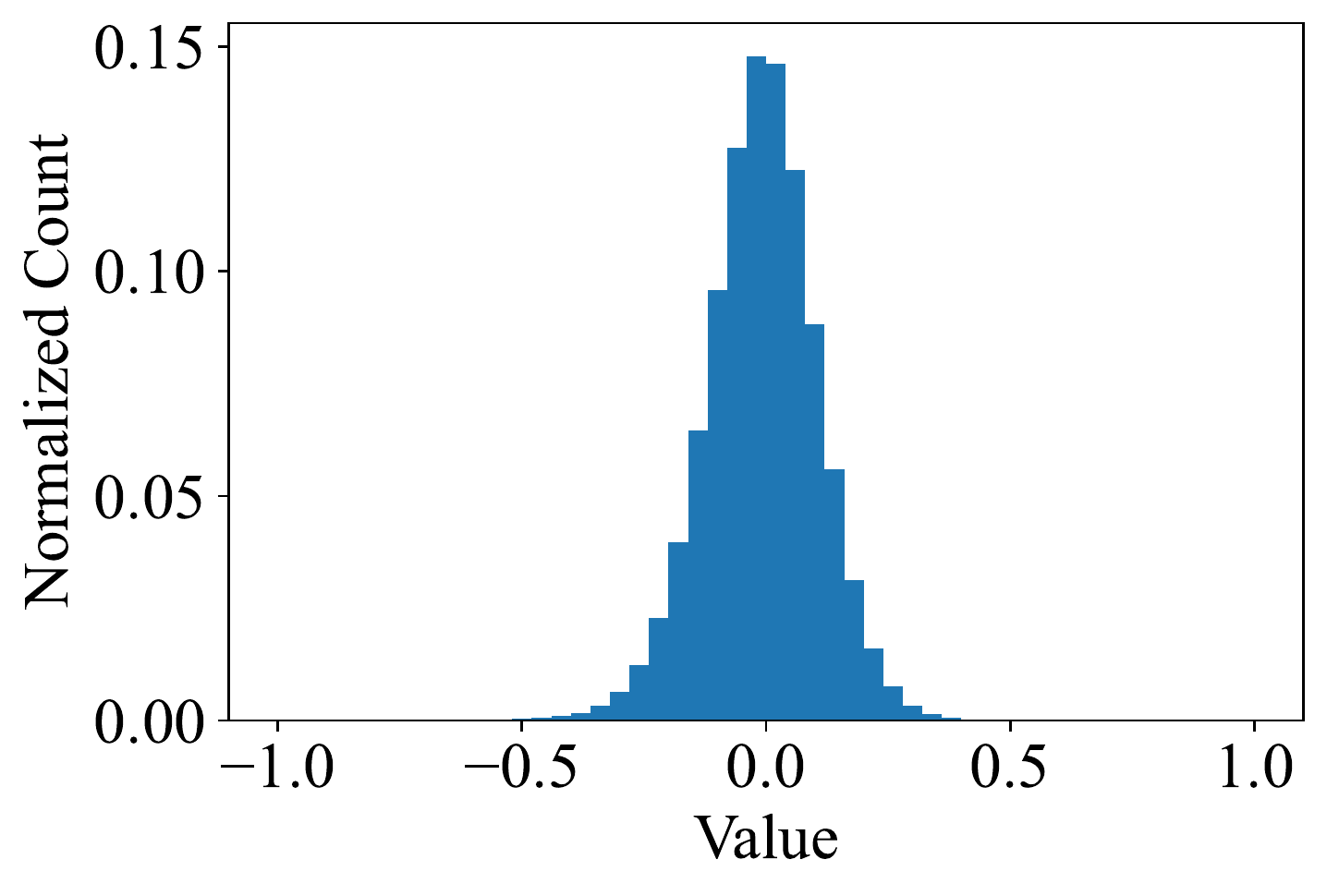}
	}
	\subfloat[Four secret images.]{
		\includegraphics[width=0.5\textwidth]{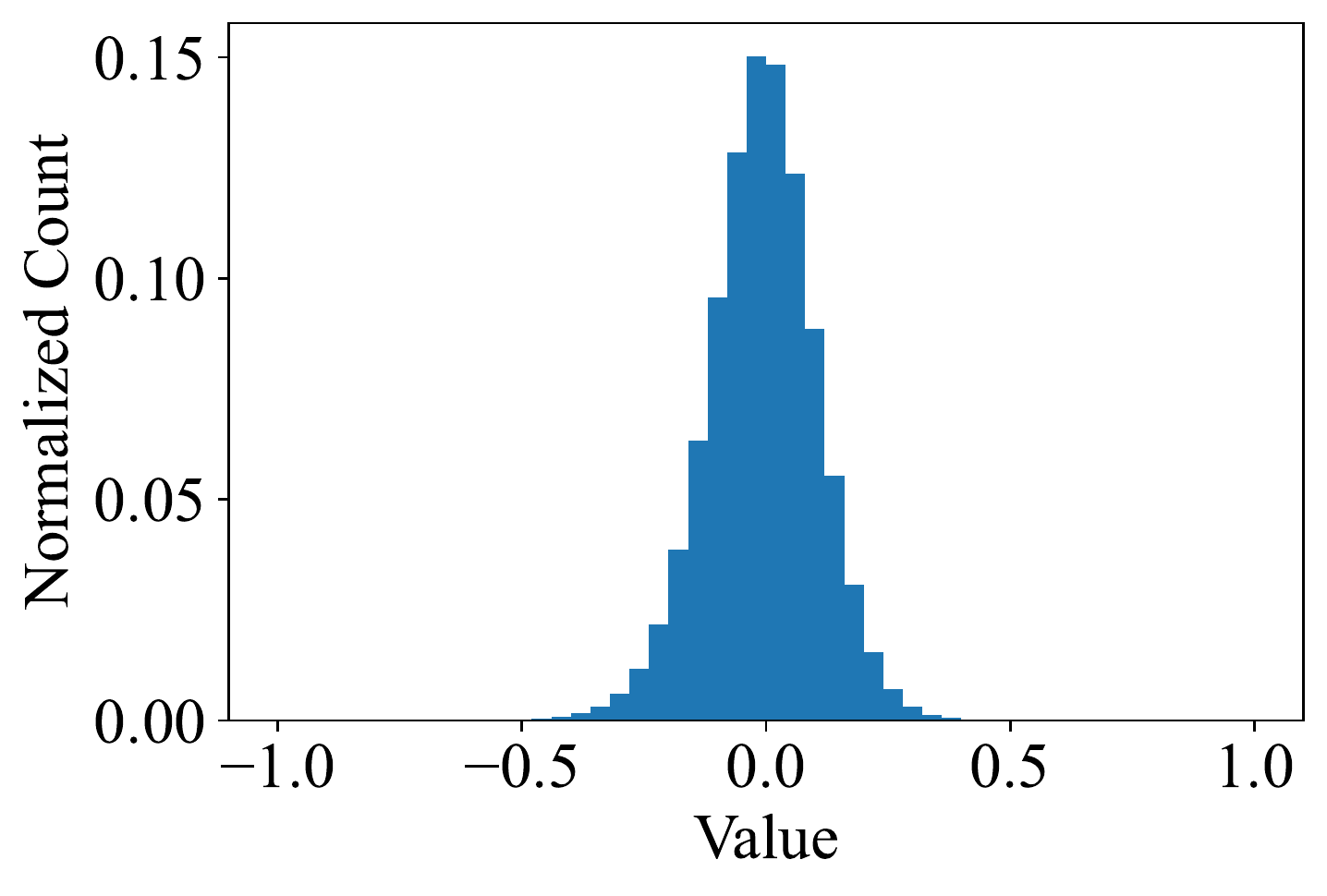}
	}

	\caption{Visual comparison of histograms of the fifth-stage weights.}
	\label{fig:s5}
\end{figure}

\begin{figure}[!h]

    \centering

    \subfloat[No secret image (original).]{
	\includegraphics[width=0.5\textwidth]{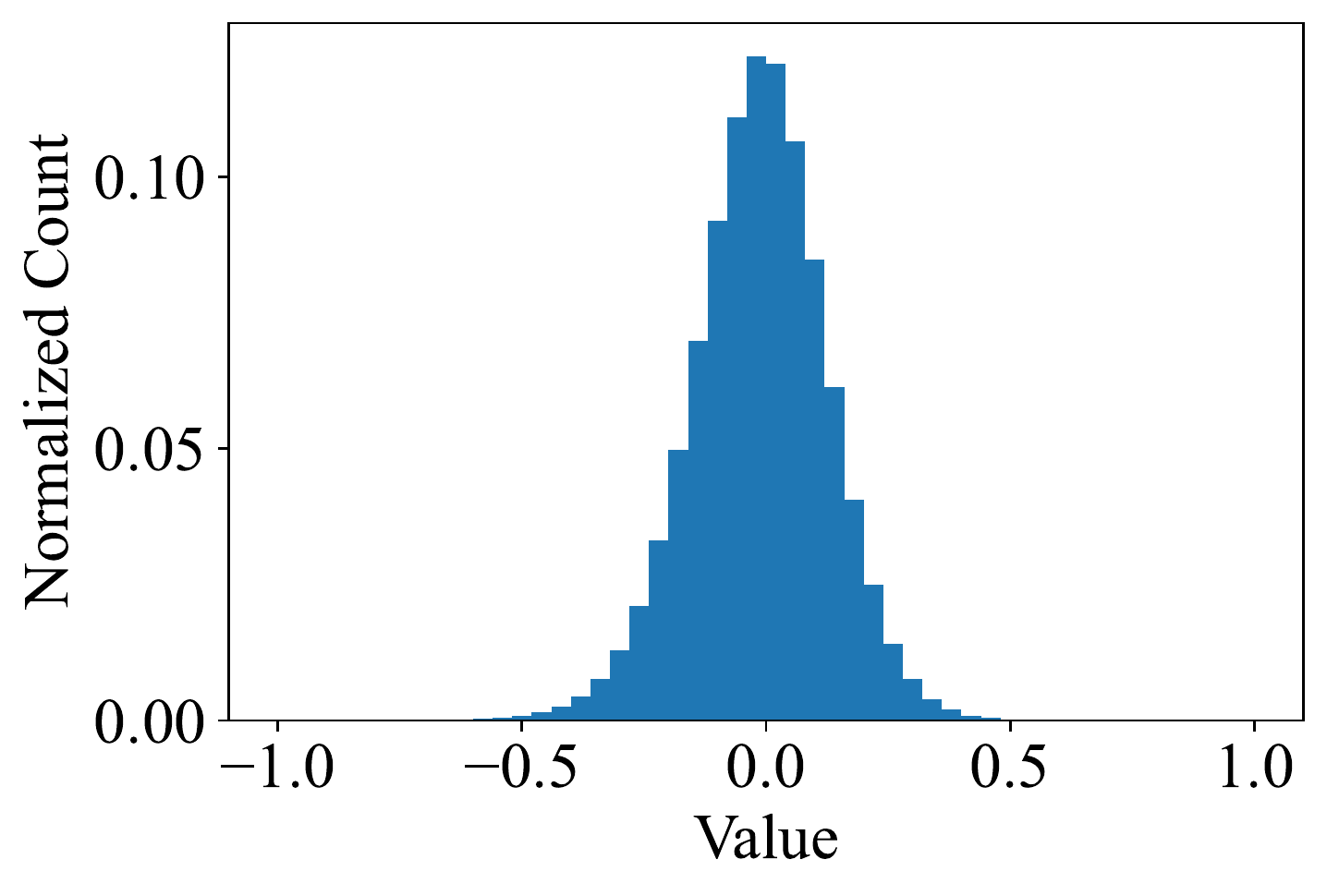}
	}
	
	\subfloat[One secret image.]{
		\includegraphics[width=0.5\textwidth]{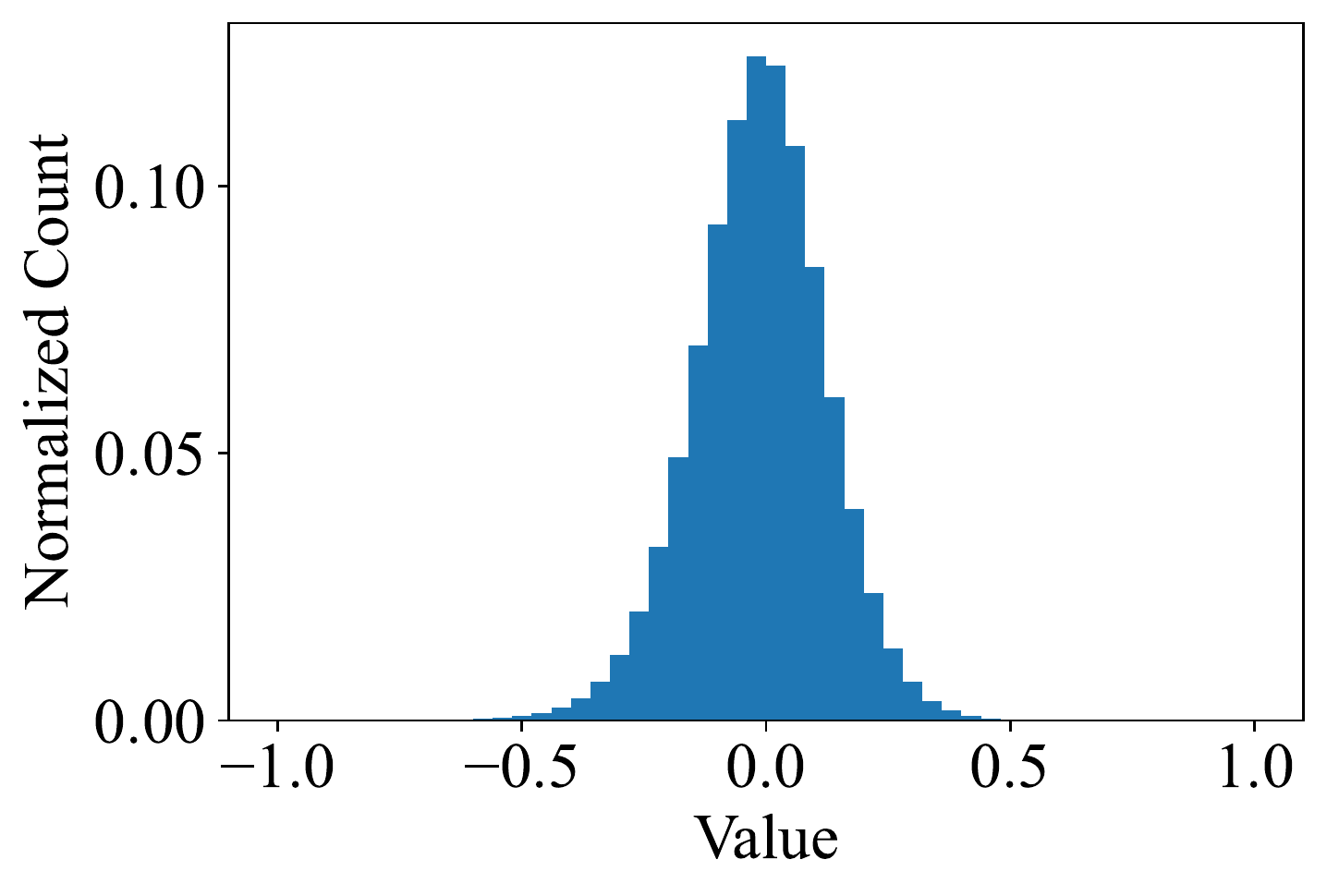}
	}
	\subfloat[Two secret images.]{
		\includegraphics[width=0.5\textwidth]{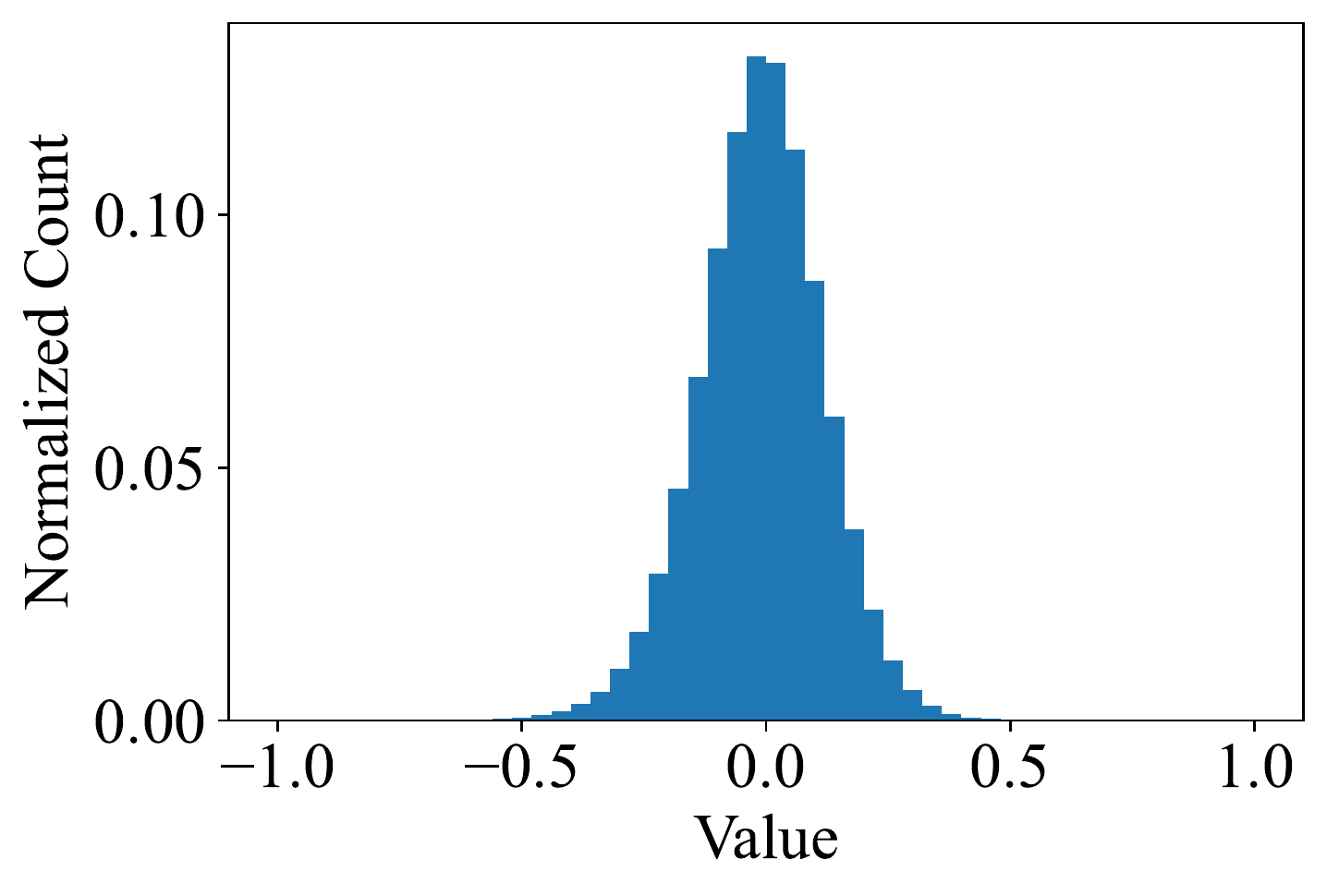}
	}
	
	\subfloat[Three secret images.]{
	    \centering
		\includegraphics[width=0.5\textwidth]{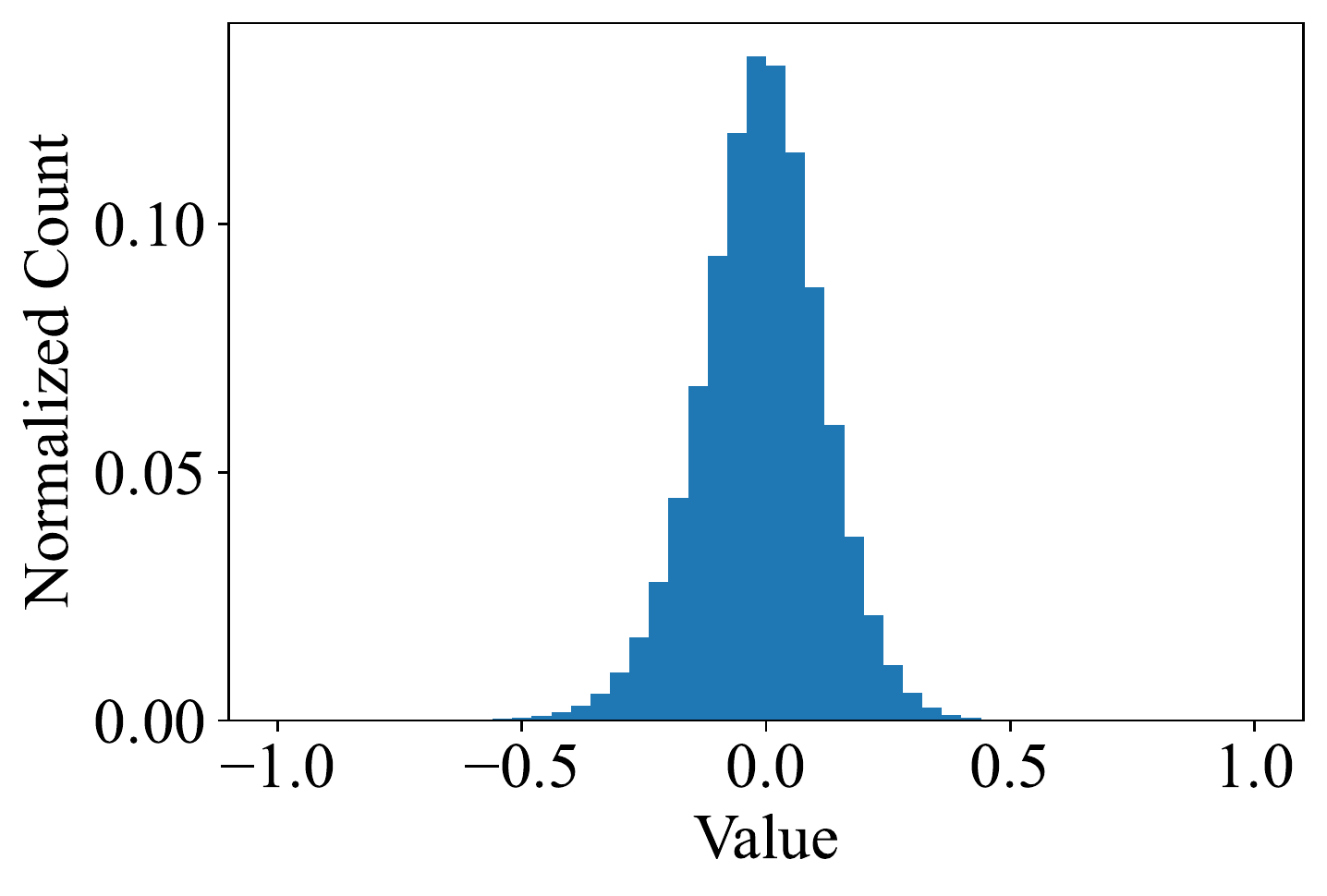}
	}
	\subfloat[Four secret images.]{
		\includegraphics[width=0.5\textwidth]{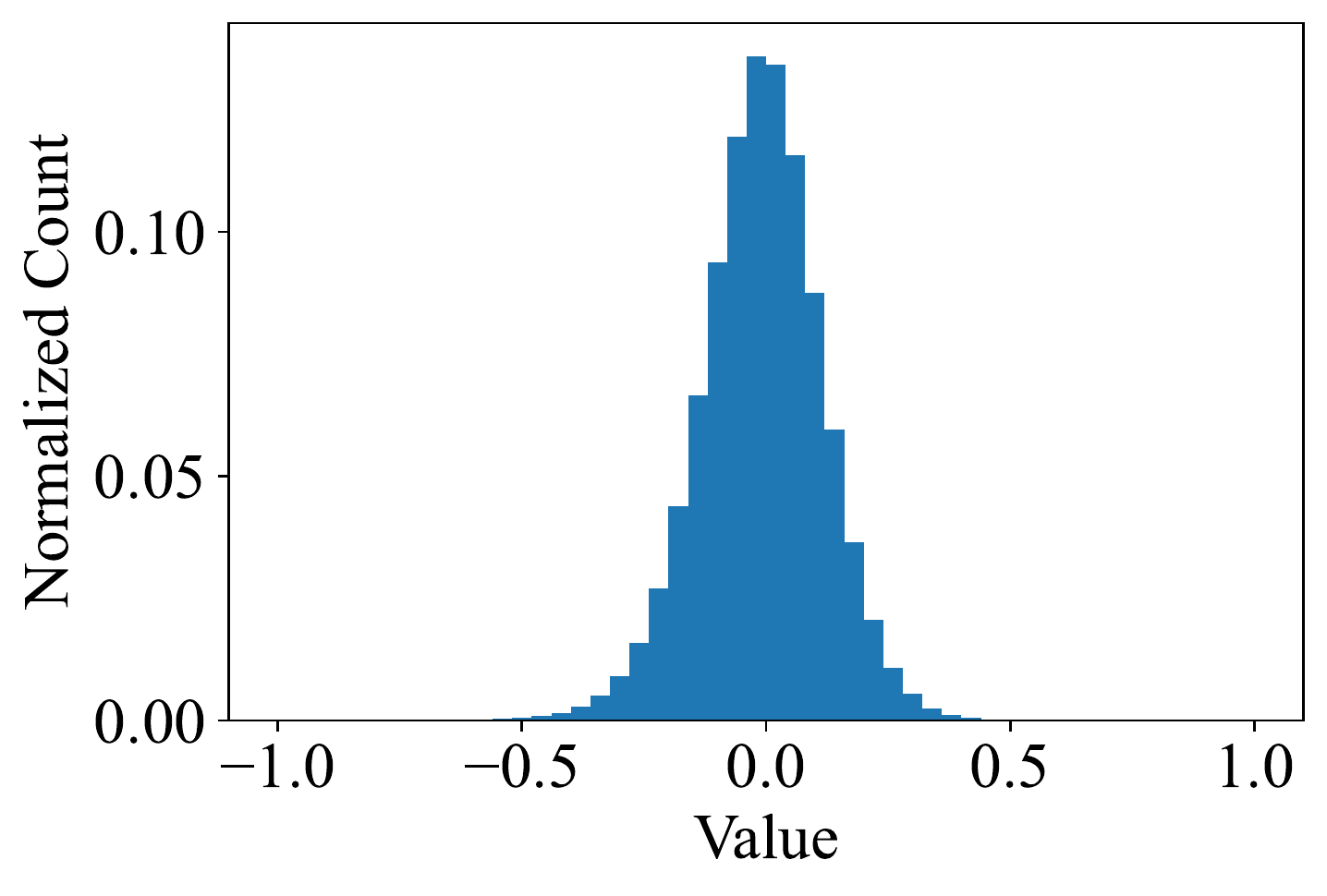}
	}

	\caption{Visual comparison of histograms of the sixth-stage weights.}
	\label{fig:s6}
\end{figure}

In addition to total weight distribution, the comparison of per-stage weight distribution is also provided. Here, one ``stage'' refers to a convolution block of the SinGAN generator, consisting of three convolution layers, as shown in Fig~\ref{fig:arc}. Per-stage comparison gives us a finer view of the weight distributions of the original and stego SinGAN generators. As our final generator has six stages (\ie, six convolution blocks), we show six sets of per-stage weight distribution comparison results in Figs.~\ref{fig:s1} to~\ref{fig:s6}. Here, the calculation of normalized count is with respect to one stage of weights across $200$ SinGANs. Again, no visually noticeable differences can be observed.

\end{document}